\documentclass[aps,floatfix]{revtex4}

\usepackage{amssymb}
\usepackage{url}
\usepackage{graphics}
\usepackage{graphicx}
\newcommand{\s}{\sum}
\newcommand{\rro}{\right)}
\newcommand{\lro}{\left( }

\begin{document}
   
\title{What's in a crowd?\\
Analysis of face-to-face behavioral networks}

\author{Lorenzo Isella}
\affiliation{Complex Networks and Systems Group, Institute for
  Scientific Interchange (ISI) Foundation, Turin, Italy}
\author{Juliette Stehl\'e} 
\affiliation{Centre de Physique Th\'eorique, CNRS UMR 6207, Marseille, France}
\author{Alain Barrat}
\affiliation{Centre de Physique Th\'eorique, CNRS UMR 6207, Marseille, France}
\author{Ciro Cattuto}
\affiliation{Complex Networks and Systems Group, Institute for
  Scientific Interchange (ISI) Foundation, Turin, Italy}
\author{Jean-Fran{\c c}ois Pinton}
\affiliation{Laboratoire de Physique de l'ENS Lyon, CNRS UMR 5672, Lyon, France}
\author{Wouter~Van den Broeck}
\affiliation{Complex Networks and Systems Group, Institute for
  Scientific Interchange (ISI) Foundation, Turin, Italy}

\begin{abstract}
The availability of new data sources on human mobility is opening
new avenues for investigating the interplay of social networks,
human mobility and dynamical processes such as epidemic spreading.
Here we analyze data on the time-resolved face-to-face proximity
of individuals in large-scale real-world scenarios. We compare two
settings with very different properties, a scientific conference
and a long-running museum exhibition.
We track the behavioral networks of face-to-face proximity,
and characterize them from both a static and a dynamic point of view,
exposing differences and similarities.
We use our data to investigate the dynamics
of a susceptible-infected model for epidemic spreading that unfolds
on the dynamical networks of human proximity.
The spreading patterns are markedly different for the conference
and the museum case, and they are strongly impacted by the causal
structure of the network data. A deeper study of the spreading paths
shows that the mere knowledge of static aggregated networks
would lead to erroneous conclusions about the transmission paths
on the dynamical networks.
\end{abstract}

\maketitle

\section{Introduction}
\label{intro}
Access to large data sets on human activities and interactions has long
been limited by the difficulty and cost of gathering such
information. Recently, the ever increasing availability of digital
traces of human actions is widely enabling the representation and the
analysis of massive amounts of information on human behavior. The
representation of this information in terms of complex
networks~\cite{science,Dorogovtsev:2003,Newman:2003,Pastor:2004,Caldarelli:2007,Barrat:2008,Wasserman:1994,watts-short}
has led to many research efforts because of the naturally interlinked
nature of these new data sources.

Tracing human behavior in a variety of contexts has become possible at very
different spatial and temporal scales: from mobility of individuals inside a
city~\cite{Chowell:2003} and between cities~\cite{Montis:2007}, to mobility
and transportation in an entire country~\cite{brockmann}, all the way to
planetary-scale travel~\cite{alain-vespi,Balcan:2009}. Mobile devices such as
cell phones make it possible to investigate mobility patterns and their
predictability~\cite{Gonzalez:2008,Song:2010}. On-line interactions occurring
between individuals can be monitored by logging instant messaging or email
exchange~\cite{Eckmann:2004,Kossinets:2006,Golder:2007,Leskovec:2008,Makse:2009,Amaral:2009}.
Recent technological advances further support mining real-world interactions
by means of mobile devices and wearable sensors, opening up new avenues for
gathering data on human and social interactions. Bluetooth and Wifi
technologies give access to proximity
patterns~\cite{Hui:2005,Eagle:2006,Kostakos,Pentland:2008,persistence}, and
even face-to-face presence can be resolved with high spatial and temporal
resolution~\cite{Sociopatterns,Cattuto:2010,alani,percol}. The combination of
these technological 
advances and of heterogeneous data sources allow researchers to gather
longitudinal data that have been traditionally scarce in social network
analysis~\cite{Padgett:1993,Lubbers:2010}. A dynamical perspective on
interaction networks paves the way to investigating interesting problems such
as the interplay of the network dynamics with dynamical processes taking place
on these networks.

In this paper, we capitalize on recent efforts~
\cite{Sociopatterns,Cattuto:2010,alani,percol} that made possible to
mine behavioral networks of face-to-face interactions between
individuals, in a variety of real-world settings and in a
time-resolved fashion. We present an in-depth analysis of the data we
collected at two widely different events. The first event was the
INFECTIOUS exhibition~\cite{infectious} held at the Science Gallery in
Dublin, Ireland, from April $17^{th}$ to July $17^{th}$, 2009. The
second event was the ACM Hypertext 2009 conference~\cite{ht2009}
hosted by the Institute for Scientific Interchange Foundation in
Turin, Italy, from June $29^{th}$ to July $1^{st}$, 2009. In the
following, we will refer to these events as SG and HT09,
respectively. Intuitively, interactions among conference participants
differ from interactions among museum visitors, and the concerned
individuals have very different goals in both settings. The study of
the corresponding networks of proximity and interactions, both static
and dynamic, reveals indeed strong differences but also interesting
similarities.
We take advantage of the availability of time-resolved data to show how
dynamical processes that can unfold on the close proximity network --- such as
the propagation of a piece of information or the spreading of an infectious
agent --- unfold in very different ways in the investigated settings. In the
epidemiological literature, traditionally, processes of this kind have been
studied using either aggregated data or under assumptions of stationarity for
the interaction networks: here we leverage the time-resolved nature of our
data to assess the role of network dynamics on the outcome of spreading
processes. At a more fundamental level, simulating simple spreading processes
over the recorded interaction networks allows us to expose several properties of
their dynamical structure as well as to probe their causal structure.

The paper is organized as follows: first, we briefly describe the data
collection platform and our data sets in Section~\ref{data}; in
Section~\ref{static-properties} we discuss the salient features of
the networks of interactions aggregated on time windows of one day. These
networks are static objects, carrying only information about the cumulative
time that -- daily -- each pair of individuals has spent in face-to-face
proximity. Section~\ref{dynamic-properties} analyzes the dynamical properties
of face-to-face interactions between conference participants and museum
visitors. Section~\ref{resilience} further characterizes the aggregated
network structures by investigating the effect of incremental link removal.
Finally, Section~\ref{information-diffusion} investigates the role played by
causality in information spreading along the proximity network, and
Section~\ref{conclusions} concludes the paper and defines a number of open
questions.

%
%
\section{Data}
\label{data}
The data collection infrastructure uses active Radio-Frequency Identification
Devices (RFID) embedded in conference badges to mine face-to-face proximity 
relations of persons wearing the badges. RFID devices exchange
ultra-low power radio packets in a peer-to-peer fashion, as described in
Refs.~\cite{Sociopatterns,Cattuto:2010,alani,percol}. Exchange of radio packets
between badges is only possible when two persons are at close range ($1$ to $1.5$m) and facing each other, as the human body acts as a RF shield at the
carrier frequency used for communication. The operating parameters of the
devices are programmed so that the face-to-face proximity of two individuals
wearing the RFID tags can be assessed with a probability in excess of $99\%$
over an interval of $20$ seconds, which is a fine enough time scale to resolve
human mobility and proximity at social gatherings. False positives are
exceedingly unlikely, as the ultra-low power radio packets used
for proximity sensing cannot propagate farther than $1.5$-$2$m,
and a sustained excess of packets is needed in order to signal
a proximity event. When a relation of
face-to-face proximity (or ``contact'', as we will refer to it in the
following) is detected, the RFID devices report this information to receivers
installed in the environment (RFID readers). The readers are connected to a
central computer system by means of a Local Area Network. Once a contact has
been established, it is considered ongoing as long as the involved devices
continue to exchange at least one radio packet for every subsequent interval
of $20$ seconds. Conversely, a contact is considered terminated if an interval
of $20$ seconds elapses with no packets exchanged. For a detailed description of
the sensing platform and some of its deployments, see
Refs.\cite{Sociopatterns,Cattuto:2010,alani,percol}.

The deployments at the Science Gallery in Dublin~\cite{infectious} and at the
HT09 conference in Turin~\cite{ht2009} involved vastly different numbers of
individuals and stretched along different time scales. The former lasted for
about three months and recorded the interactions of more than
$14\mathpunct{,}000$ visitors (more than $230\mathpunct{,}000$ face-to-face
contacts recorded), whereas the latter took place over the course of three
days and involved about $100$ conference participants (about
$10\mathpunct{,}000$ contacts). Behaviors are also very different: in a
museum, visitors typically spend a limited amount of time on site,
well below the maximum duration permitted by the museum opening hours,
they are not likely to return, and they follow a rather pre-defined path, 
touching different locations that host the exhibits. In a conference setting,
on the other hand, most attendees stay on-site for the entire duration of
the conference (a few days), and move at will between different areas
such as conference room, areas for coffee breaks and so on.
The coverage of the community was different in both settings. At the Science
Gallery, visitors were equipped with a RFID tag upon entering the venue, as
part of an interactive exhibit, and therefore almost the totality of them were
tracked. On the other hand, at HT09, about 75\% of the participants
volunteered to being tracked. This sampling may introduce some biases in the
results. Sampling issues are also commonly encountered in the study of static
complex networks
\cite{Willinger:2002,Petermann:2004,Clauset:2005,Dallasta:2005}.
Reference~\cite{Cattuto:2010} has shown that for a broad variety of real-world
deployments of the RFID proximity-sensing platform used in this study, the
behavior of the statistical distributions of quantities such as contact
durations is not altered by unbiased sampling of individuals. On the other
hand, we cannot completely rule out that a systematic bias is introduced by
the selection of volunteers, if volunteers and non-volunteers have different
behavioral patterns. Accurately checking this point would require monitoring
an independent data source for face-to-face contacts, and because of
scalability issues this would be feasible only for small control groups.
Issues regarding the effect of missing data and incomplete sampling on the
properties of dynamical processes unfolding on the networks also deserve
attention and will be the subject of future investigations.

%
%
\section{The static interaction network}
\label{static-properties}
We start by analyzing aggregated networks of interaction obtained by
aggregating the raw proximity data over one day. This aggregation yields a
social graph where nodes represent individuals, and an edge is drawn between
two nodes if at least one contact was detected between those nodes during the
interval of aggregation. Therefore, every edge is naturally weighted by the
total duration of the contact events that occurred between the tags involved,
i.e., by the total time during which the corresponding individuals have been
in face-to-face proximity.

The choice of daily time windows seems quite natural in our settings. It would
represent, for instance, a typical time scale for a description of articulated
social networks based on surveys, in which each participant would (ideally)
declare who s/he has encountered during the course of the day. Such a choice
for the duration of the time-window, albeit natural, is by no means
unique~\cite{persistence}. For instance, it is possible to aggregate the data
over longer periods of time (weeks or months) to investigate the stationarity
of the collected data~\cite{Cattuto:2010}. Shorter aggregation times of the
order of a few minutes are also useful, for instance, to resolve circadian
activity patterns at the venue under investigation.

\begin{figure}
\includegraphics[width=0.45\columnwidth]{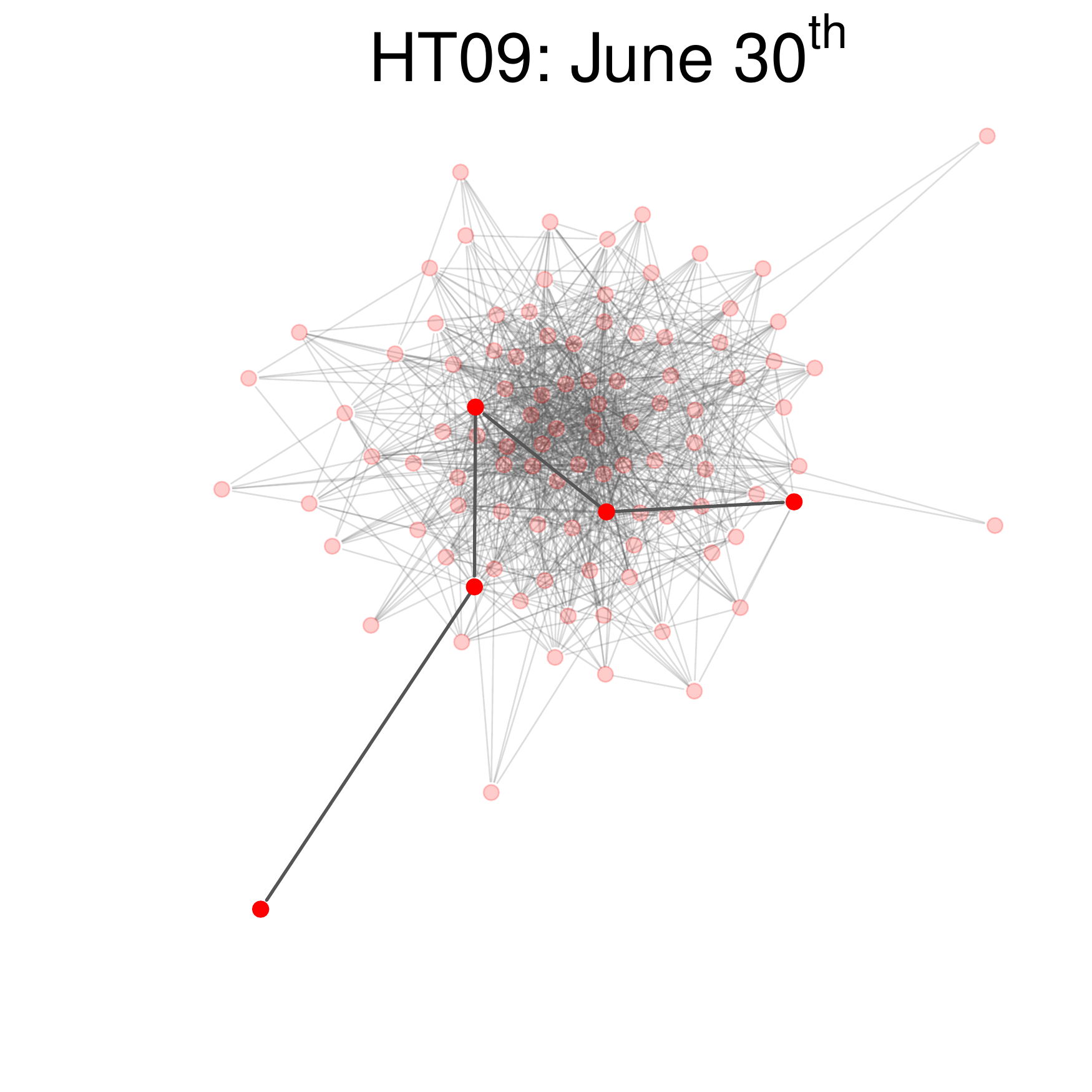}
\includegraphics[width=0.45\columnwidth]{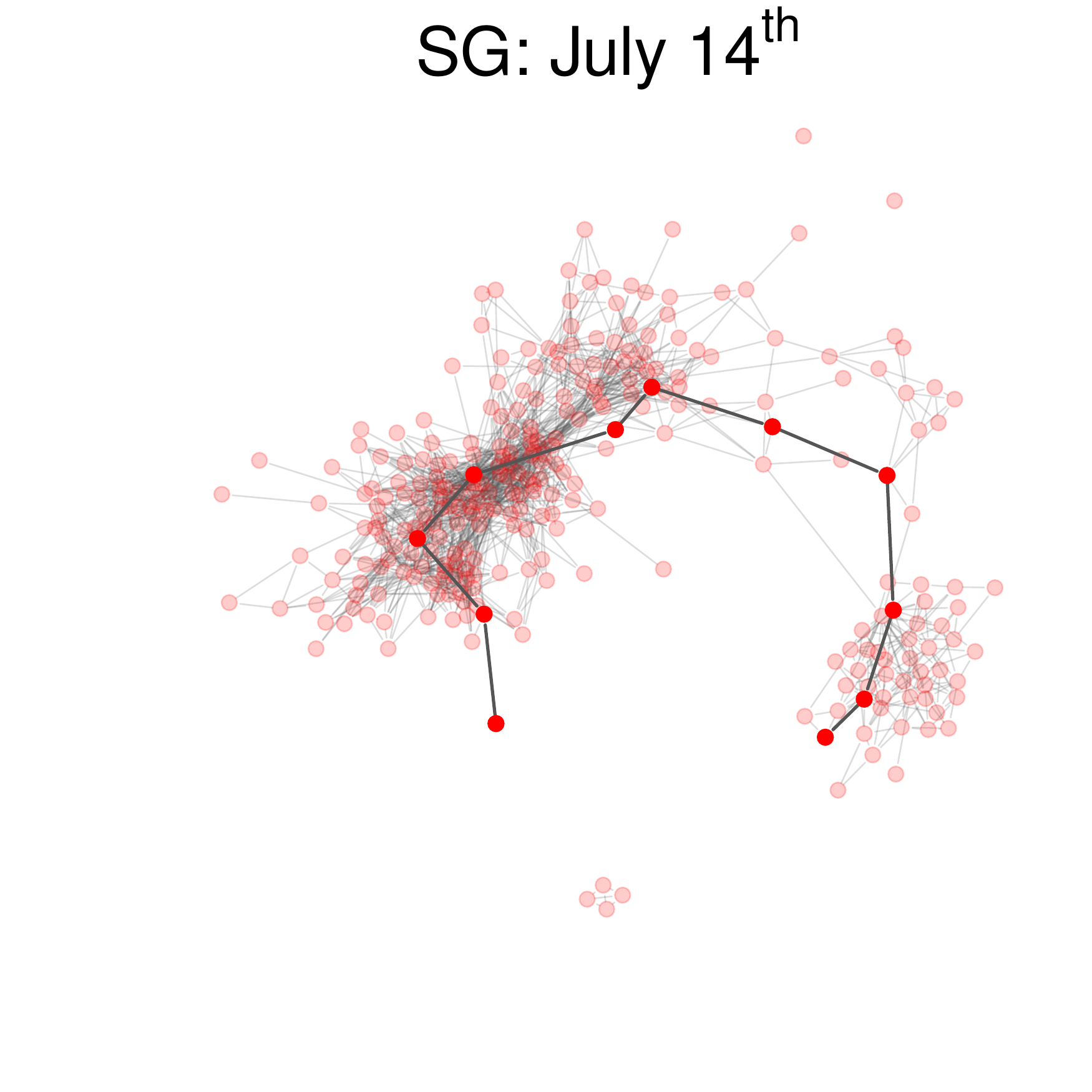}
\includegraphics[width=0.45\columnwidth]{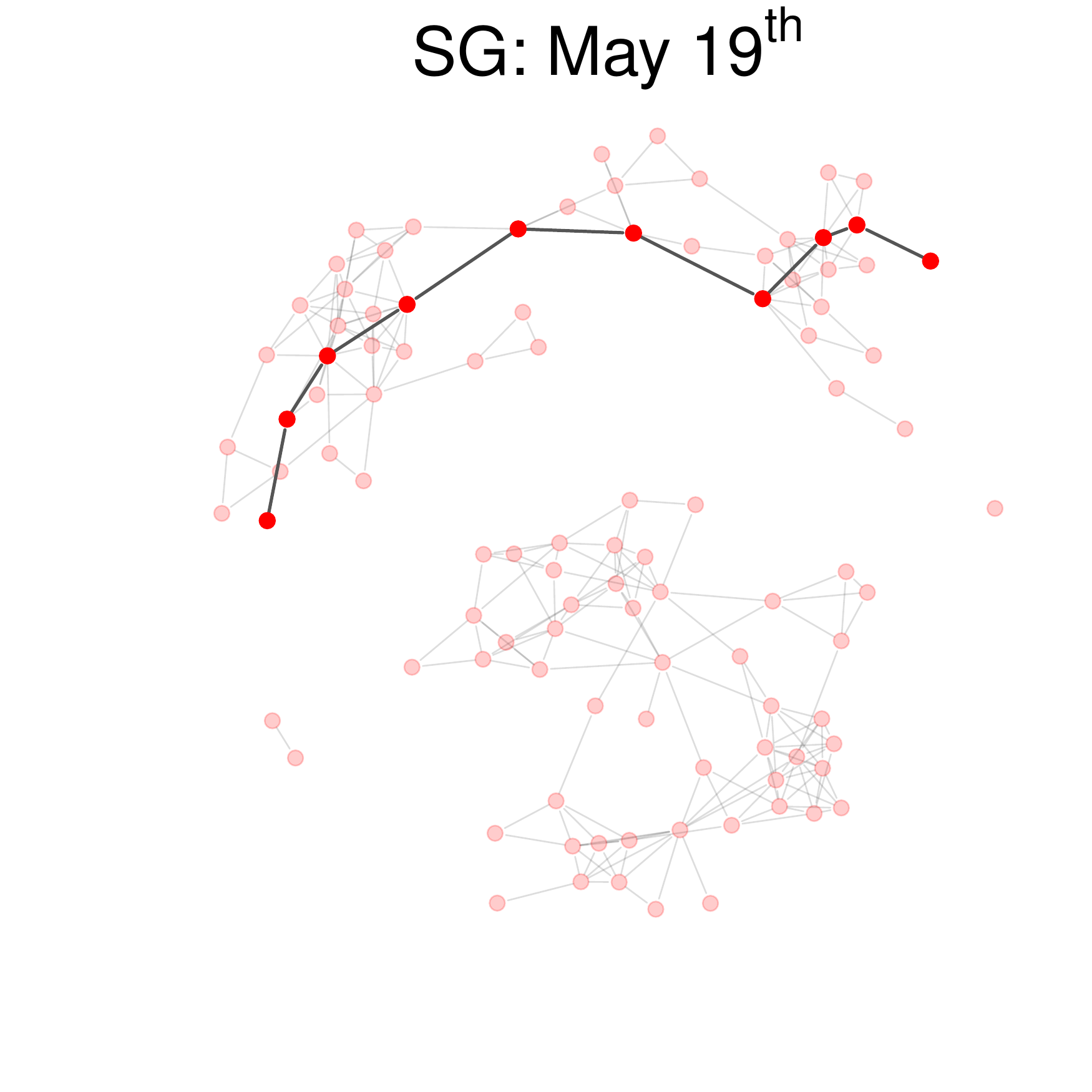}
\includegraphics[width=0.45\columnwidth]{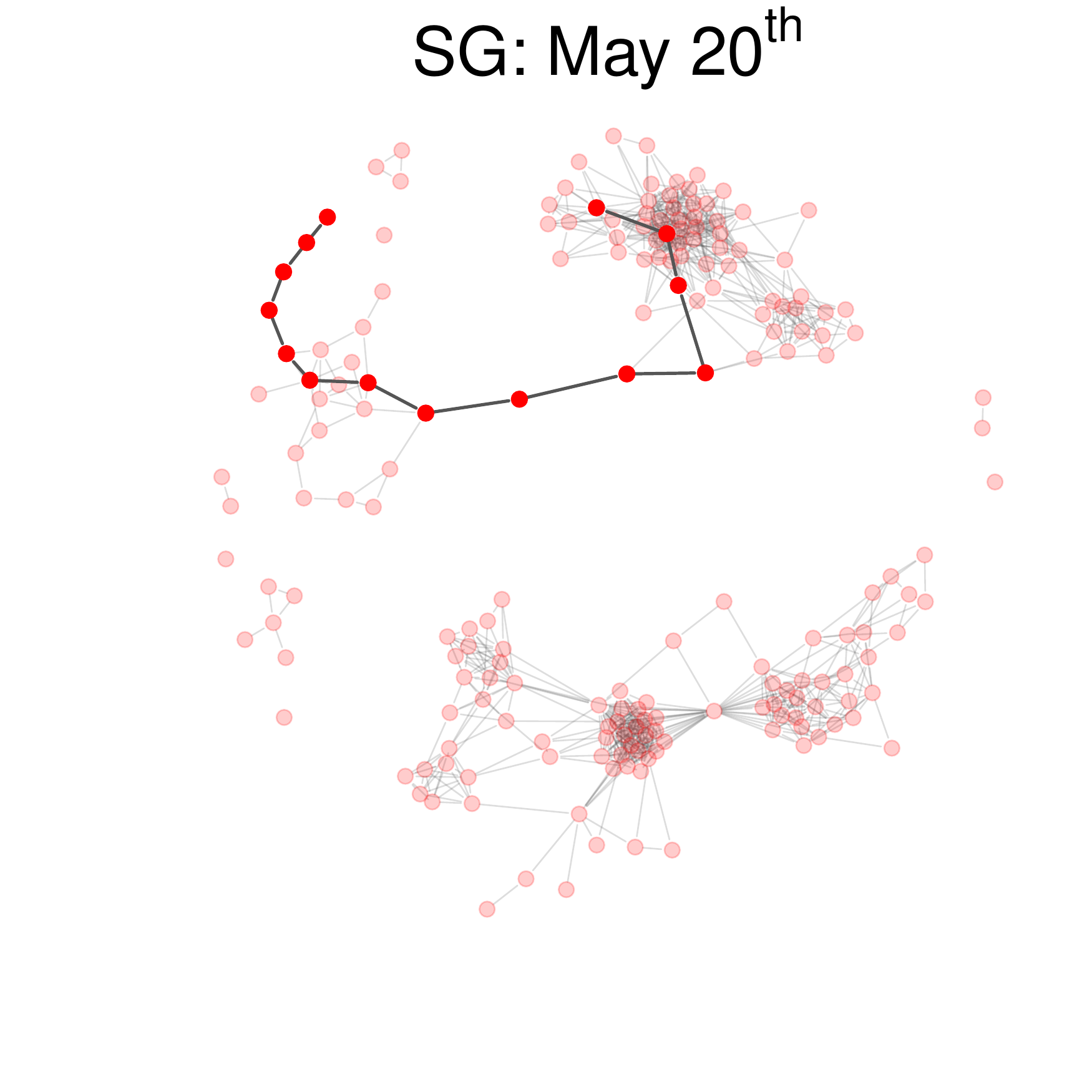}
\caption{Daily aggregated networks in the HT09 and SG deployments.
  Nodes represent individuals and edges are drawn between nodes if at
  least one contact event was detected during the aggregation
  interval. Clockwise from top: aggregated network for one day of the
  HT09 conference, and for three representative days at the SG
  deployment. In each case, the network diameter is highlighted.  All
  the network visualizations in this study were produced using the
  igraph library \cite{igraph}.  }\label{aggregated-networks}
\end{figure}

\begin{figure}
\includegraphics[width=0.45\columnwidth]{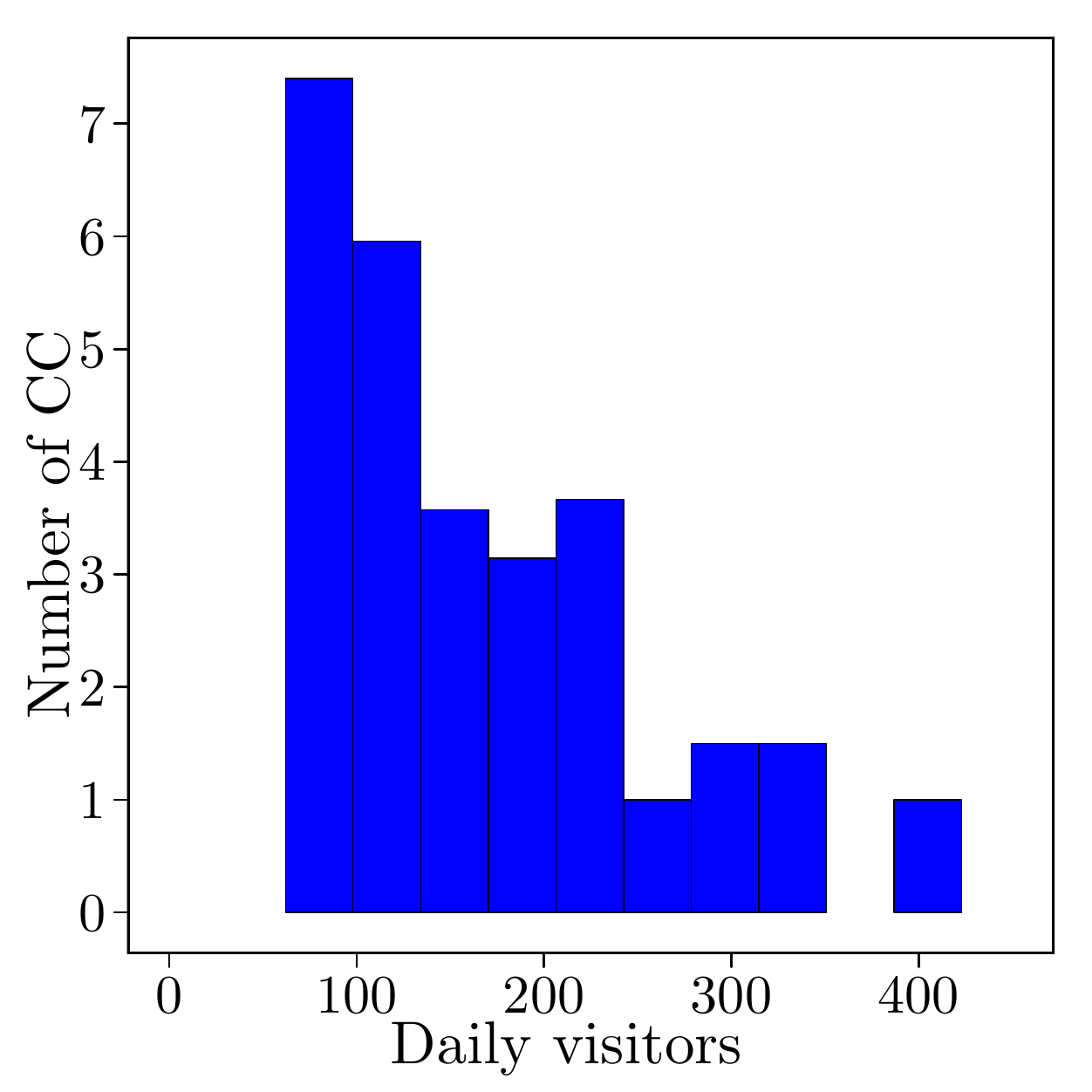}
\caption{Number of connected components (CCs) in the daily aggregated
  networks of the SG deployment as a function of the number
of visitors.}\label{nclusters}
\end{figure}

Figure~\ref{aggregated-networks} displays the aggregated contact
networks for June 30$^{\rm th}$ at the HT09 conference (top left), and
for three representative days for the SG museum deployment.
Despite the large variation in the number of daily museum
visitors, ranging from about $60$ to $400$, the
chosen days illustrate many features of the SG aggregated networks,
in particular the presence of either a single or two large connected
components (CC) in the network. Days with smaller numbers of visitors
can also give rise to aggregated networks made of a larger number of
small isolated clusters. As shown in Fig.~\ref{nclusters}, depending
on the number of visitors the number of CC can in fact vary
substantially. For a large number of visitors, typically only one CC
is observed. For a low number of visitors, on the other hand, many
clusters are formed. Overall one also notices that the network
diameter (highlighted in all the plots of
Fig.~\ref{aggregated-networks}) is considerably longer for SG 
than for HT09 aggregated networks, reflecting the different behavioral
patterns in these settings.

\begin{figure}
\includegraphics[width=0.45\columnwidth]{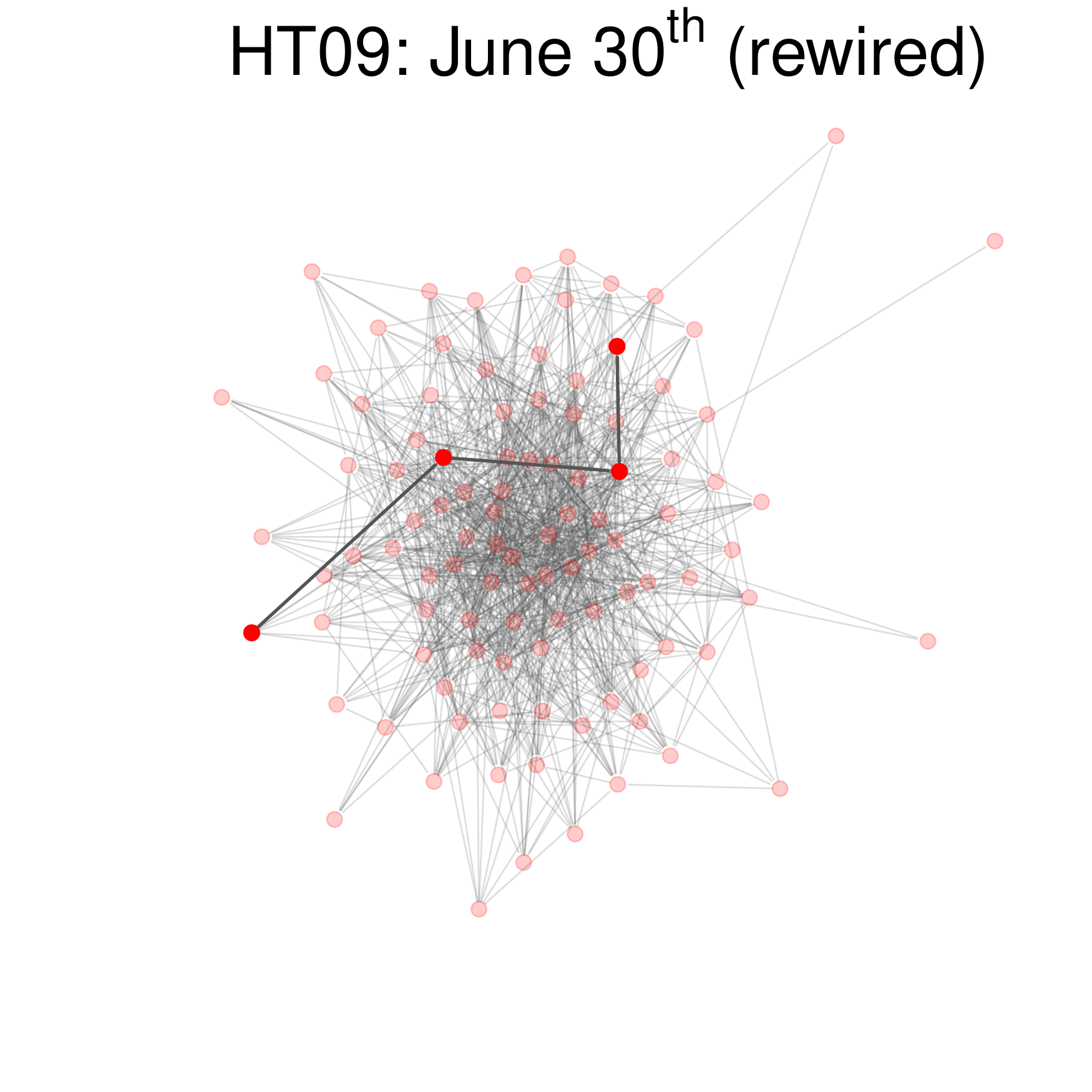}
\includegraphics[width=0.45\columnwidth]{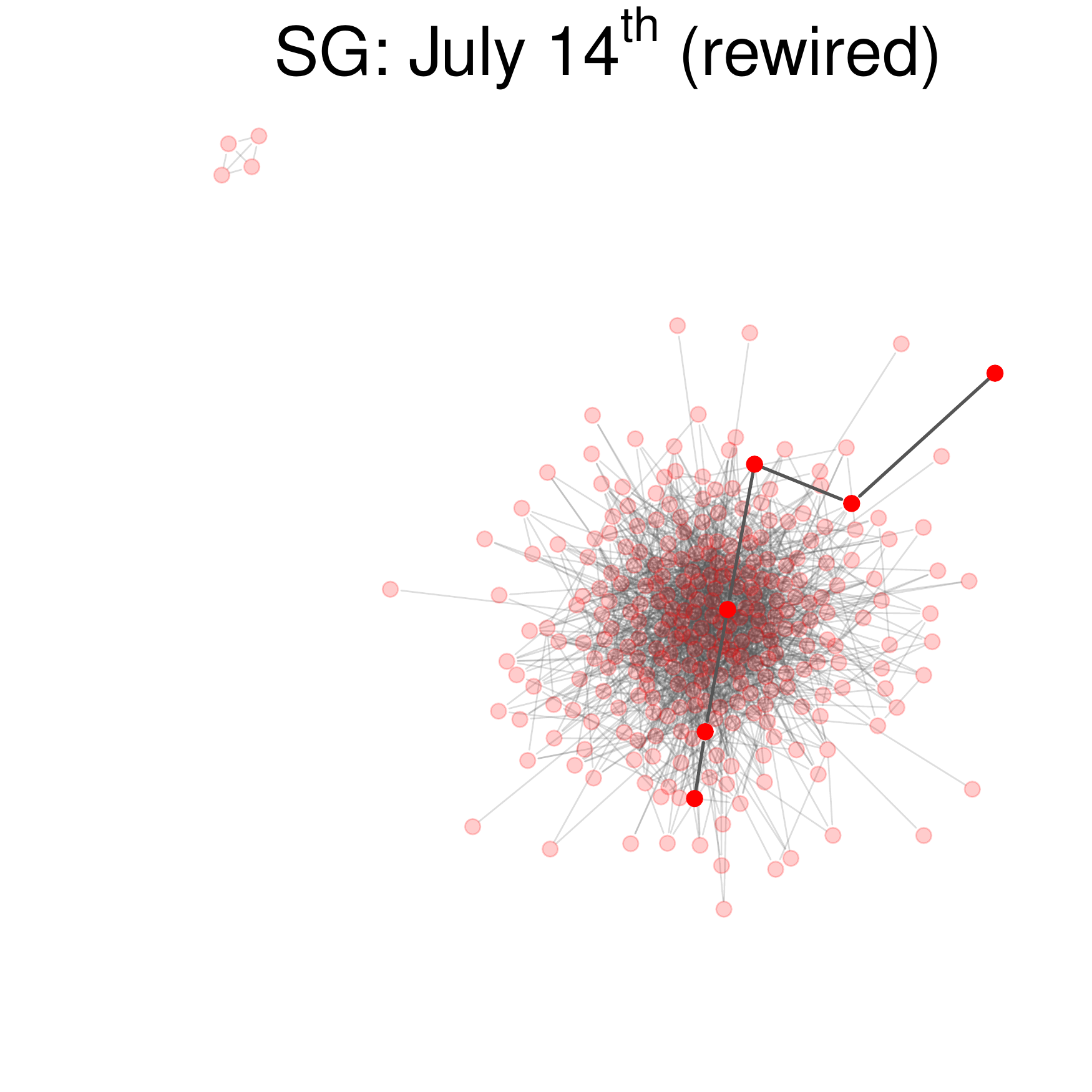}
\caption{Randomized versions of the daily aggregated networks in
  the top row of Fig.~\ref{aggregated-networks}. Left: HT09
  deployment, June 30$^{\rm th}$. Right: SG deployment, July
  14$^{\rm th}$.  The network diameters are highlighted as in
  Fig.~\ref{aggregated-networks}. In the SG case, the randomized
  network is much more ``compact'' than the original one, with a much
  shorter diameter.}\label{rewired-networks}
\end{figure}

The small-world nature --- or lack thereof --- of the aggregated
networks can be investigated statistically by introducing a proper
null model. To this end, we construct a randomized network using the
rewiring procedure described in Ref.~\cite{rewiring}. The procedure
consists in taking random pairs of links $(i,j)$ and $(l,m)$ involving
$4$ distinct nodes, and rewiring them as $(i,m)$ and $(j,l)$. This
procedure preserves the degree of each node and the degree
distribution $P(k)$, while destroying the degree correlations between
neighboring nodes, as well as any other correlations linked to node
properties. The procedure is carried out so that initially distinct
CCs do not get merged. Since the rewiring procedure cannot be
implemented for the rare CCs with less than four nodes, these small
CCs are removed from the aggregated networks before rewiring.
Figure~\ref{rewired-networks} displays a single realization of the
null model for the networks in the top row of
Fig.~\ref{aggregated-networks}. We notice that the rewired version of
the aggregated HT09 network is very similar to the original version,
whereas the null model for the aggregated network of the SG data on
July 14$^{\rm th}$ is more ``compact'' than the original network and
exhibits a much shorter diameter. Similar considerations hold for the
other aggregated networks of the SG deployment.

\begin{figure}
\includegraphics[width=0.45\columnwidth, height=0.45\columnwidth]{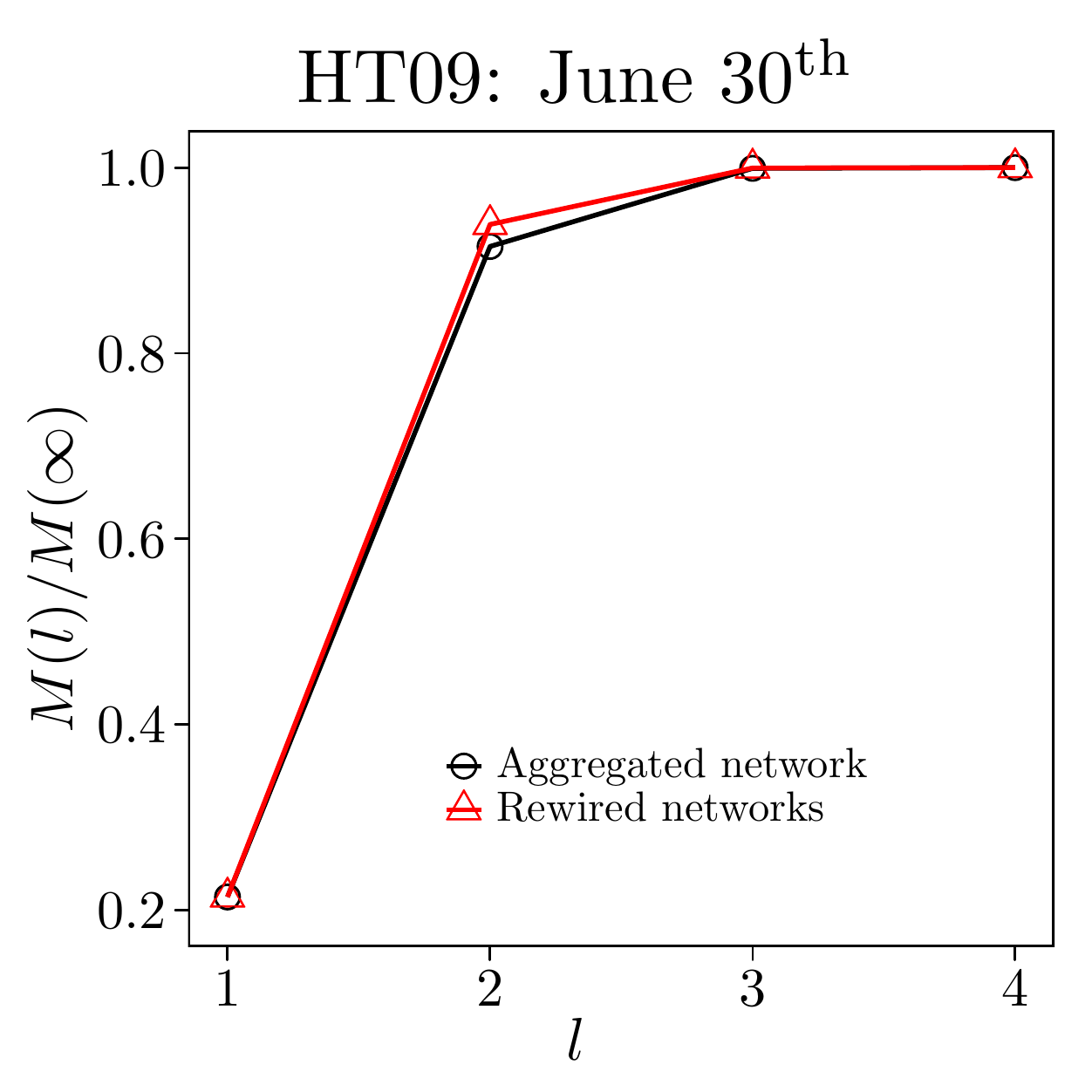}
\includegraphics[width=0.45\columnwidth, height=0.45\columnwidth]{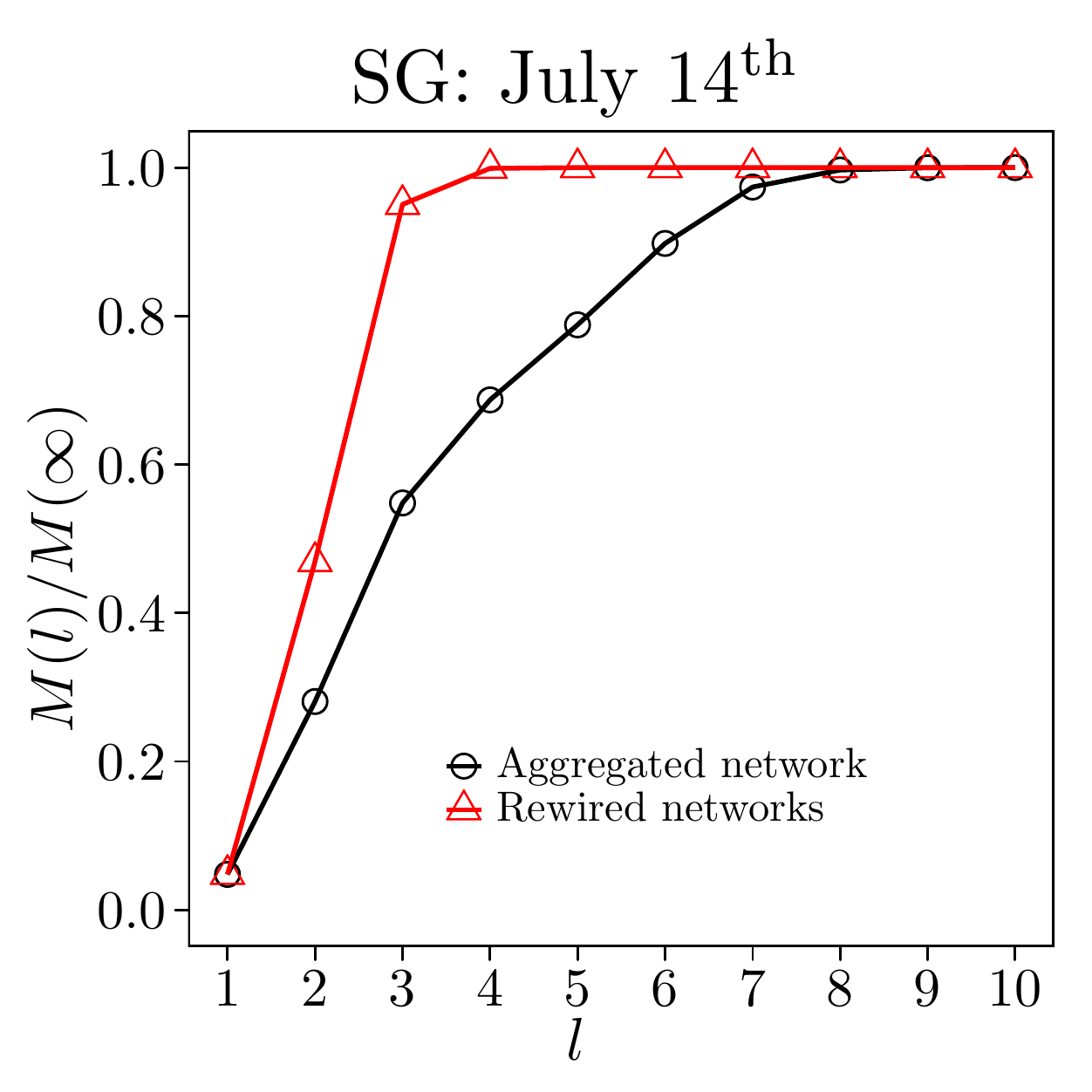}
\caption{Average number of nodes reachable from a randomly chosen node
  by making $l$ steps on the network, $M(l)$, divided by its
  saturation limit $M(\infty)$, for daily aggregated networks
  (circles) and their randomized versions (triangles). For the
  randomized case, data are averaged on $100$ realizations. Left:
  network aggregated on June 30$^{\rm th}$ for the HT09 case. Right:
  SG deployment, July 14$^{\rm th}$.  The solid lines are only guides
  for the eye.}\label{M-l-plots}
\end{figure}

More quantitatively, we measure the mean number of nodes one can reach
from a randomly chosen node by making $l$ steps on the network, a
quantity hereafter called $M(l)$. For a network consisting of a single
connected component, the definition of $M(l)$ implies that
\begin{equation}
\label{M-l-single-component} M(1)=\langle k \rangle +1\;\;\;\; {\rm
  and} \;\;\;\; M(\infty)=N \, , 
\end{equation} 
where $\langle k \rangle$ is the average node degree, $N$ is the total
number of nodes in the network and $M(\infty)$ the saturation value of
$M$ on the network. The saturation value $M(\infty)$ is reached when
$l$ is equal to the length of the network diameter, and
may vary for different realizations of the random networks.  For a
network consisting of several CCs one has to take into account
the probability $N_{i}/N$ that the chosen node belongs to a given CC,
where $N_{i}$ is the number of nodes in the $i$-th CC.
As a consequence, Eq.~(\ref{M-l-single-component}) generalizes to
\begin{equation}\label{M-l-multi-component} 
M(1)=\frac{1}{N}\s_{i}{N_{i}\lro\langle k
  \rangle_{i} +1\rro}\;\;\;\; {\rm and} \;\;\;\;
M(\infty)=\frac{1}{N}\s_{i}N_{i}^{2} \, , 
\end{equation} 
where $\langle k \rangle_{i}$ is the average node degree on the $i$-th
CC.  This ensures that the quantity $M(l)/M(\infty)$, regardless on
the number of CC, assumes the same value when $l=1$, and saturates to
unity for both the aggregated and rewired network.
Figure~\ref{M-l-plots} displays $M(l)/M(\infty)$ for the aggregated
networks on the top row of Fig.~\ref{aggregated-networks}, as well as
its value averaged on $100$ randomized networks (the average value of
$M(l)$ converges rapidly already when calculated on a few tens of
randomized networks). We notice the striking similarity between the
results for the HT09 original and randomized networks, where about
$90{\%}$ of the individuals lie, in both cases, within two degrees of
separation. In the SG case, conversely, the same $90{\%}$ percentage
is reached with six degrees of separation for the original network,
but with only three degrees of separation on the corresponding
randomized networks.  The same calculation, performed on other
aggregated SG networks, yields qualitatively similar results, always
exposing a dramatic difference from the null model. 

\begin{figure}
\includegraphics[width=0.45\columnwidth, height=0.45\columnwidth]{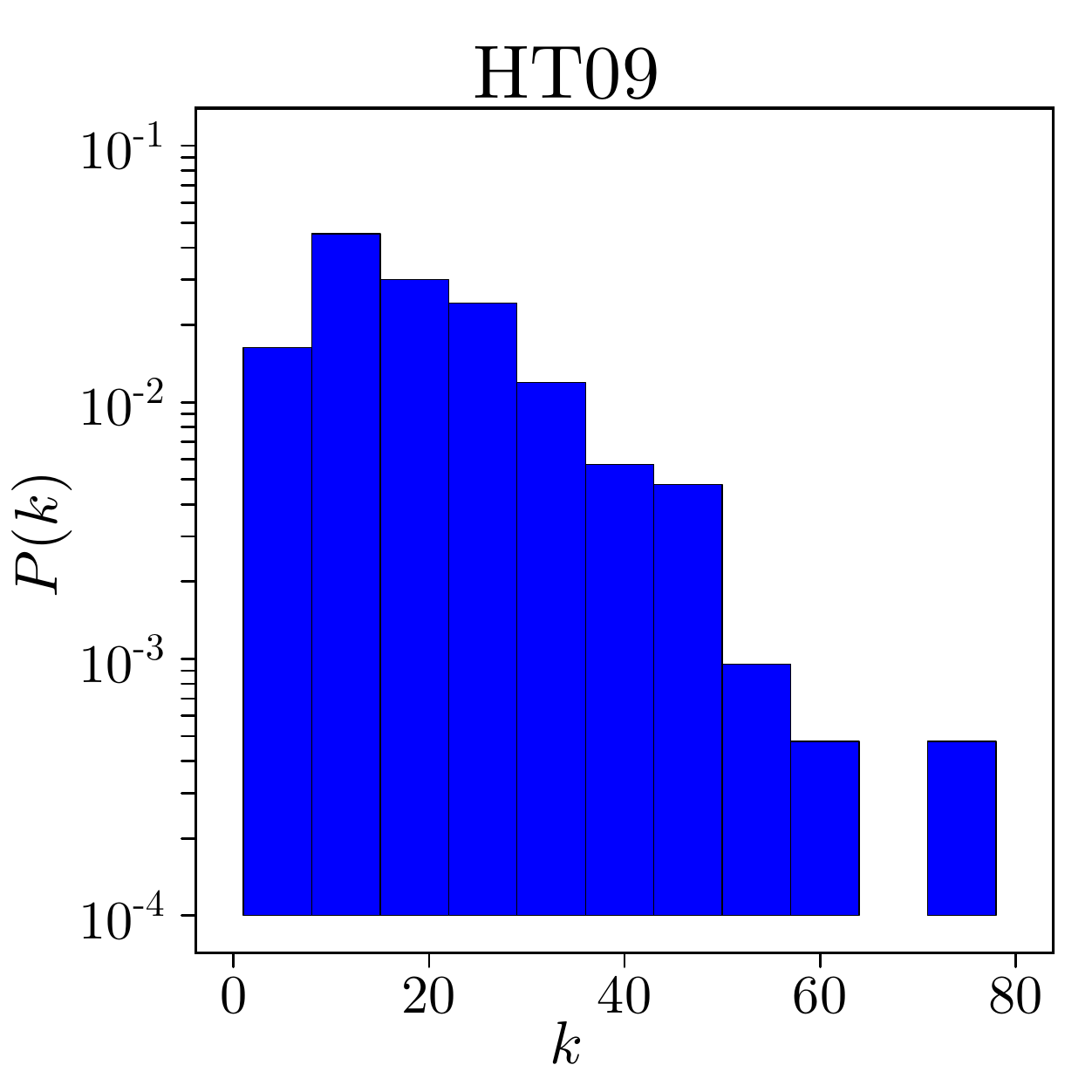}
\includegraphics[width=0.45\columnwidth,height=0.45\columnwidth]{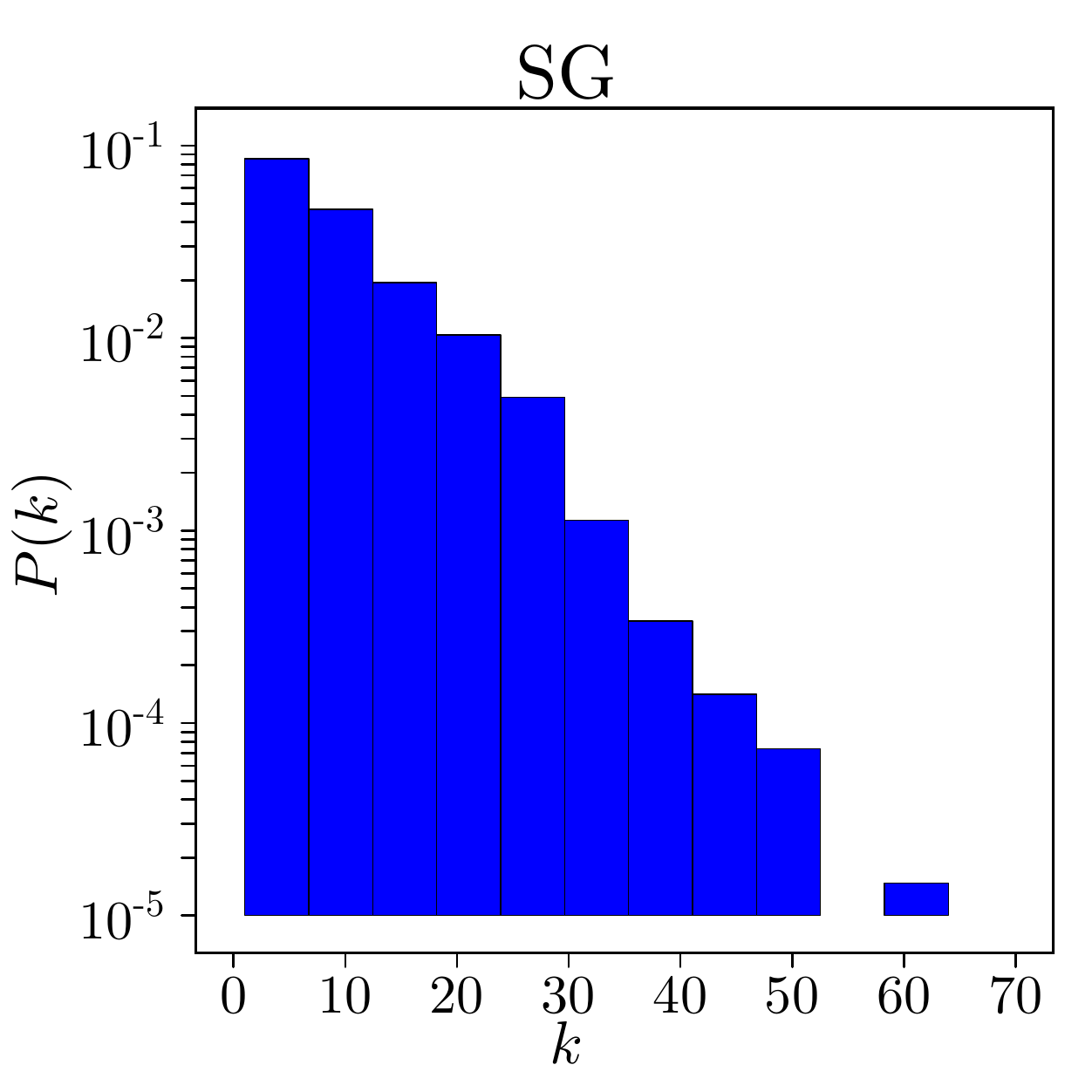}
\caption{Degree distributions $P(k)$ averaged over all daily aggregated networks,
  for the HT09 (left) and the SG (right) cases.  }\label{P-k}
\end{figure}

One of the standard observables used to characterize a network
topology is the degree distribution $P(k)$, i.e., the probability that
a randomly chosen node has $k$ neighbors. Figure~\ref{P-k} reports the
degree distributions of the daily aggregated networks, averaged over
the whole duration of the HT09 deployment (left) and SG deployment
(right).  For the SG case, we left out the few isolated nodes that
contribute to the degree distribution for $k=0$ only.  The $P(k)$
distributions are short-tailed in all cases: $P(k)$ decreases
exponentially in the SG case, and even faster for HT09. We notice that
the HT09 degree distribution exhibits a peak at $k$ around $15-20$, pointing
to a characteristic number of contacts established during the
conference. Moreover, the average degree in the HT09 case, $\langle k
\rangle$, close to $20$, is more than twice as high as that for the SG
networks, $\langle k\rangle$ which is close to $8$.  This represents another clear
indication of the behavioral difference of conference participants
versus museum visitors (the fact that the average degree is high for
conference attendees can be regarded as a goal of the conference
itself).  Finally, we observe that a large fraction of the recorded
contacts are sustained for a short time: for instance, removing all
the contacts with a cumulated duration below one minute yields
$\langle k\rangle$ about $7.5$ for HT09 and $\langle k\rangle$ about $3.5$
for SG.

\section{Temporal features}
\label{dynamic-properties}

The availability of time-resolved data allows one to gain much more
insight into the salient features of the social interactions taking place
during the deployments than what could be possible by the only knowledge
of ``who has been in face-to-face proximity of whom''.

\begin{figure}
\includegraphics[width=0.45\columnwidth,height=0.45\columnwidth]{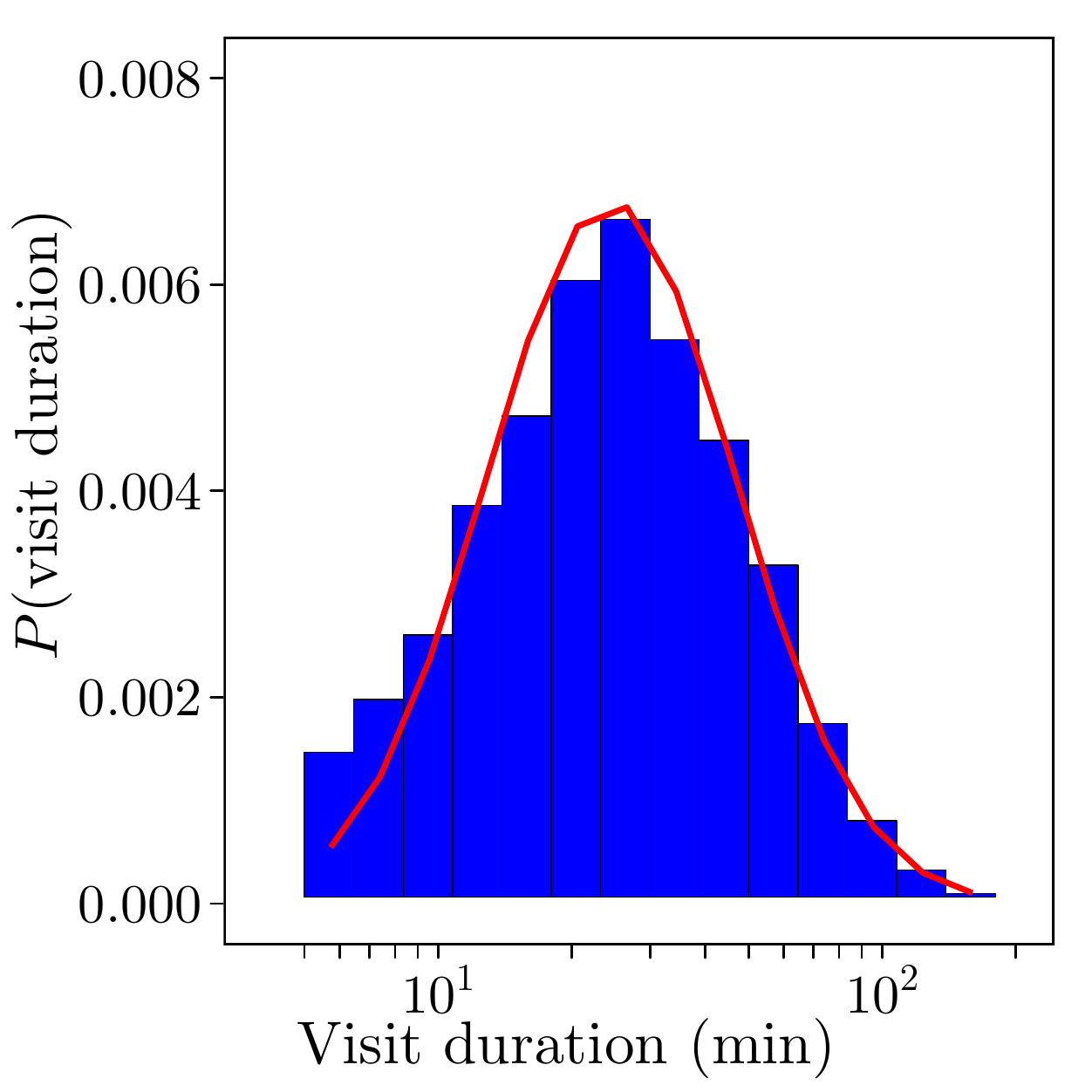}
\caption{Visit duration distribution at the SG museum (histogram) and
  fit to a lognormal distribution (red line). }\label{P-visits}
\end{figure}

\begin{figure}
\includegraphics[width=\columnwidth]{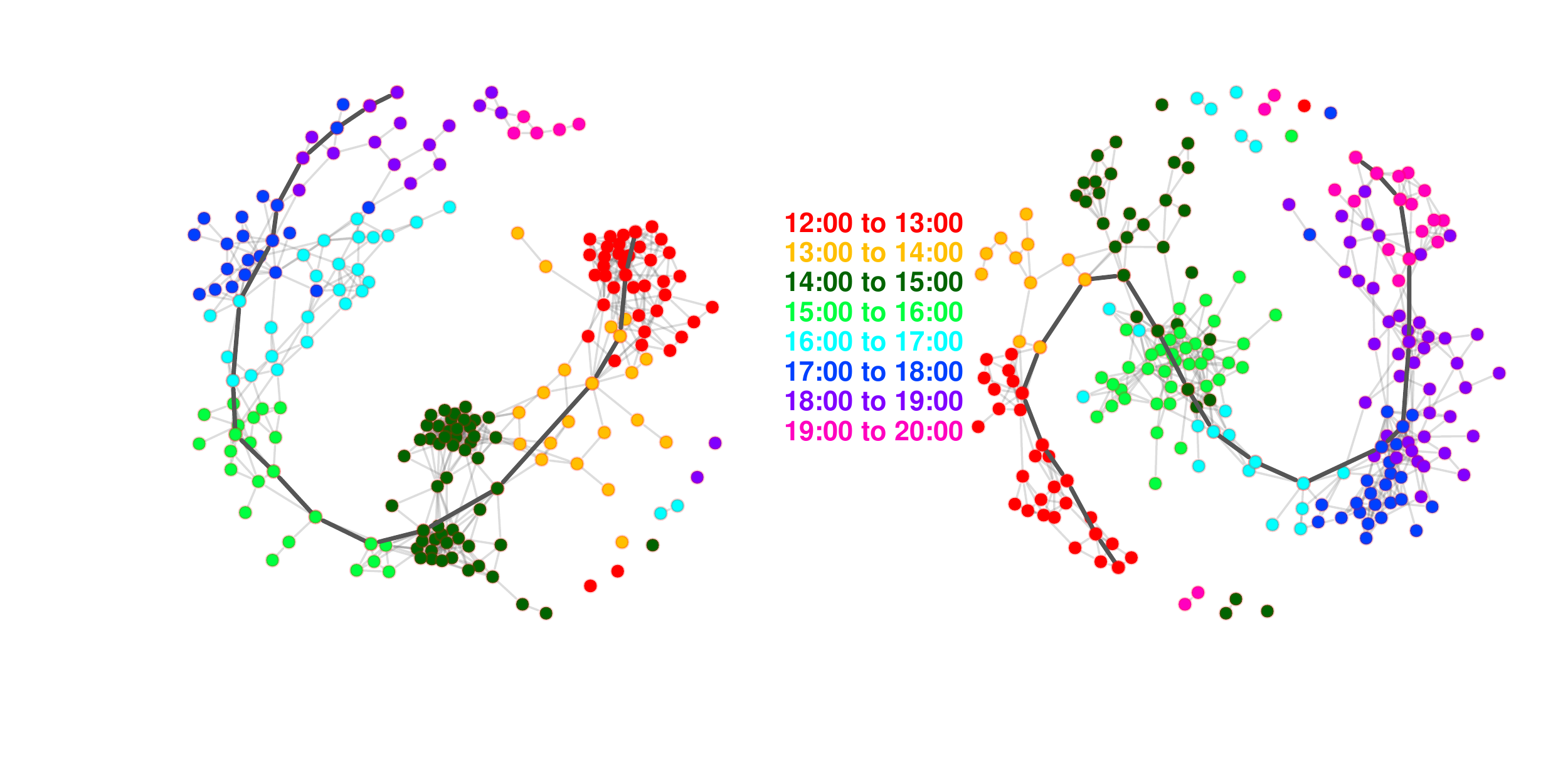}
\caption{(Color online) Aggregated networks for two different days of
  the SG museum deployment. Nodes are colored according to the
  corresponding visitor's entry time slot. The network diameter is
  highlighted in each case.  }\label{node-color-code}
\end{figure}

We first investigated the presence duration distribution in both
settings. For the conference case, the distribution is rather trivial,
as it essentially counts the number of conference participants
spending one, two or three days at the conference. The visit duration
distribution for the museum, instead, can be fitted to a lognormal
distribution (see Fig.~\ref{P-visits}), with geometric mean
around $35$ minutes. This shows that, unlike the case of
the conference, here one can meaningfully introduce the concept of a
characteristic visit duration that turns out to be well below the
cutoff imposed by museum opening hours.
The existence of a characteristic visit duration sheds light on the
elongated aspect of the aggregated networks of visitor interactions
(see Fig.~\ref{aggregated-networks}). Indeed museum visitors are
unlikely to interact directly with other visitors entering the venue
more than one hour after them, thus preventing the aggregated network
from exhibiting small-world properties.
Figure~\ref{node-color-code} reports the SG aggregated networks
for two different days, where the network diameter
is highlighted and each node is colored according to
the arrival time of the corresponding visitor.
One notices that, as expected from the aforementioned
properties of the visit duration distribution, there is limited
interaction among visitors entering the museum at different times.
Furthermore, the network diameter clearly defines a path connecting
visitors that enter the venue at subsequent times, mirroring the longitudinal
dimension of the network. These findings show that aggregated 
network topology and
longitudinal/temporal properties are deeply interwoven.

\begin{figure}
\includegraphics[width=0.45\columnwidth,height=0.45\columnwidth]{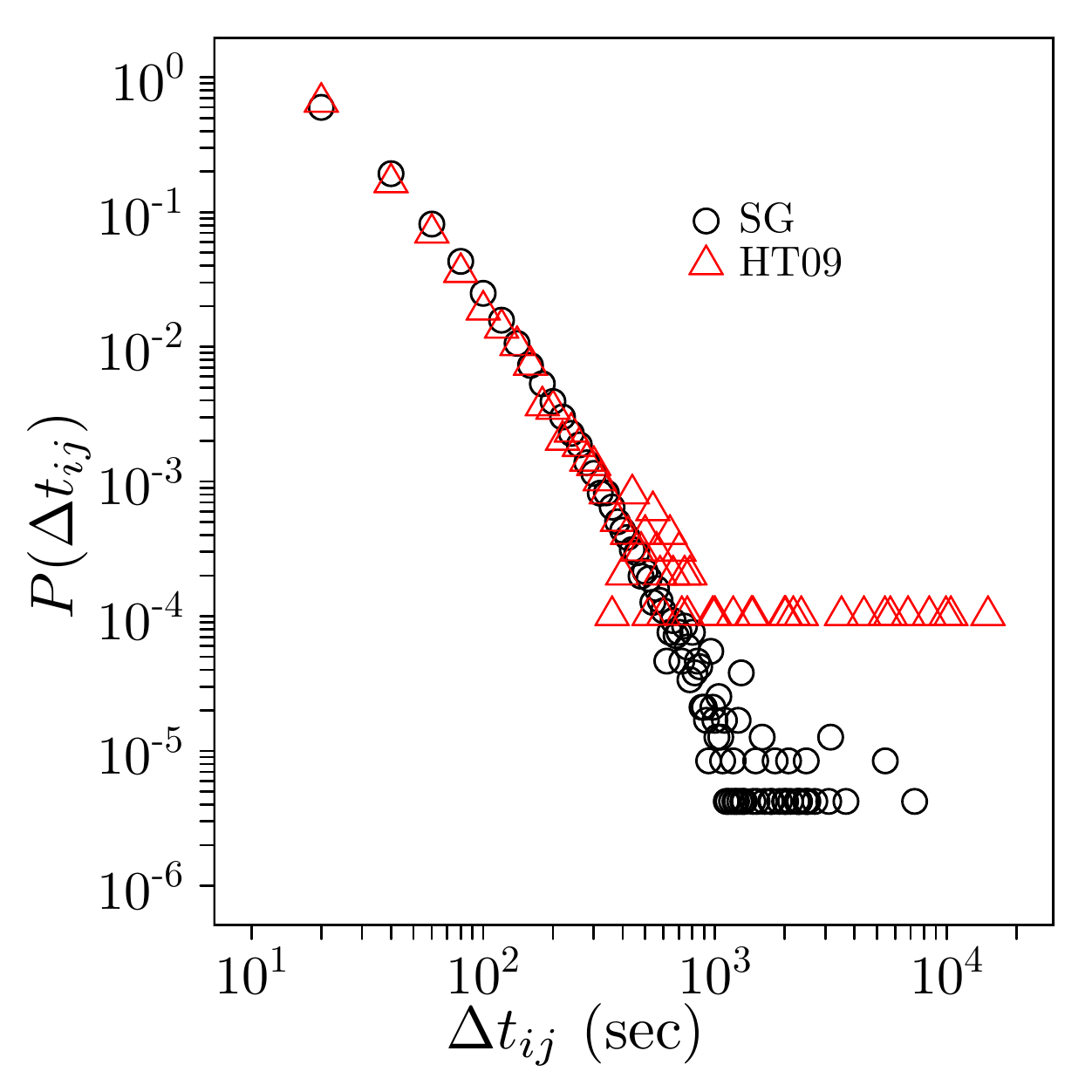}
\caption{Distributions of the contact durations for the HT09
  (triangles) and SG (circles) deployments, averaged over all
  days. Despite the differences in the measurement contexts, the
  distributions are superimposed.}\label{P-contact}
\end{figure}

\begin{figure}
\includegraphics[width=0.45\columnwidth, height=0.45\columnwidth]{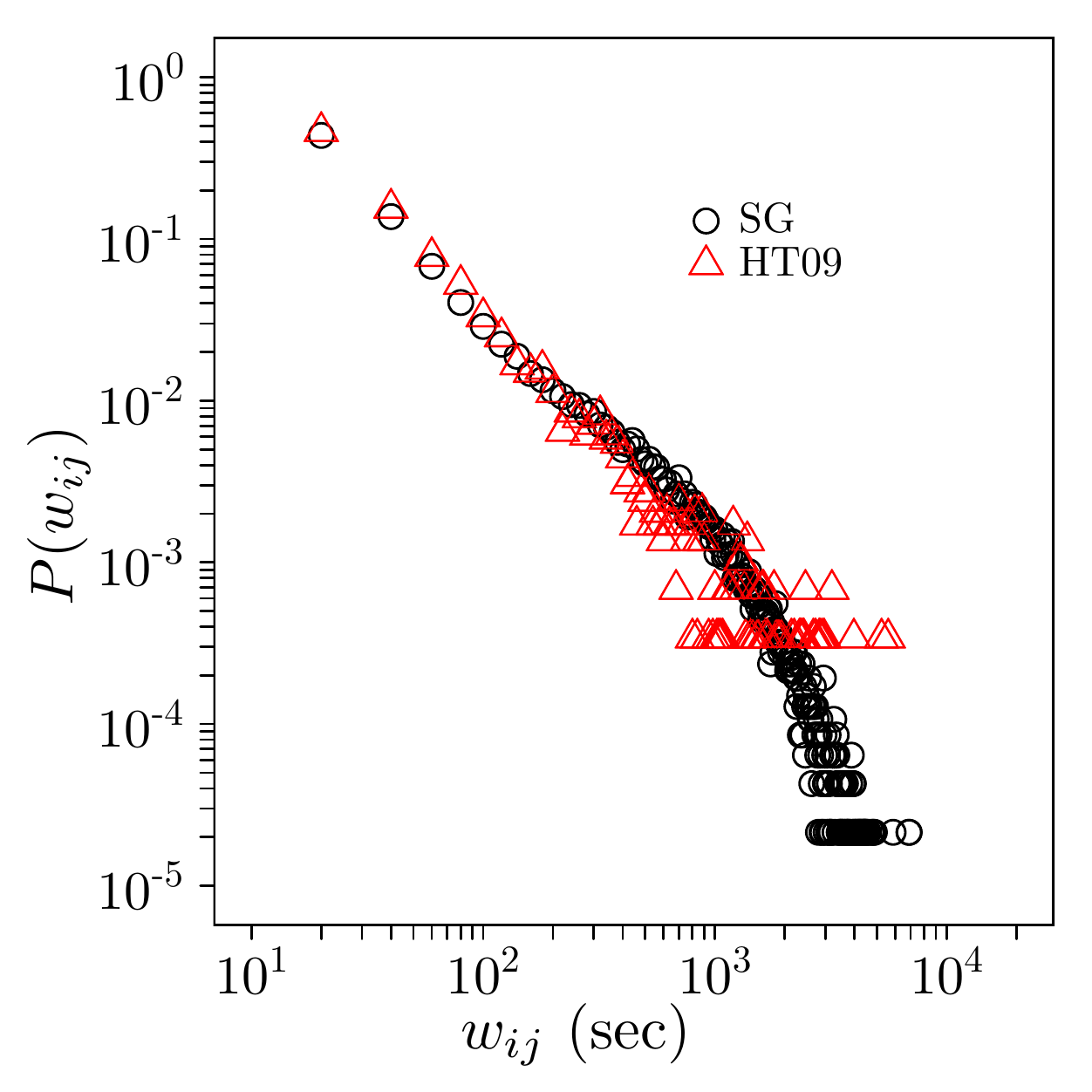}
\caption{Weight distributions for the daily aggregated networks of one
  HT09 conference day (triangles) and for the SG aggregated networks
  (circles), averaged over all daily aggregated networks. The weight
  of a link represents the total time spent in face-to-face proximity
  by the two linked individuals during the aggregation interval (here
  one day).}\label{P-w}
\end{figure}

Let us now focus on the temporal properties of social interactions.
At the most detailed level, each contact between two individuals is
characterized by its duration. The corresponding distributions are
shown in Fig.~\ref{P-contact}. As noted before, in both the HT09 and
SG cases most of the recorded interactions amount to shortly-sustained
contacts lasting less than one minute. However, both distributions
show broad tails -- they decay only slightly faster than a power
law. This behavior does not come as a surprise, as it has been
observed in social sciences in a variety of context ranging from human
mobility to email or mobile phone calls networks
\cite{Eckmann:2004,Hui:2005,Cattuto:2010,Oliveira:2005}.  More
interestingly, the distributions are very close (except in the noisy
tail, due to the different number of contributing events), showing
that the statistics of contact durations are robust across two very
different settings. This robustness has been observed in
Ref.~\cite{Cattuto:2010} across different scientific conferences, but
the museum setting corresponds to a situation in which a flux of
individuals follows a predefined path, and this strong similarity
between distributions was therefore not expected a priori.  At a
coarser level, aggregated networks are characterized by weights on the
links, that quantify for how long two individuals have been in
face-to-face proximity during the aggregation interval.
Figure~\ref{P-w} displays the distributions of these weights $w$.
These distributions are very broad~\cite{Cattuto:2010}: while most
links correspond to very short contacts, some correspond to very long
cumulated durations, and all time scales are represented, that is, no
characteristic interaction timescale (except for obvious cutoffs) can
be determined.  We note that at this coarser level of analysis the
distributions are again very similar.

\begin{figure}
\includegraphics[width=0.45\columnwidth, height=0.45\columnwidth]{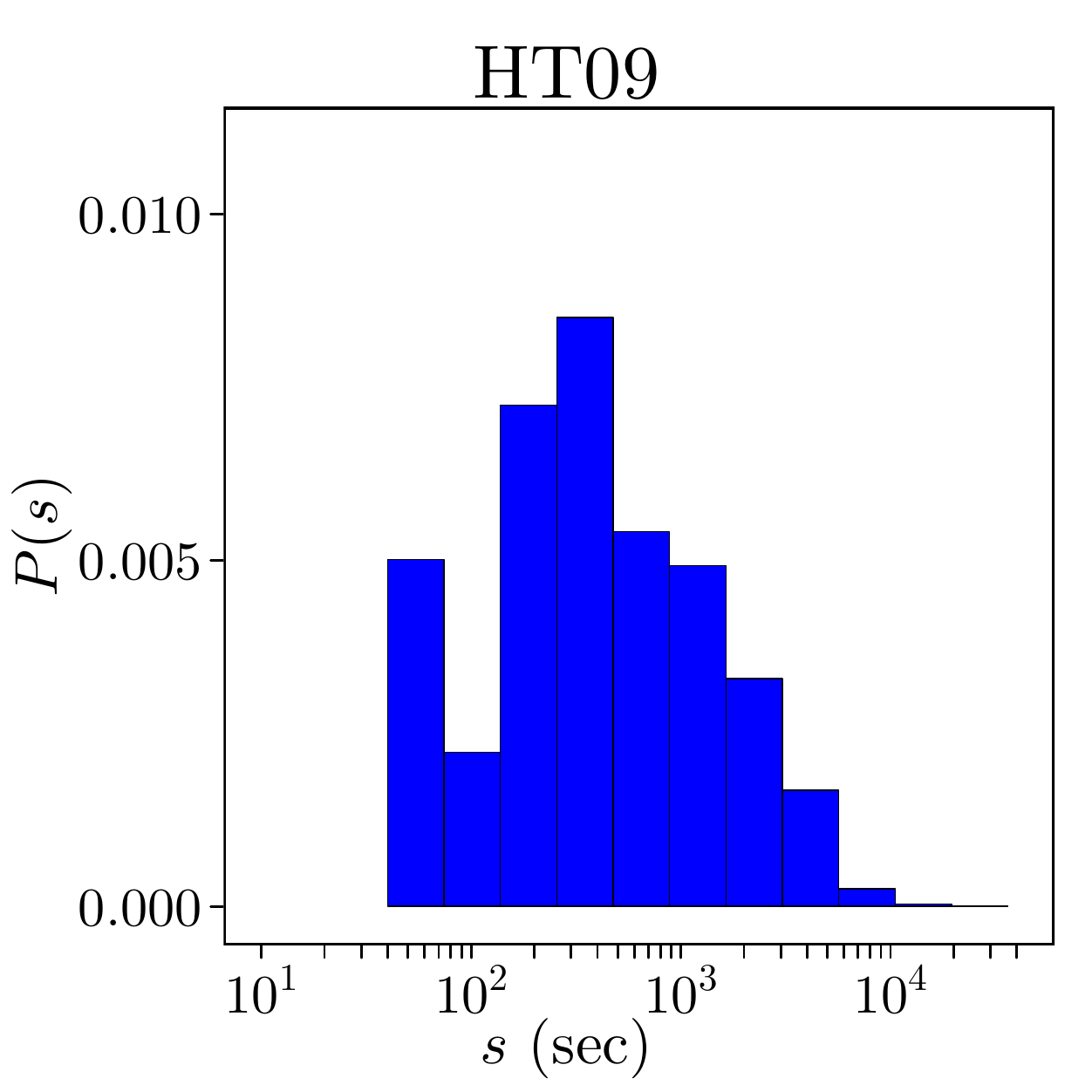}
\includegraphics[width=0.45\columnwidth,height=0.45\columnwidth]{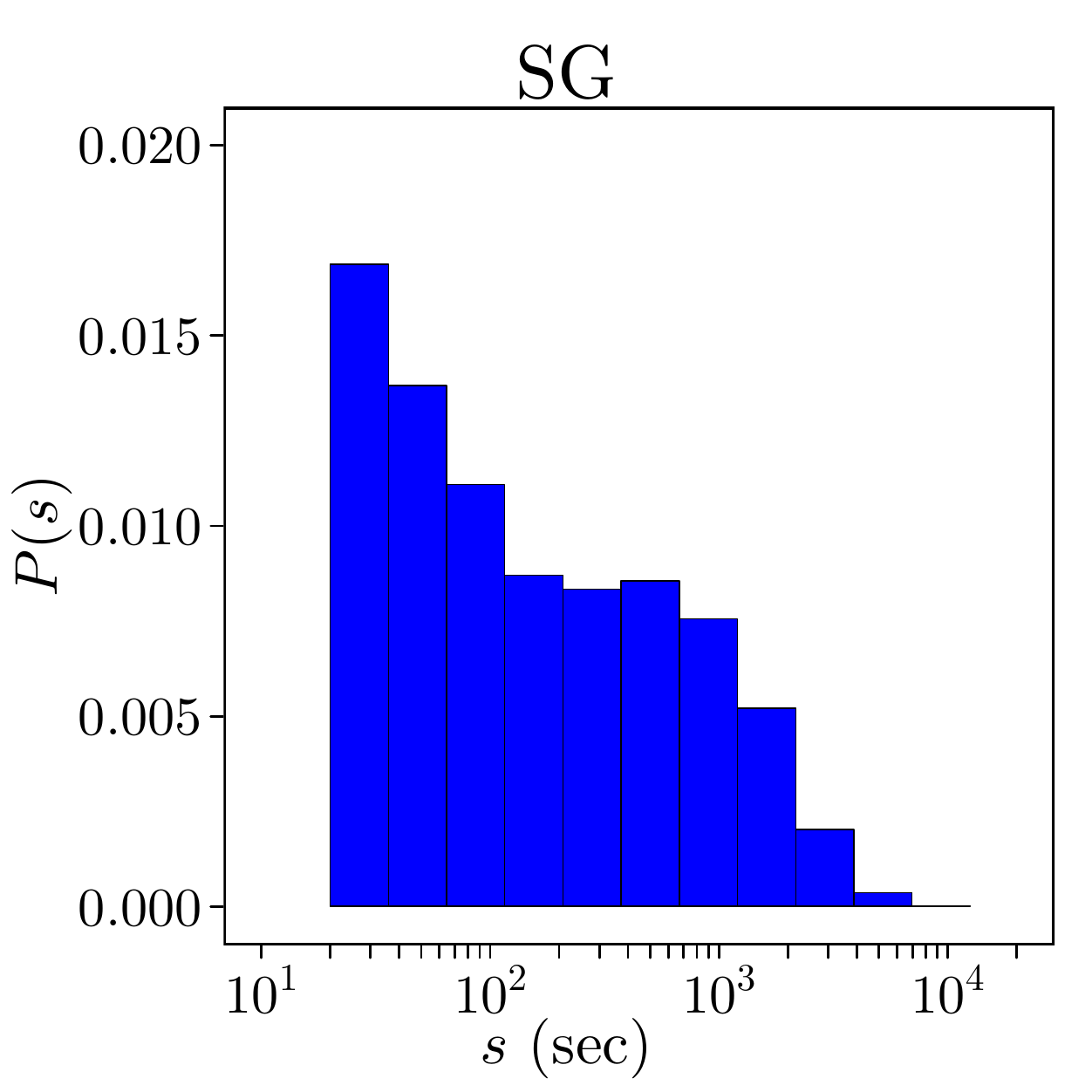}
\caption{Strength distributions $P(s)$ in the HT09 (left) and SG
  (right) aggregated networks (data for all daily networks). The
  strength of a node quantifies the cumulated time of interaction of
  the corresponding individual with other individuals. }\label{P-s}
\end{figure}

\begin{figure}
\begin{center}
\includegraphics[width=0.45\columnwidth, height=0.45\columnwidth]{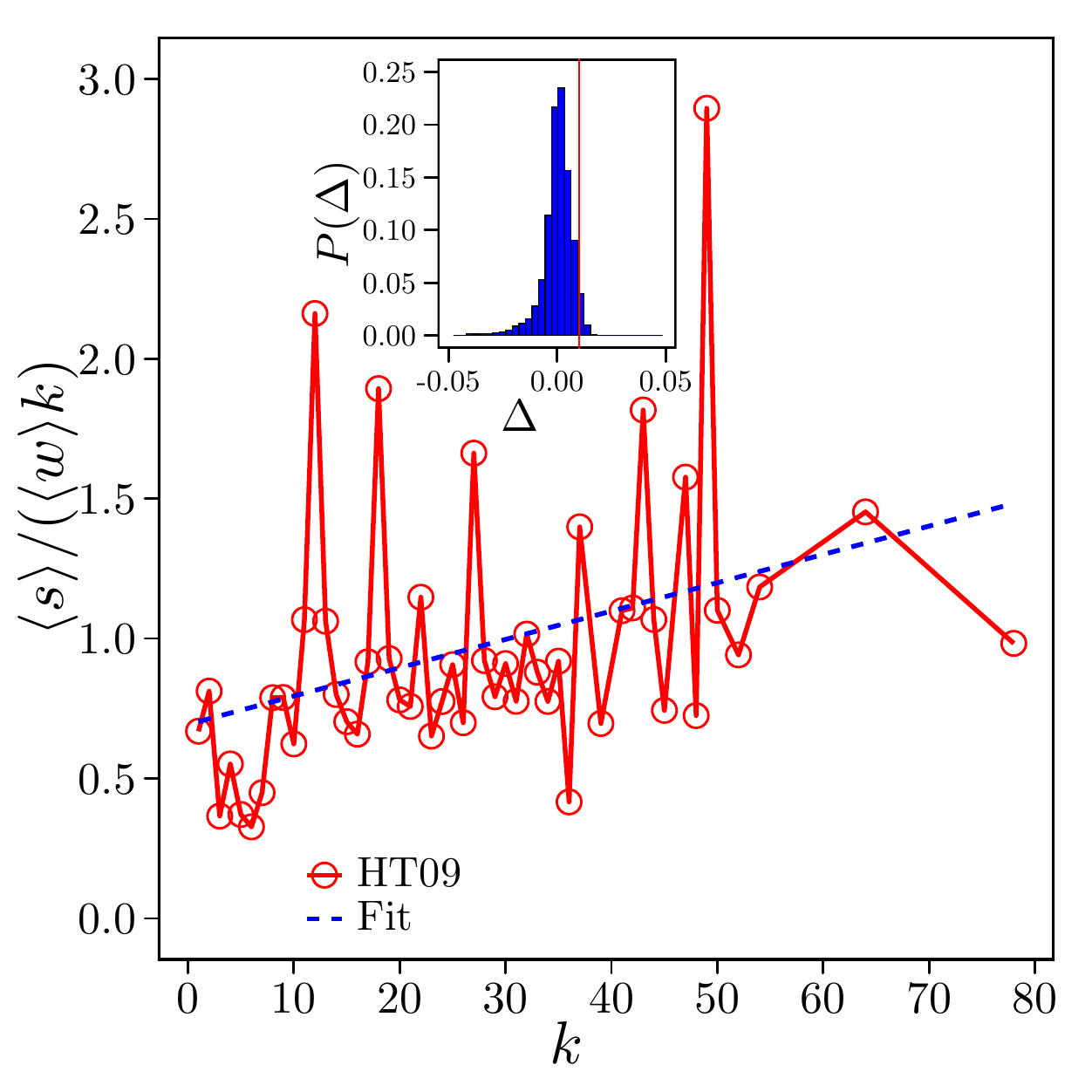}
\includegraphics[width=0.45\columnwidth,height=0.45\columnwidth]{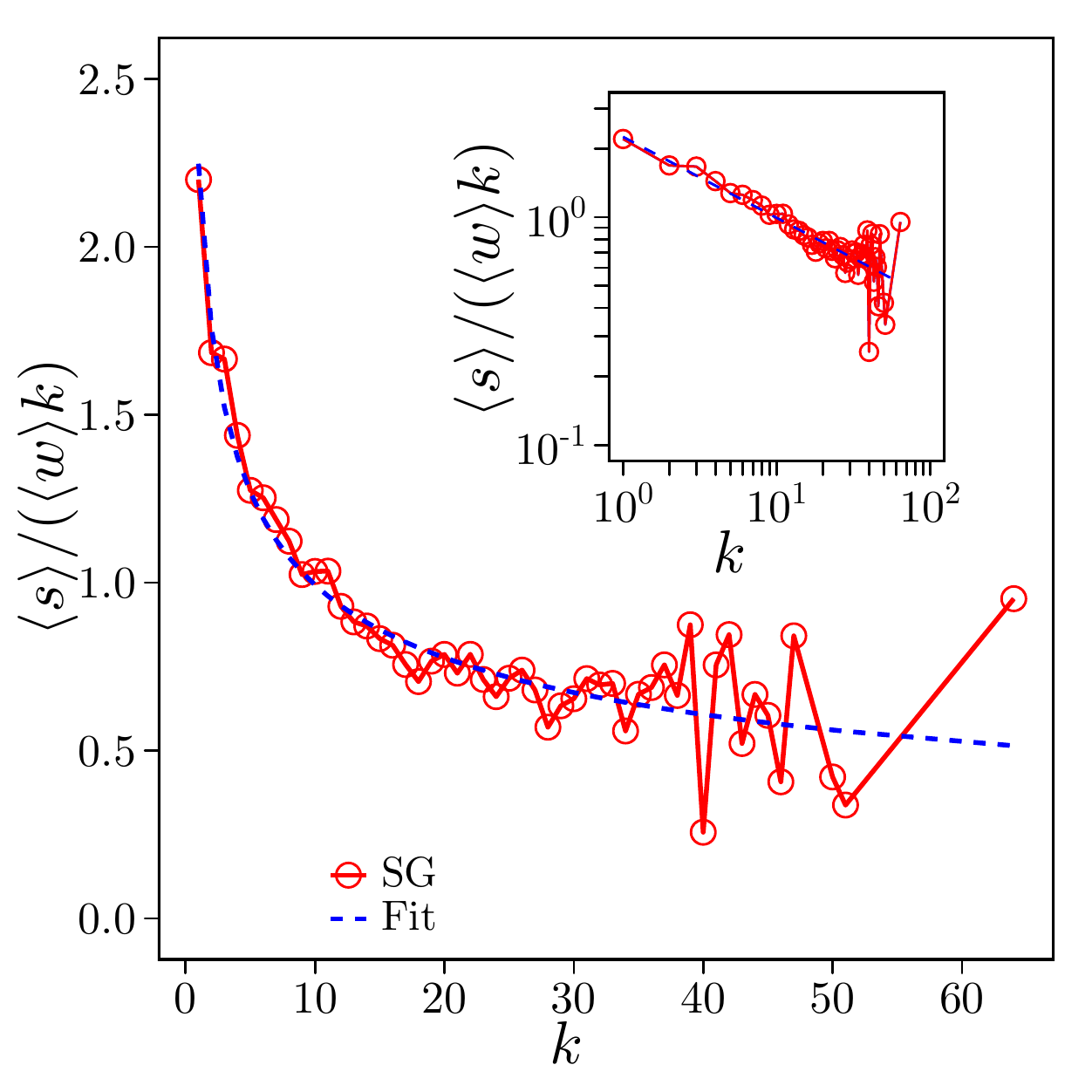}
\caption{(Color online) Correlation between node's strength and
  degree, as measured by the average strength $\langle s(k) \rangle$
  of nodes of degree $k$. The figures show $\langle s(k) \rangle
  /(\langle w \rangle k)$ (circles), for the HT09 (left) and SG
  (right) deployments (the solid line is only a guide for the
  eye). The dashed lines stand for a linear fit and a power law fit to
  the data for the HT09 and SG deployments, respectively.  Distinct
  increasing and decreasing trends are respectively observed.  The
  inset for the HT09 deployment shows a distribution of linear
  coefficients $\Delta$ calculated for 4000 reshufflings of the
  network weights and the fitted value from the data collected at HT09
  (vertical line). The inset for the SG deployment shows $\langle s(k)
  \rangle /(\langle w \rangle k)$ on a doubly logarithmic scale
  (circles) together with the power law fit to the data (dashed line).
}\label{s-over-k}
\end{center}
\end{figure}

For each individual, the cumulated time of interaction with other
individuals is moreover given by the strength $s$ of the corresponding
node \cite{alain-vespi}, i.e., by the sum of the weights of all links
inciding on it.
The strength distributions $P(s)$ are displayed in Fig.~\ref{P-s} for
the aggregated networks of the HT09 conference (left) and of the SG museum case
(right). Unlike $k$, the node strength $s$ spans several orders of
magnitude, ranging from a few tens of seconds to well above one hour.
The node strength $s$ can be correlated with
the node degree $k$ by computing the average strength
$\langle s(k)\rangle$ of nodes of degree $k$ \cite{alain-vespi}.
While a completely random assignment of weights yields a linear
dependency with $\langle s(k)\rangle$ proportional to $\langle w\rangle k$, where
$\langle w\rangle$ is the average link weight,
super-linear or sub-linear behaviors have been observed in various
contexts~\cite{alain-vespi,Cattuto:2010,Onnela:2007}.
A super-linear dependence such as the one observed in some conference settings~\cite{Cattuto:2010} hints at the presence of super-spreader nodes
that play a prominent role in processes such as information diffusion~\cite{vespi-classic,anderson-may}. On the other hand, the sub-linear
dependence observed for large-scale phone call networks~\cite{Onnela:2007}
corresponds to the fact that more active individuals spend on average
less time in each call. Figure \ref{s-over-k} displays the ratio
$\langle s(k) \rangle/(\langle w \rangle k)$ for the SG and HT09 daily
aggregated networks. Two different trends appear despite the
large fluctuations: a slightly increasing trend in the conference setting,
and a clearly decreasing one in the museum setting.
In particular, the behavior of  $\langle s(k) \rangle/(\langle w
\rangle k)$  for the HT09 case
(left plot in Figure \ref{s-over-k}) can be fitted linearly yielding a linear
coefficient $\Delta=0.01$ (p-value $=0.007$). By reshuffling
$4000$ times the weights of the network links and performing the same
linear fit for each reshuffling, we obtain a distribution of linear
coefficients $\Delta$. Such distribution, whose mean is zero,  is shown in the inset of the left plot in Figure
\ref{s-over-k} together with the value of $\Delta$ from the HT09
 daily aggregated networks (vertical line). The observed value of $\Delta$ at
the HT09 is an outlier of the distribution ($96^{th}$ percentile), thus
showing  that the observed behavior of $\langle s(k) \rangle/(\langle w
\rangle k)$ can hardly arise by a random assignment of link weights.
On the other hand, the observed behavior of $\langle s(k) \rangle/(\langle w
\rangle k)$  at the SG can be fitted to a power law with a
negative exponent  
i.e.  it decreases linearly on a double
logarithmic scale such as the one shown in the inset of the right plot
in  Figure \ref{s-over-k}. 
These results indicate that individuals who encountered the same number
of distinct persons can have different spreading potentials,
depending on the setting. It also gives a warning about characterizing
spreading by only measuring the number of encounters,
which can yield a rather misleading view.

\section{Percolation analysis}
\label{resilience}

The issue of network vulnerability to successive node removal has
attracted a lot of interest in recent years starting from the
pioneering works of Refs. \cite{barabasi-resilience,cohen-resilience},
that have shown how complex networks typically retain their integrity
when nodes are removed randomly, while they are very fragile with
respect to targeted removal of the most connected nodes. While the
concepts of node failures and targeted attacks are pertinent for
infrastructure networks, successive removals of nodes or links is more
generally a way to study network
structures~\cite{Holme:2002a,Girvan:2002,pippo-resilience,Dallasta:2006}.  For
instance, detecting efficient strategies for dismantling the network
sheds light on the network community structure, as it amounts to
finding the links that act as bridges between different
communities~\cite{Girvan:2002,pippo-resilience}. Moreover, in the context of
information or disease spreading, the size of the largest connected
component gives an upper bound on the number of nodes affected by the
spreading. Identifying ways to reduce this size, by removing
particular links, in order to break and disconnect the network as much
as possible, is analogous in terms of disease spreading to finding
efficient intervention and containment strategies.

In order to test different link removal strategies, we consider
different definitions of weight for a link connecting nodes
$i$ and $j$ in the aggregated contact network:\\
-- The simplest definition of link weight is given by the 
cumulated contact duration $w_{ij}$ between $i$ and $j$. In the following, we will refer to this weight as ``contact weight''.\\
-- The topological overlap $O_{ij}$, introduced in Ref.~\cite{onnela},
is defined as
\begin{equation}
  \label{eq:onnela-weight}
  O_{ij}=\frac{n_{ij}}{(k_{i}-1)+(k_{j}-1)-n_{ij}} \in [0,1] \, ,
\end{equation}
where $k_{i(j)}$ is the degree of node $i(j)$ and $n_{ij}$ measures
the number of neighbors shared by nodes $i$ and $j$.
This measure is reminiscent of the edge
clustering coefficient~\cite{pippo-resilience},
and evaluates the ratio of the number of triangles leaning upon the $ij$
edge with the maximum possible number of such triangles 
given that $i$ and $j$ have degrees $k_{i}$ and $k_{j}$, respectively.
Edges between different communities are expected to have a low number
of common neighbors, hence a low value of $O_{ij}$. \\
-- Finally, the structural similarity
of two nodes is defined as the cosine similarity
\begin{equation}
  \label{eq:cos-simil}
{\rm sim}_{ij}=\frac{\s_{l\in
    {\cal V}}w_{il}w_{jl}}{\sqrt{\s_{l}w_{il}^{2}\s_{l}w_{jl}^{2}}} \in [0,1] \, ,
\end{equation}
where ${\cal V}$ is the set of neighbors shared by nodes $i$ and $j$,
and the sums at the denominator are computed over all the neighbors
of $i$ and $j$. Cosine similarity, which is one of the simplest 
similarity measures used in the field of
information retrieval~\cite{newman-cosine-similarity,book-retrieval,book-baeza},
takes into account not only the number of shared neighbors
of $i$ and $j$, but also the similarity of the corresponding edge strengths,
i.e. the similarity of individuals in terms of the time they spent
with their neighbors. Once again, edges connecting different communities
are expected to have a low value of ${\rm sim}_{ij}$.

Based on these three weight definitions, we consider four different
strategies for link removal, namely: removing the links in
increasing/decreasing order of contact weight, in increasing order of
topological overlap, and in increasing order of cosine similarity. The
former two strategies are the simplest one can devise, as they do not
consider the neighborhoods' topology. The latter two
strategies were implemented in an incremental fashion,
by recomputing the lists of links ranked in order of increasing overlap
or cosine similarity whenever a link was removed, and then removing
the links in the updated list order\footnote{As shown in 
Ref.~\cite{Holme:2002a,Dallasta:2006}
and verified numerically (not shown) for the present case,
a procedure that does not update the link ranking upon every link removal,
based on the quantities (\ref{eq:onnela-weight}) and (\ref{eq:cos-simil}),
leads to sub-optimal results.
The deviation from the updating strategy becomes
apparent only when more than $20{\%}$  of links have been removed 
since $O_{ij}$ and ${\rm sim}_{ij}$  deal with local
quantities only. As a consequence, each link removal amounts
to a local perturbation of the network, contrary  
to what happens with non-local quantities such as the betweenness
centrality~\cite{Holme:2002a,Dallasta:2006}.}.
An issue also arises from the fact that all
the generalized weights mentioned above produce a certain amount of
link degeneracy (in particular when using the contact weight): for
instance, many links may have the same (small) value $w_{ij}$,
or exactly $0$ overlap or similarity. Each link removal procedure carries
therefore a certain ambiguity, and the results may depend on which
links, among those with the same contact weight/overlap/similarity, are
removed first.

\begin{figure}
\includegraphics[width=0.45\columnwidth]{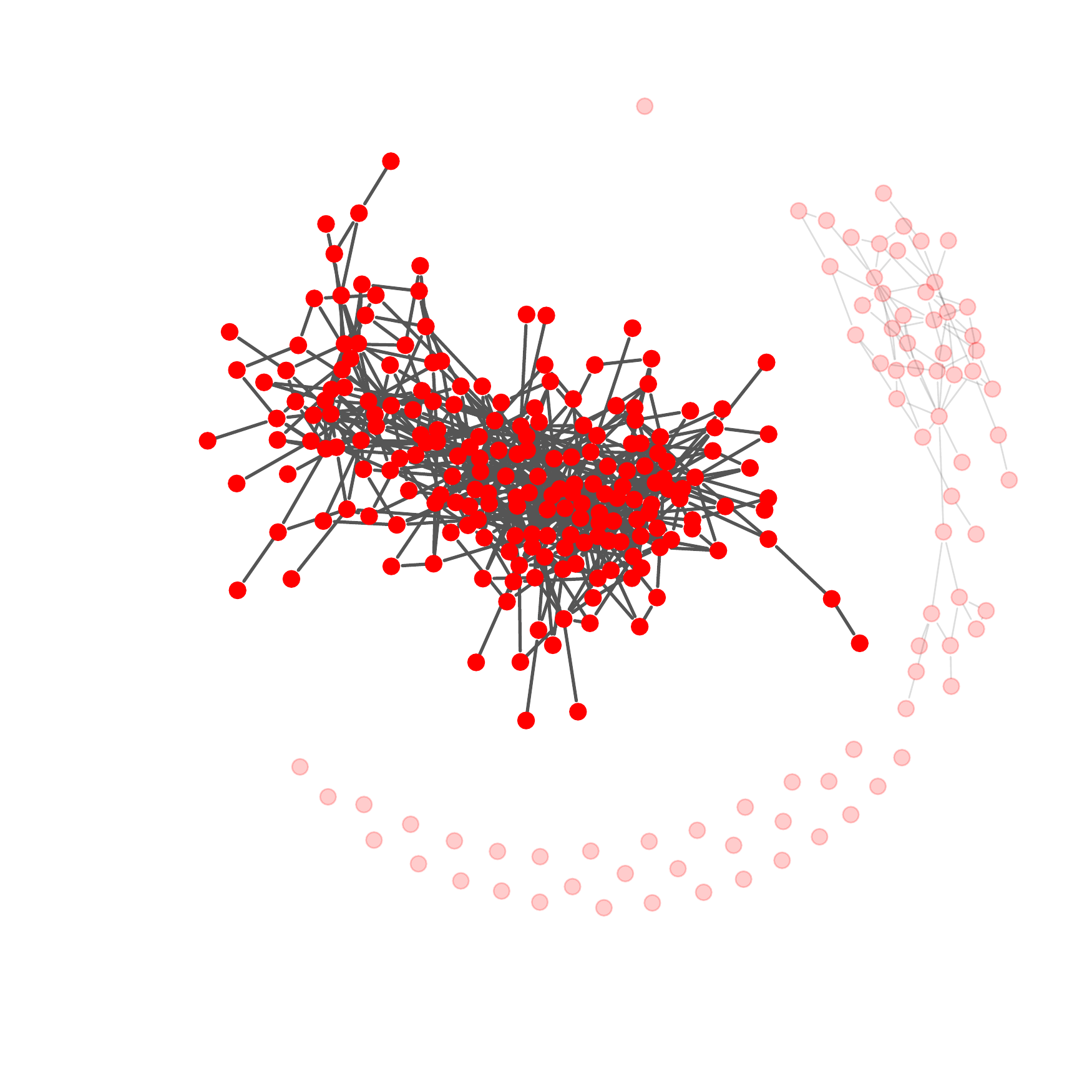}
\includegraphics[width=0.45\columnwidth]{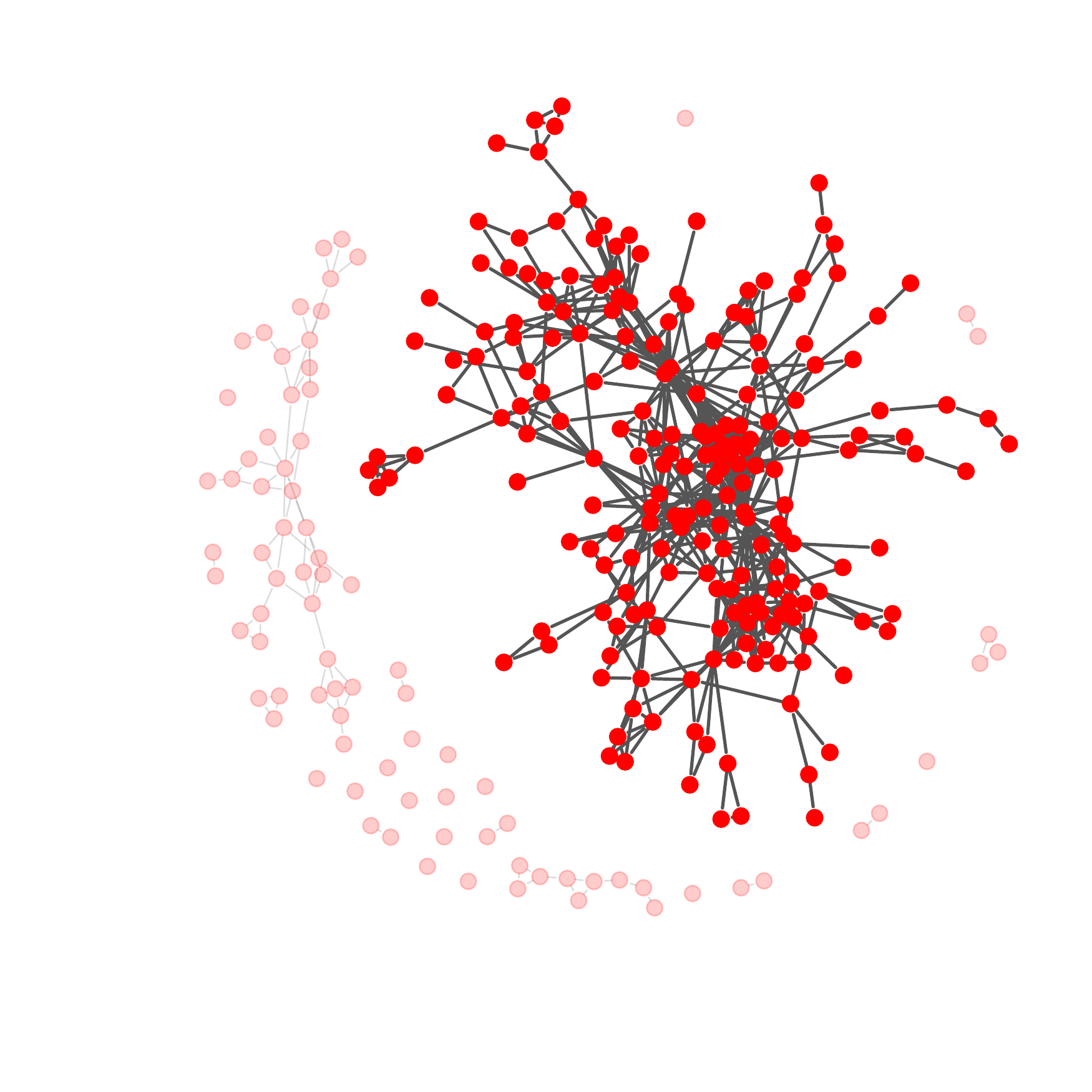}
\includegraphics[width=0.45\columnwidth]{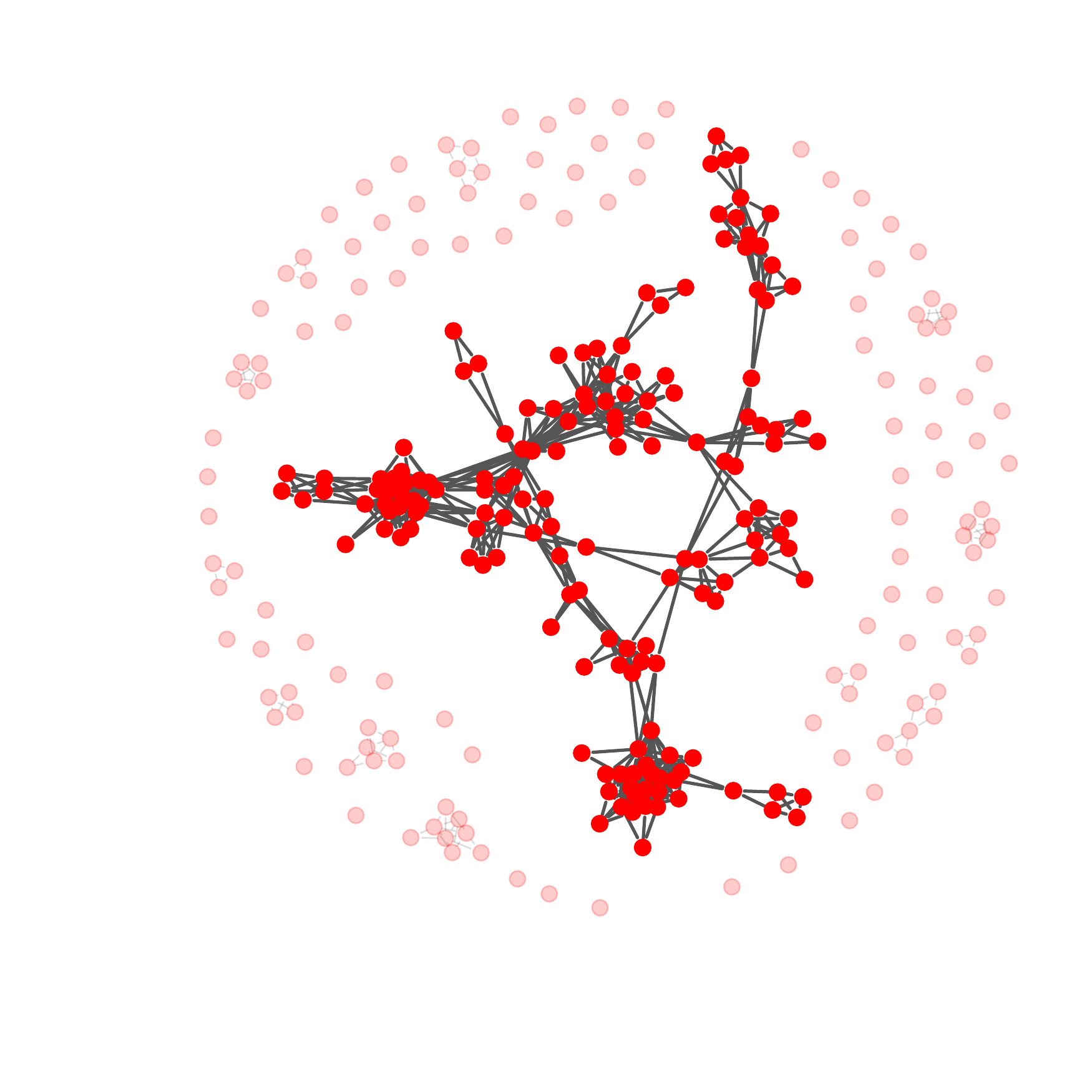}
\includegraphics[width=0.45\columnwidth]{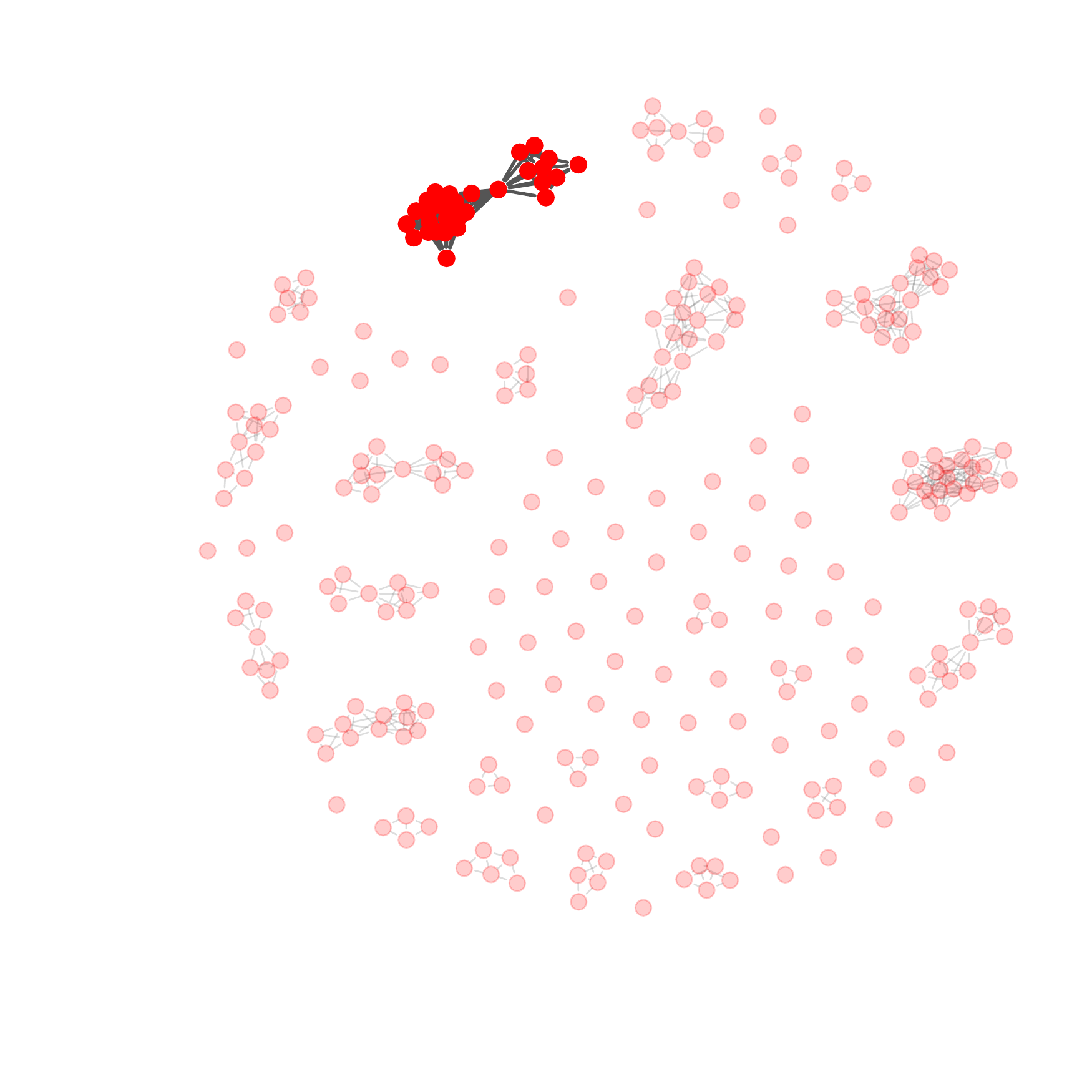}
\caption{Effect of four different ways of removing $60\%$ of the
  links on the SG museum daily aggregated network of July 14$^{\rm
    th}$.  Clockwise from top: links removed in decreasing contact
  weight order, increasing contact weight order, increasing
  topological overlap order and increasing cosine similarity
  order. The largest CC is highlighted in each
  case.}\label{dismantled-networks}
\end{figure}

The impact of link removal on network fragmentation can be measured by
monitoring the variations of the size of the largest CC, hereafter called
$N_{1}$, as a function of link removal. If the network is initially divided
into two CCs, labeled $C_{1}^{0}$ and $C_{2}^{0}$, of similar initial sizes
$N_{1}^{0}\ge N_{2}^{0}$, we call $N_{1}$ the size of the largest CC surviving
in the network (which does not need to be a subnetwork of $C_{1}^{0}$). We
used the apex ``$0$'' to denote quantities expressed for the original network,
before any link removal. In order to alleviate the problems arising from link
degeneracy, we averaged $N_{1}$ on $100$ different link orderings
(i.e. we reshuffled the list of links of equal generalized weight
before removing them).

\begin{figure}
\includegraphics[width=0.45\columnwidth, height=0.5\columnwidth]{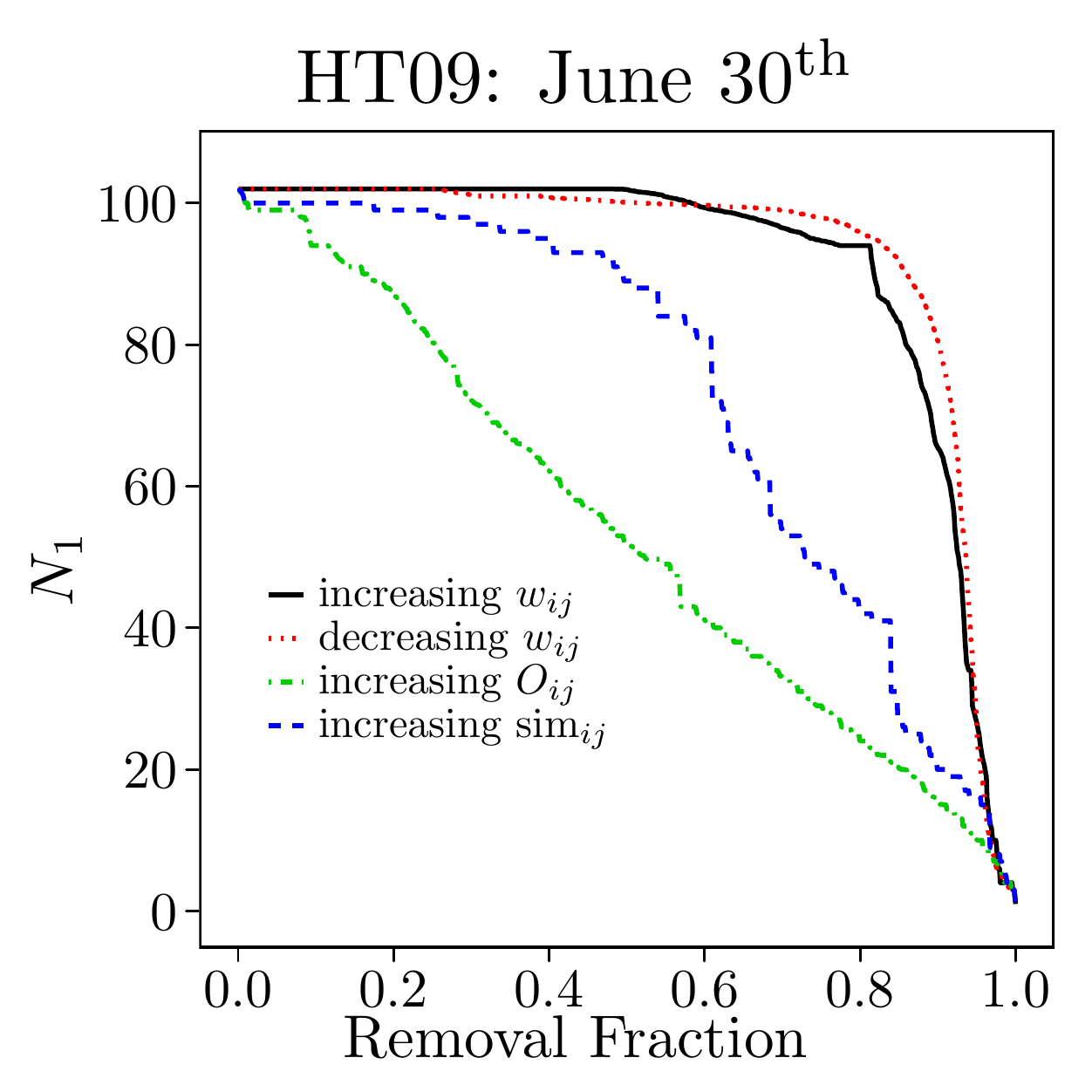}
\includegraphics[width=0.45\columnwidth, height=0.5\columnwidth]{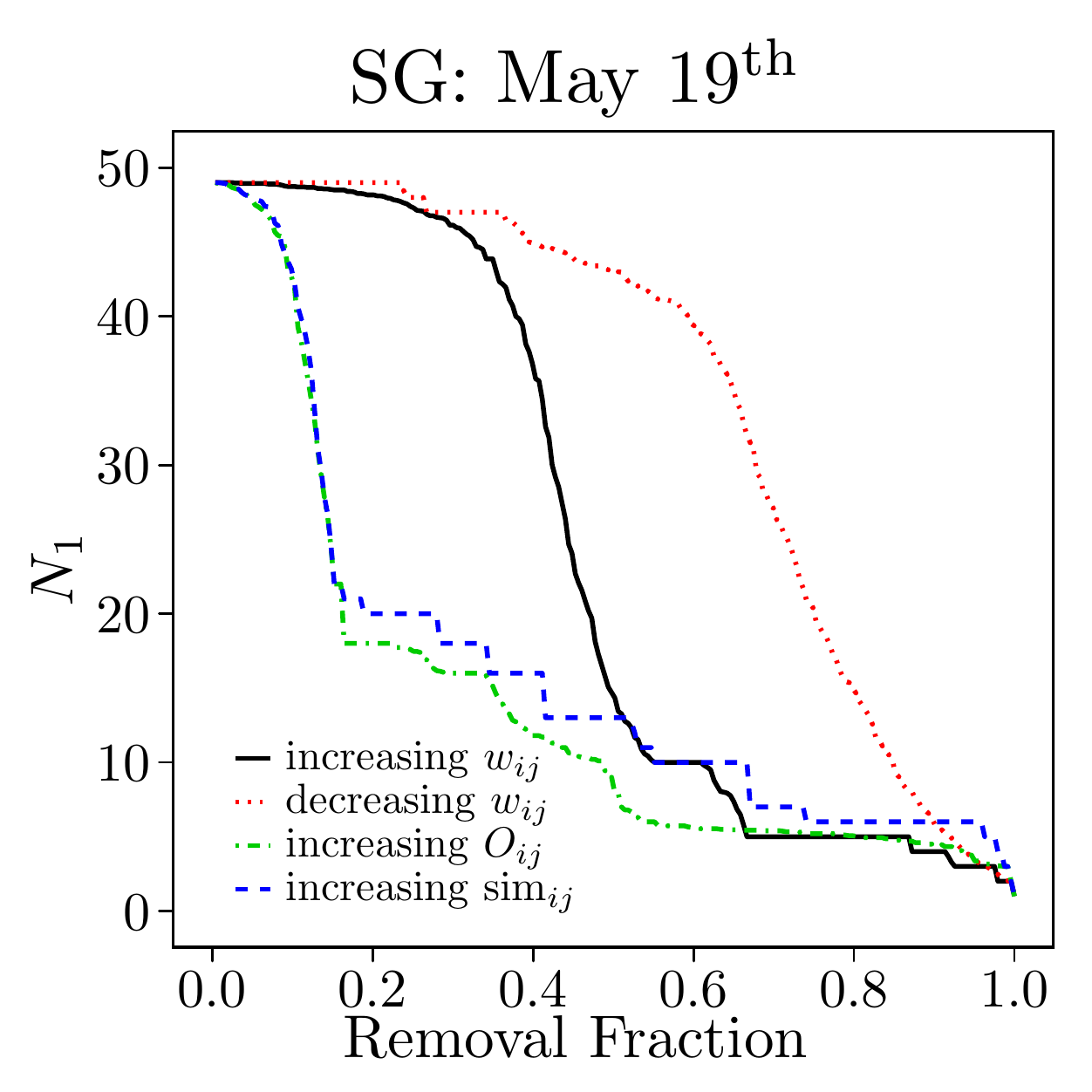}
\includegraphics[width=0.45\columnwidth, height=0.5\columnwidth]{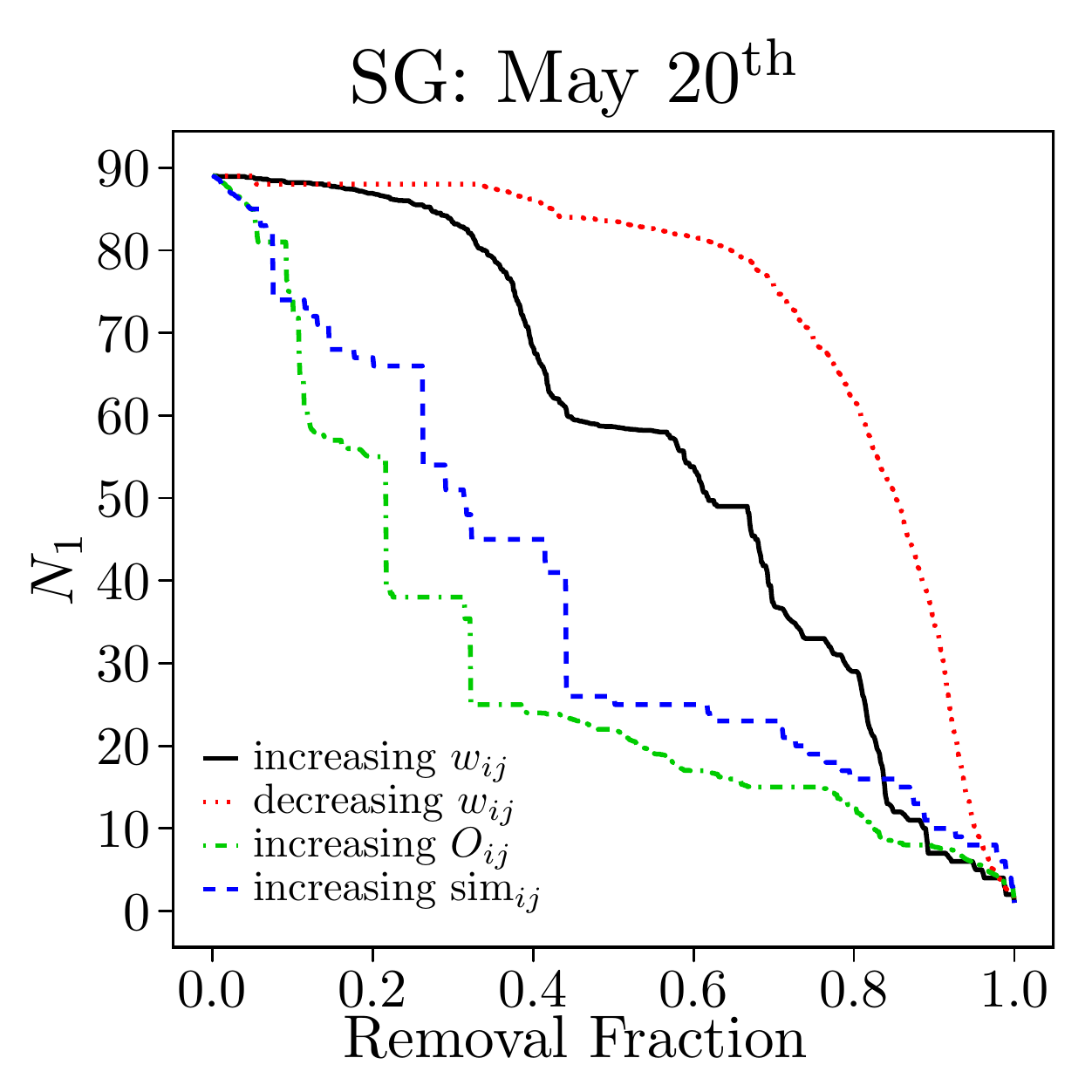}
\includegraphics[width=0.45\columnwidth, height=0.5\columnwidth]{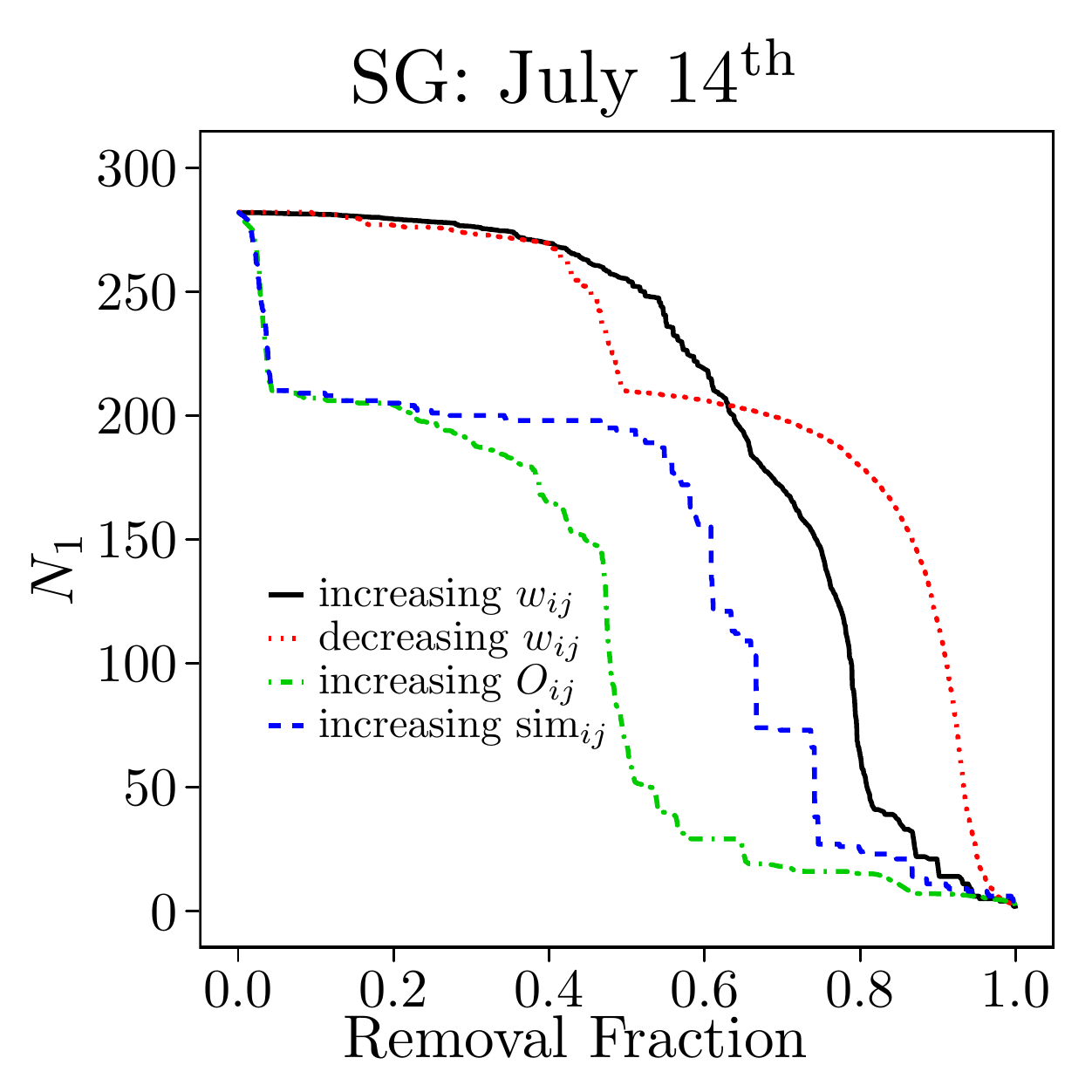}
\caption{(Color online) Size $N_1$ of the largest CC as a function of the fraction
  of removed links, for several removal strategies, and for different
  daily aggregated networks in the HT09 and SG deployments.  For all
  networks, removing links in increasing topological overlap order and
  increasing cosine similarity order have the most disruptive
  effects. The HT09 aggregated network is in all cases more
  resilient than the SG aggregated
  networks.}\label{dismantled-networks-delta-n1}
\end{figure}

An example of a single realization of the removal strategies for the
SG aggregated network of July 14$^{\rm th}$ is shown in
Fig.~\ref{dismantled-networks}. We observe that a removal of $60\%$ of
the network links has a far deeper impact on the network when the
removal is based on the topological overlap (the size of the largest
CC is $N_{1}=30$) or cosine similarity ($N_{1}=155$) rather than on
decreasing (increasing) contact weight ($N_{1}=204$ $(205)$). More
quantitatively, Fig.~\ref{dismantled-networks-delta-n1} shows that
removing links according to their topological overlap is the most
efficient strategy. This is in agreement with previous
results~\cite{Holme:2002a,Girvan:2002,pippo-resilience,Dallasta:2006,onnela} that
have shown that topological criteria detect efficiently the links that
act as bridges between communities. Due to their high degeneracy,
removing first the links with small contact weights approximates a
random removal strategy that is far from optimal. Despite this
limitation, removing the links with small contact weights
can outperform the removal of links with high contact weight since the
latter are usually found within dense communities, while links between
communities have typically small contact weights.

The strategy based on link topological overlap proves slightly
more effective
than the strategy based on link similarity:
the information on the link contact weights incorporated
in the definition of ${\rm sim}_{ij}$ (Eq.~\ref{eq:cos-simil})
does not enhance the decrease of $N_{1}$. This can be explained through the
following argument: topological overlap
link ranking usually leads to a higher degeneracy with respect to
similarity-based link ranking. As a consequence, for a network with similar
values of $N_{1}^{0}$ and $N_{2}^{0}$, a strategy based on topological
overlap is more likely to dismantle in parallel both $C_{1}^{0}$
and $C_{2}^{0}$ than a similarity-based strategy, as it has no bias
towards a specific component. The opposite strategy of a complete
dismantling of $C_{1}^{0}$ that leaves $C_{2}^{0}$ intact
would result in $N_{1}=N_{2}^{0}$
even after the complete disintegration of $C_{1}^{0}$.
This effect is illustrated in Fig.~\ref{dismantled-networks-delta-n1}
for the SG aggregated networks of May $19^{\rm th}-20^{\rm th}$, which are
indeed composed of two large CCs (see Fig.~\ref{aggregated-networks}).

Interestingly, and as expected from the previous comparisons,
rather different results are obtained for the HT09 and SG
aggregated networks. The conference network is more resilient to all
strategies, and significant levels of disaggregation are reached only
by removing large fractions ($\ge 40-60 {\%}$) of the links, sorted by
their topological overlap. For the SG aggregated networks, on the
other hand, targeting links with small topological overlap or cosine
similarity is a quite effective strategy, which can be intuitively
related to the modular structure visible in Fig.~\ref{aggregated-networks}.

%
%
\section{Dynamical spreading over the network}
\label{information-diffusion}
Aggregated networks often represent the most detailed information that
is available on social interactions. In the present case, they would
correspond to information obtained through ideal surveys in which
respondents remember every single person they encountered and the
overall duration of the contacts they had with that person.  While
such a static representation is already informative, it lacks
information about the time ordering of events, and it is unable to
encode causality. The data from our measurements do not suffer from
this limitation, as they comprise temporal information about every
single contact. Therefore, these data can be used to investigate the
unfolding of dynamical processes.  They also allow to study the role
of causality in diffusion processes, such as the spreading of an
infectious agent or of a piece of information on the encounter
networks of individuals.  In the following we will mainly use an
epidemiological terminology, but we may equally imagine that the RFID
devices are able to exchange some information whenever a contact is
established. Individuals will be divided into two categories,
susceptible individuals (S) or infected ones (I): susceptible
individuals have not caught the ``disease'' (or have not received the
information), while infected ones carry the disease (or have received
the information) and can propagate it to other individuals.

In order to focus on the structure of the dynamical network itself, we consider
in the following a deterministic snowball SI model~\cite{anderson-may}:
every contact between a susceptible individual and an infected one, no matter
how short, results in a transmission event in which the susceptible becomes
infected, according to $S+I \to 2I$. In this model, individuals, once
infected, do not recover. Such a deterministic model allows to isolate the
role played in the spreading process by the structure of the dynamical network
(e.g. its causality). Its role would otherwise be entangled with the
stochasticity of the transmission process and the corresponding interplay of
timescales. Of course, any realistic epidemiological model should include a
stochastic description of the infection process, since the transmission from
an infected individual to a susceptible one is a random event that depends on
their cumulative interaction time. The resulting dynamics would depend on the
interplay between contact and propagation times. We leave the study of this
interesting type of interplay to future investigations.

In our numerical experiments, for each day we select a single ``seed'', i.e.,
an individual who first introduces the infection into the network. All the
other individuals are susceptible and the infection spreads deterministically
as described above. By varying the choice of the seed over individuals, we
obtain the distribution of the number of infected individuals at the end of
each day. The transmission events can be used to define the network along
which the infection spreads (i.e., the network whose edges are given by
$S\leftrightarrow I$ contacts), hereafter called the \textit{transmission
network}.

Due to causality, the infection can only reach individuals present at
the venue after the entry of the seed. As a consequence, in the
following we will use the term \textit{partially aggregated network}
to indicate the network aggregated from the time
the seed enters the museum/conference to the end of the day.
We note that the partially aggregated network defined in this way
can be radically different from (much smaller than)
the network aggregated along the whole day.

\begin{figure}
\includegraphics[width=0.45\columnwidth, height=0.45\columnwidth]{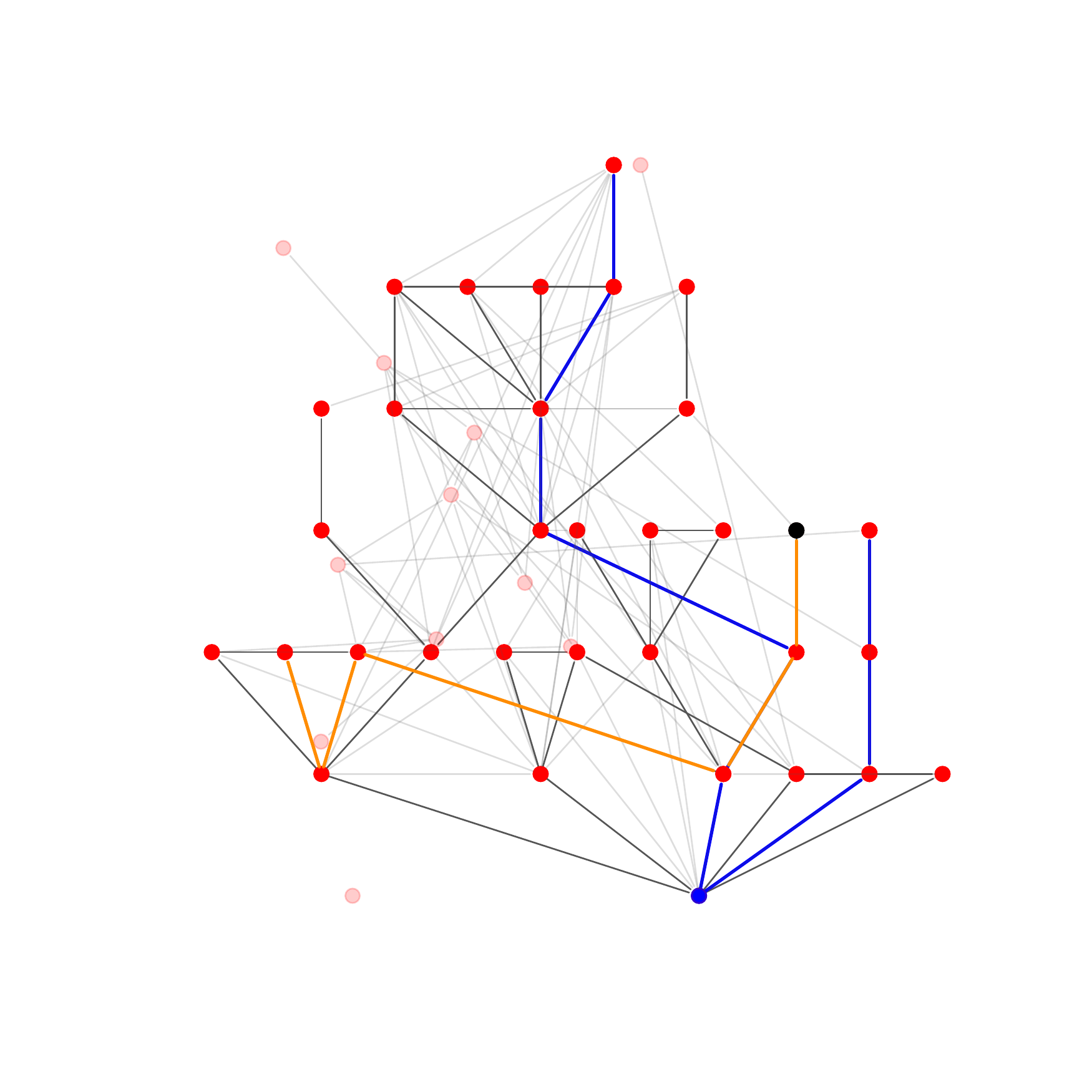}
\includegraphics[width=0.45\columnwidth,height=0.45\columnwidth]{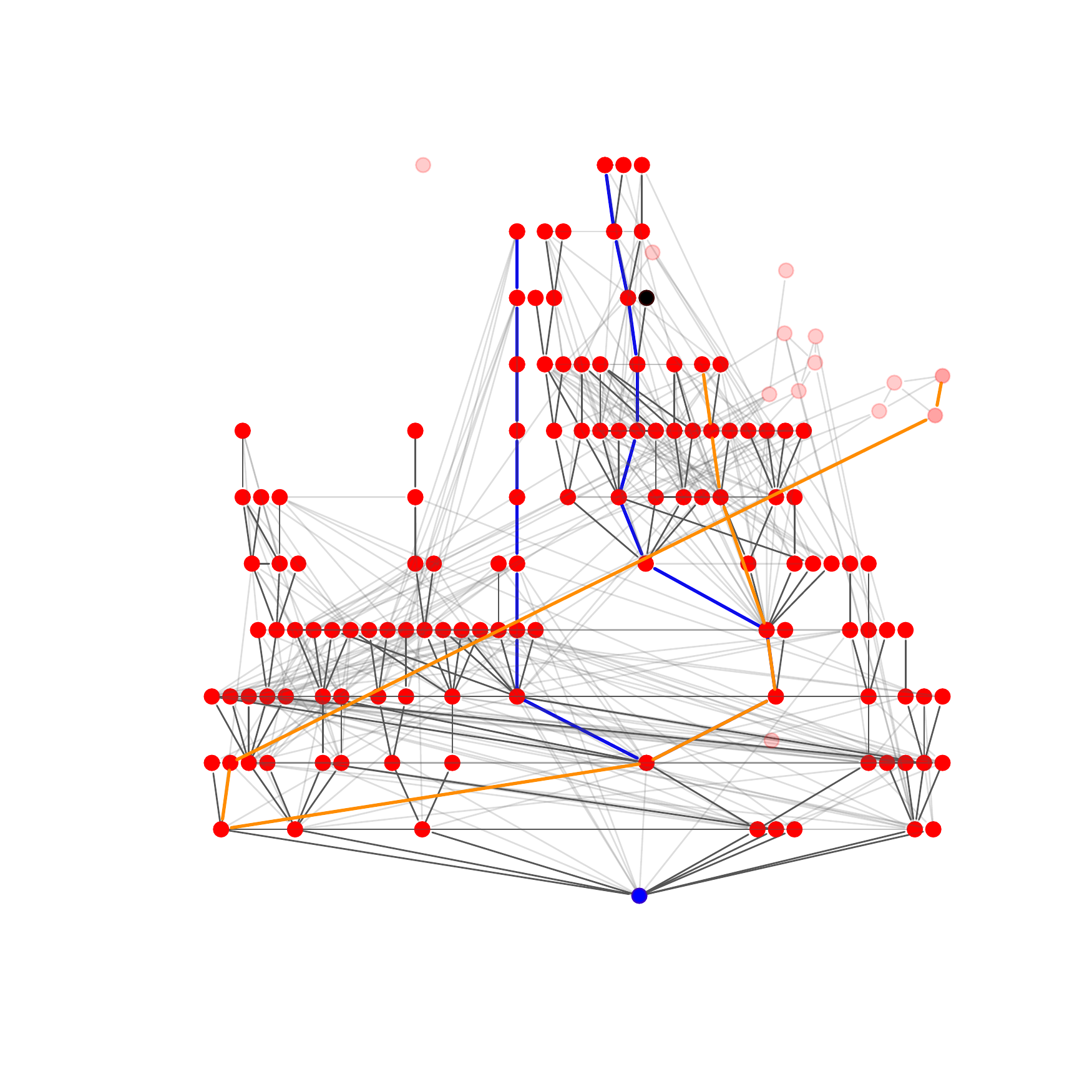}
\caption{(Color online) Partially aggregated networks on July 14$^{\rm
    th}$ at the SG museum, for two different choices of the seed (blue
  node at the bottom). Transparent nodes and light gray edges
  represent individuals not infected and contacts not spreading the
  infection, respectively. Red nodes and dark gray links represent
  infected individuals and contacts spreading the infection,
  respectively. The diameter of the transmission network and of the
  partially aggregated networks are shown respectively with blue and
  orange links. The black node represents the last infected
  individual.  }\label{infection-network}
\end{figure}

Figure~\ref{infection-network} shows two partially aggregated networks
for July 14$^{\rm th}$ at the SG museum, for two different choices of
the seed (blue node), and the corresponding transmission networks.
The transmission network is of course a subnetwork of the partially
aggregated network: not all individuals entering the premises after
the seed can be reached from the seed by a {\em causal} path, and
not all links are used for transmission events. 
In order to emphasize the branching nature of infection spreading, we
represent the transmission network with successively infected nodes 
arranged from the bottom to the top of the figure.
We notice that the diameter of both the transmission and the partially
aggregated network may not include the seed and/or the last infected
individual.

\begin{figure}
\includegraphics[width=0.45\columnwidth, height=0.45\columnwidth]{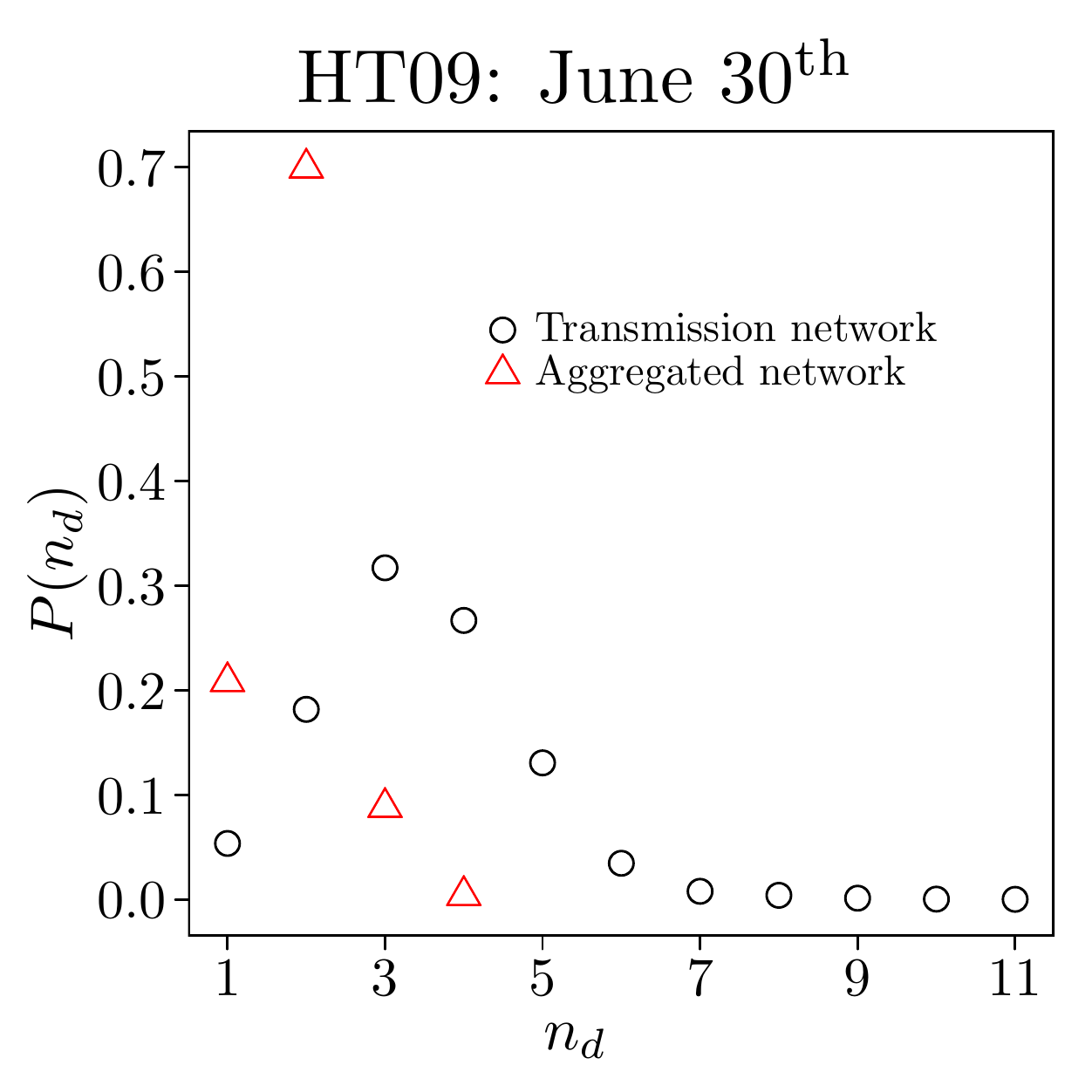}
\includegraphics[width=0.45\columnwidth,height=0.45\columnwidth]{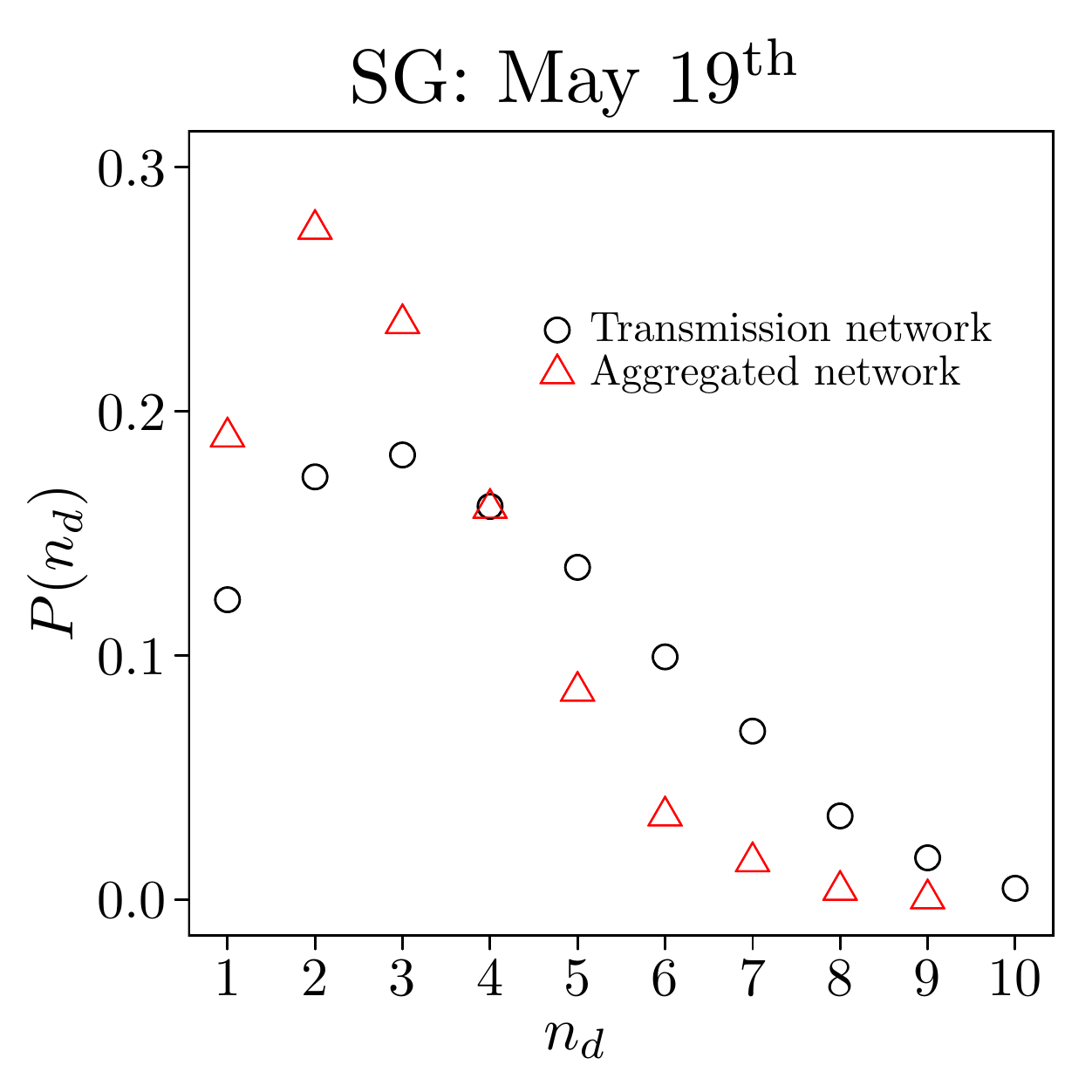}
\includegraphics[width=0.45\columnwidth,height=0.45\columnwidth]{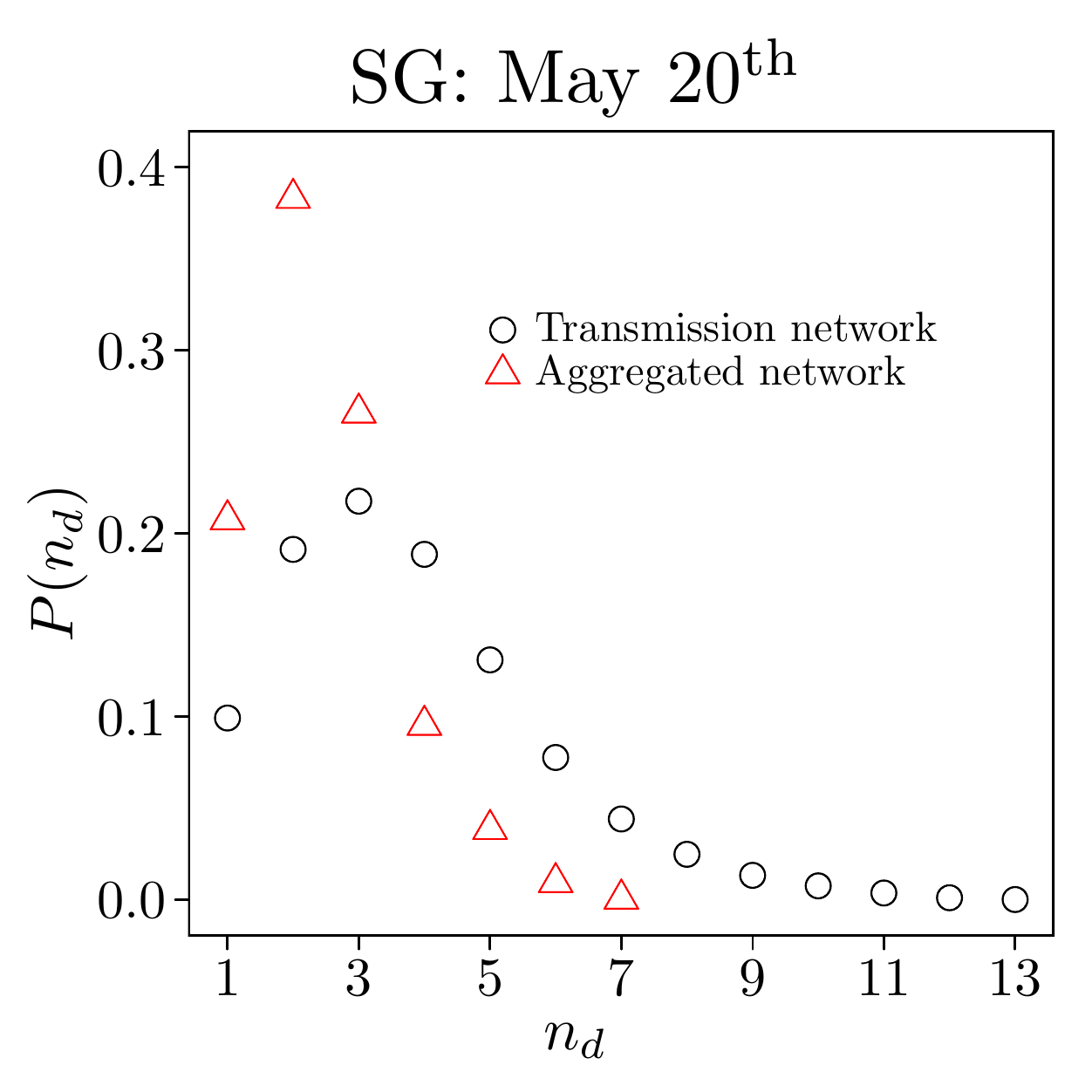}
\includegraphics[width=0.45\columnwidth,height=0.45\columnwidth]{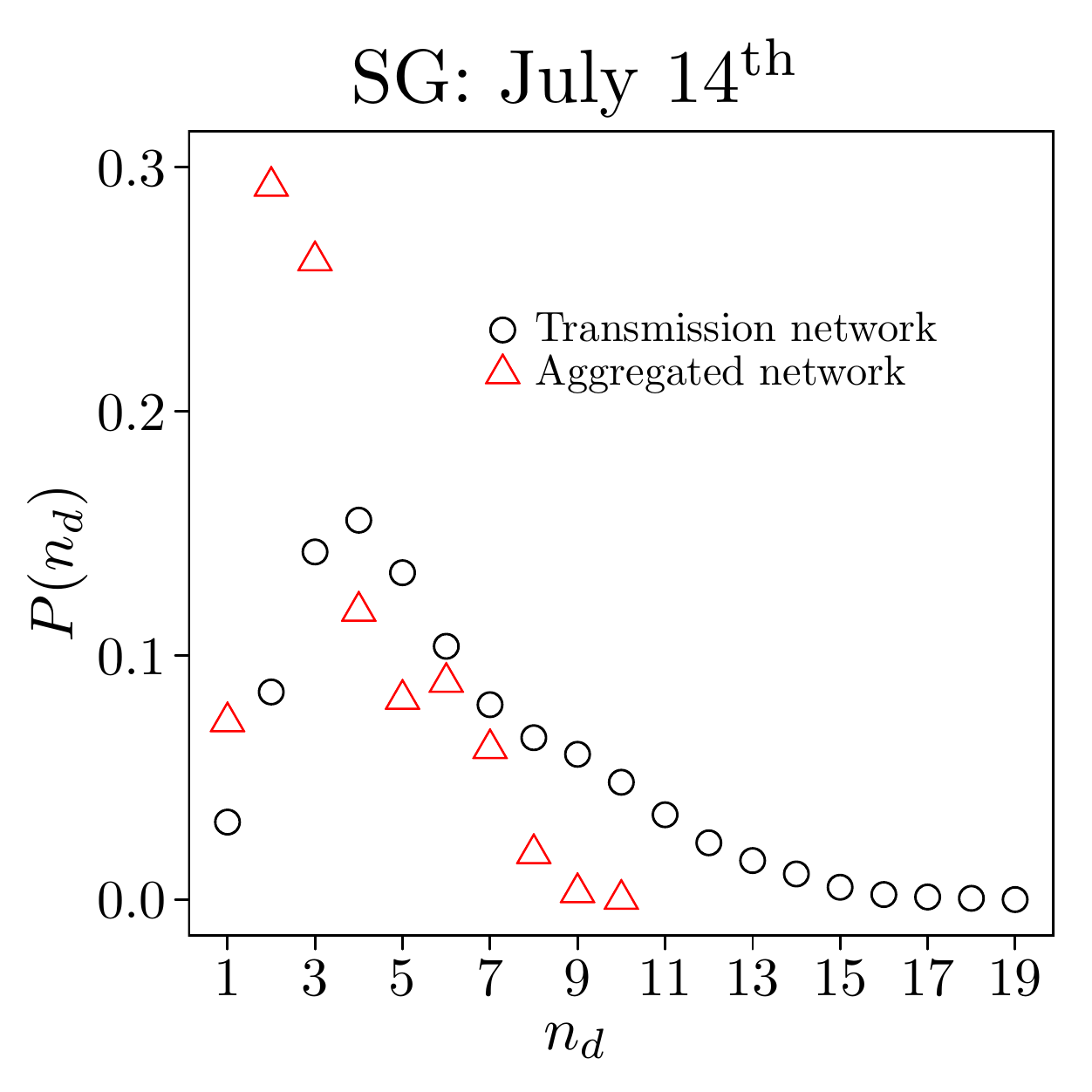}
\caption{(Color online) Distribution of the path lengths $n_{d}$ from the seed to
  all the infected individuals calculated over the transmission network
  (circles) and the partially aggregated networks (triangles).
The distributions are computed, for each day, by varying the choice
of the seed over all individuals.
}\label{path-length-distr}
\end{figure}

The presence of a few triangles in the transmission network is due to the
finite time resolution of the measurements. Let us consider, for
instance, the case of an infected visitor $A$ who infects $B$,
followed by a simultaneous contact of $A$ and $B$
with the susceptible $C$. In this case it is
impossible to attribute the infection of $C$ to either $A$ or $B$,
and both the $C\leftrightarrow A$ and the $C\leftrightarrow B$ links
are highlighted in the transmission network as admissible transmission events.
As a consequence, we slightly overestimate the number of links in the
transmission network of Fig.~\ref{infection-network}. In the case of 
Fig.~\ref{infection-network}, the number of links is between $1$ and $8\%$
larger than for a tree with the same number of nodes.
At finer time resolutions, some of the diffusion paths
of Fig.~\ref{infection-network} would actually be forbidden by causality.

A general feature exemplified by
Fig.~\ref{infection-network} is that the diameter of the transmission
network (blue path) is longer than the diameter of the partially aggregated
network (orange path), a first signature of the fact that the
fastest paths between two individuals, which are the ones followed
by the spreading process, do not coincide with the shortest path
over the partially aggregated network~\cite{kleinberg:2008}.
 
\begin{figure}
 \includegraphics[width=0.45\columnwidth, height=0.45\columnwidth]{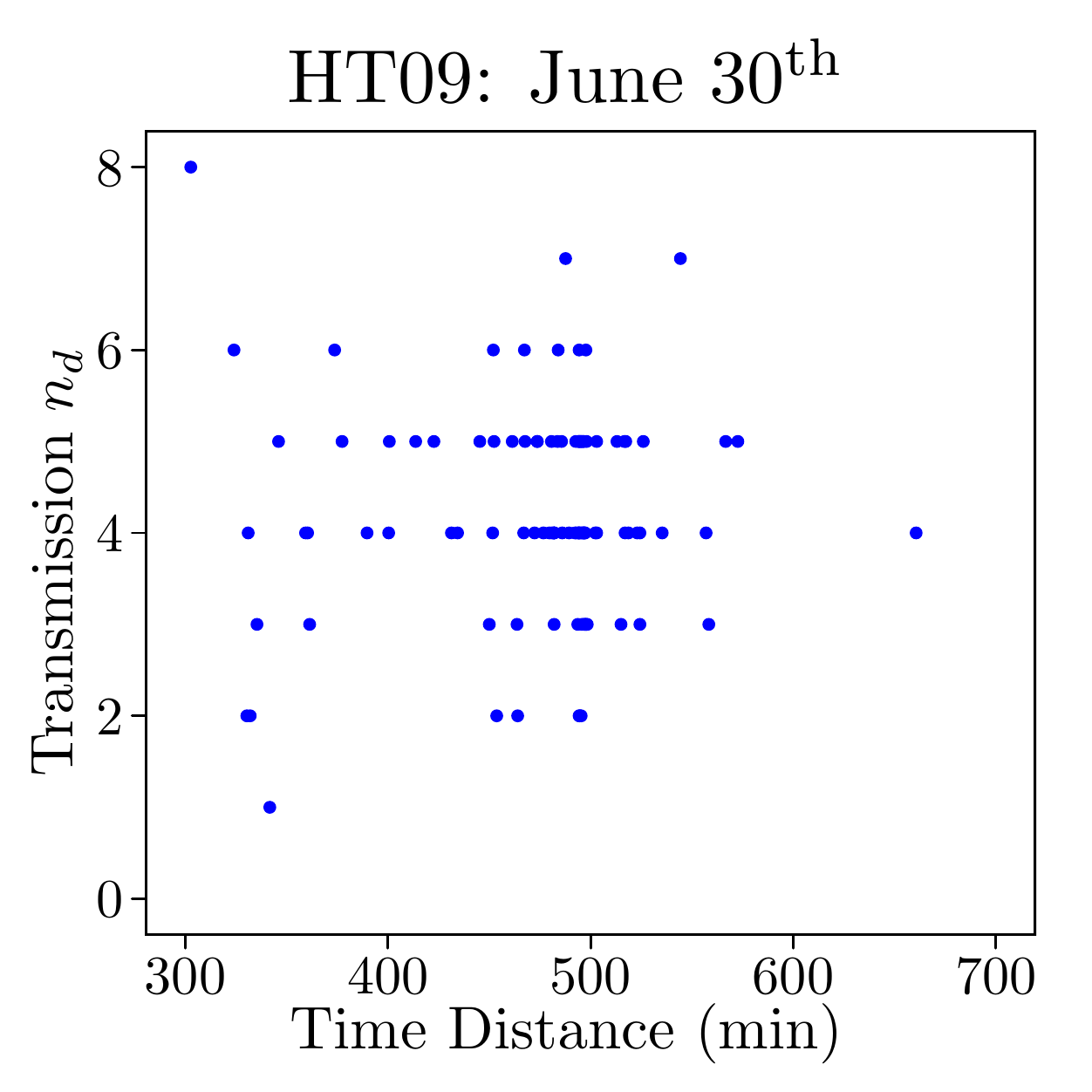}
\includegraphics[width=0.45\columnwidth,height=0.45\columnwidth]{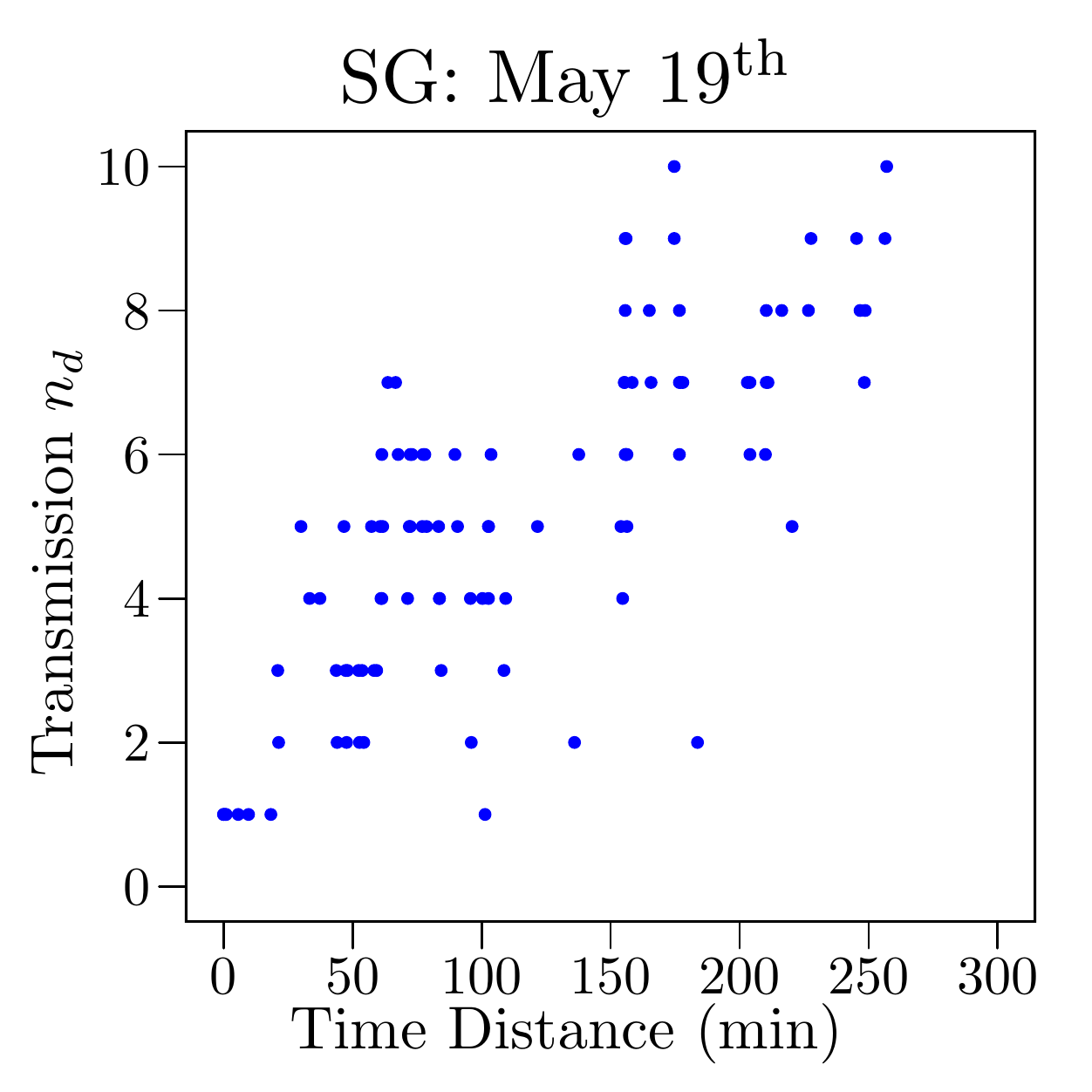}
\includegraphics[width=0.45\columnwidth,height=0.45\columnwidth]{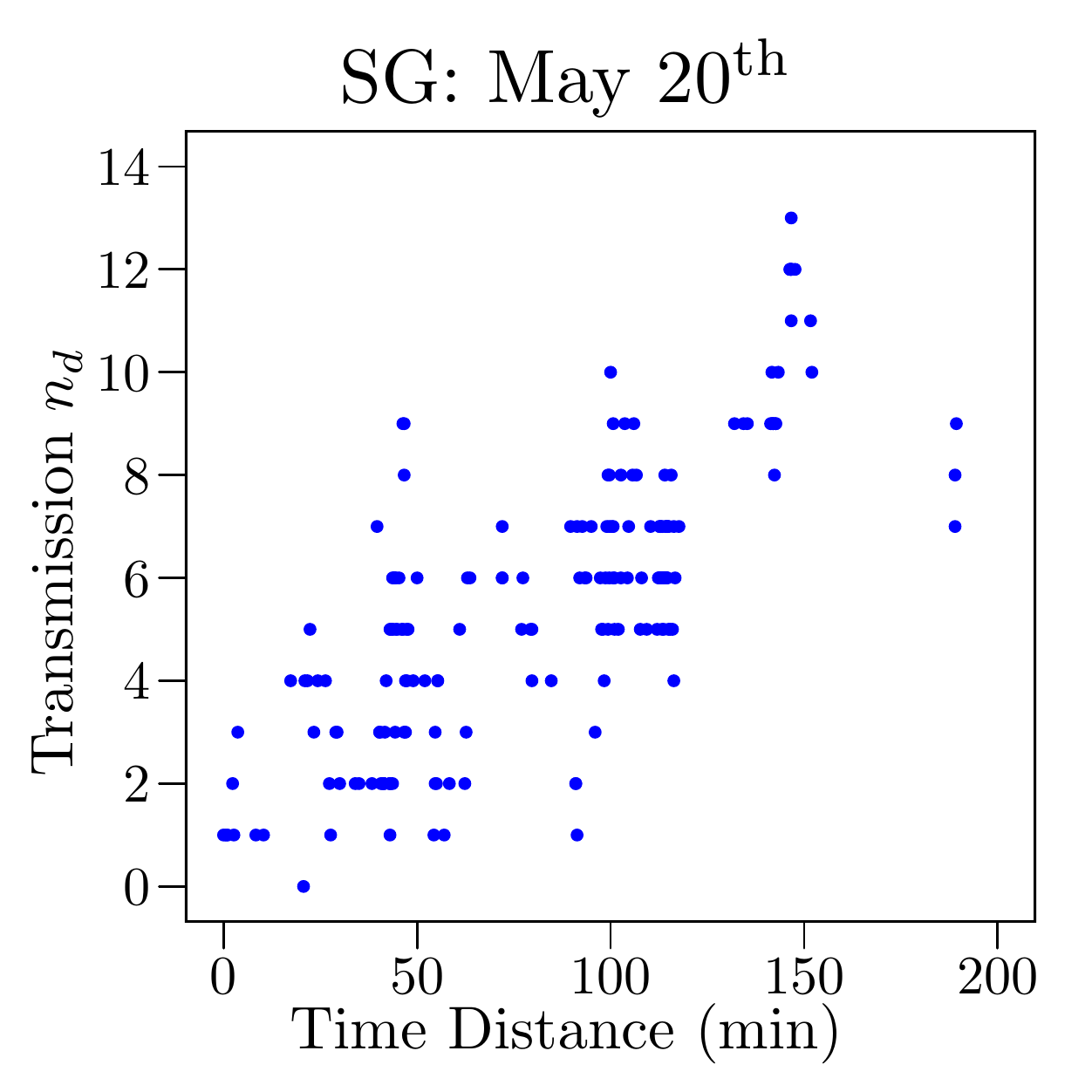}
\includegraphics[width=0.45\columnwidth,height=0.45\columnwidth]{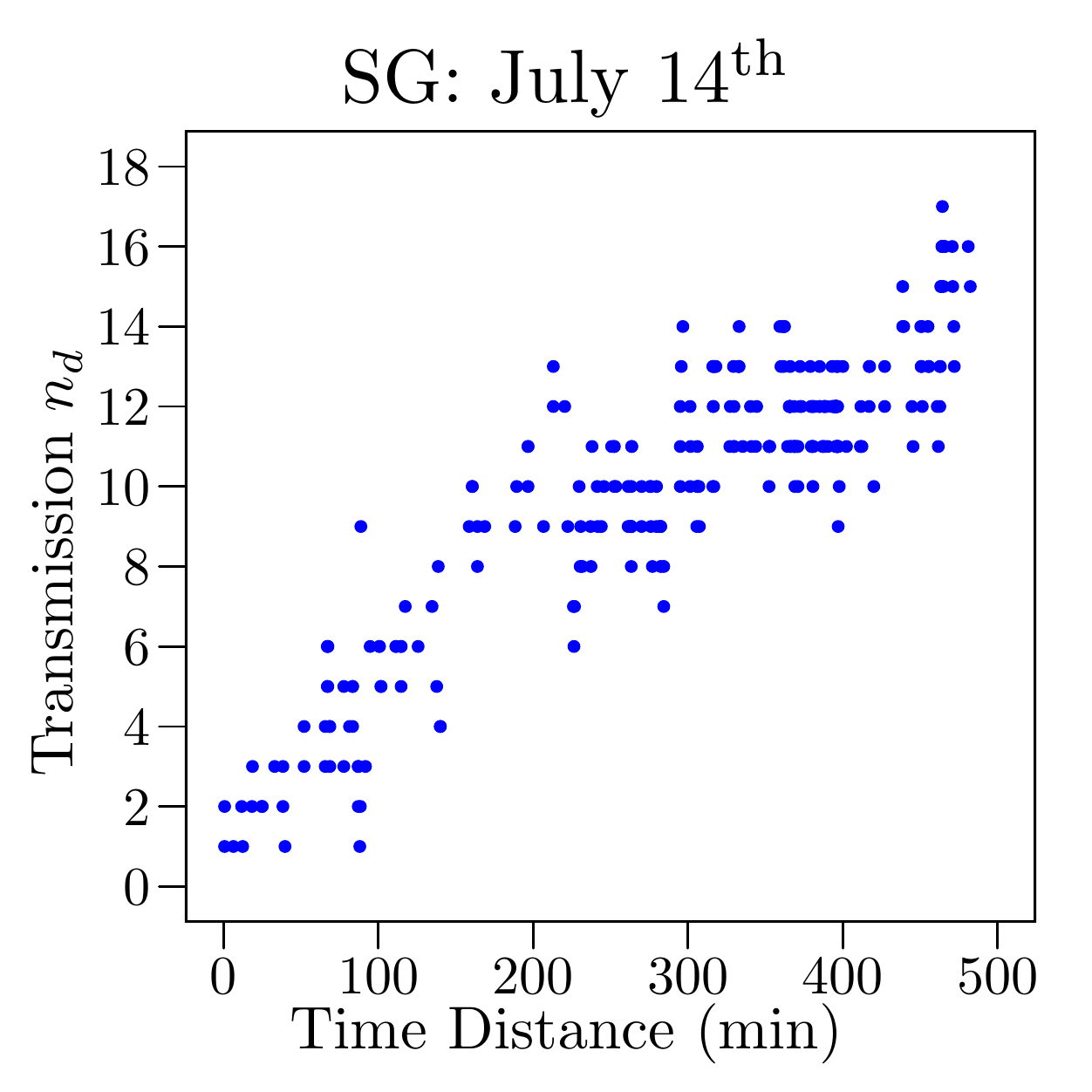}
\caption{Scatterplot of the seed-to-last-infected-individual distance
  (transmission $n_{d}$) along the transmission network,
  versus the total duration of the epidemics (time interval from the
  entry of the seed to the last infection
  event).}\label{time-length-distr}
\end{figure}

The difference between the fastest and the shortest paths for
a spreading process can be quantitatively investigated.
Figure~\ref{path-length-distr} reports the distribution of the network
distances $n_{d}$ between the seed and every other infected individual
along both the transmission networks and the aggregated networks.
When calculated on the partially aggregated network,
$n_{d}$ measures the length of the \textit{shortest}
seed-to-infected-individual path, whereas it yields the length
of the \textit{fastest} seed-to-infected-individual path
when calculated on the transmission network.
We observe that the length distribution of fastest paths,
i.e., the $P(n_{d})$ distribution for the transmission network,
always turns out to be broader and shifted toward higher values
of $n_{d}$ than the corresponding shortest path distribution,
i.e., $P(n_{d})$ for the partially aggregated network.
The difference is particularly noticeable in the
case of May $20^{\rm th}$ and July $14^{\rm th}$ for the SG deployment,
and June $30^{\rm th}$ for the HT09 conference, where the longest paths on the
transmission network are about twice as long as the longest paths
along the partially aggregated network.

These results clearly underline that in order to understand
realistic dynamical processes on contact networks,
information about the time ordering of
the contact events turns out to be essential: the
information carried by the aggregated network
may lead to erroneous conclusions on the spreading paths.

It is also possible to study the length of the path connecting the first (seed) to the last infected individual along the transmission network.
We measure the fastest seed-to-last-infected-individual path
(a quantity hereafter called ``transmission $n_{d}$'')
as a function of the duration of the spreading process,
defined as the time between the
entry of the seed and the last transmission event.
As shown by Fig.~\ref{time-length-distr}, a clear correlation
is observed between the transmission $n_{d}$
and the duration of the spreading process for the SG case
(Pearson coefficients $0.76$ for May $20^{\rm th}$ and May $19^{\rm th}$,
and $0.9$ for July $14^{\rm th}$). No significant correlation is instead
observed for the HT09 conference. This highlights the importance of the
longitudinal dimension in the SG data, and gives a first
indication of the strong differences in the spreading patterns,
that we further explore in the following.

 \begin{figure}
 \includegraphics[width=0.45\columnwidth, height=0.45\columnwidth]{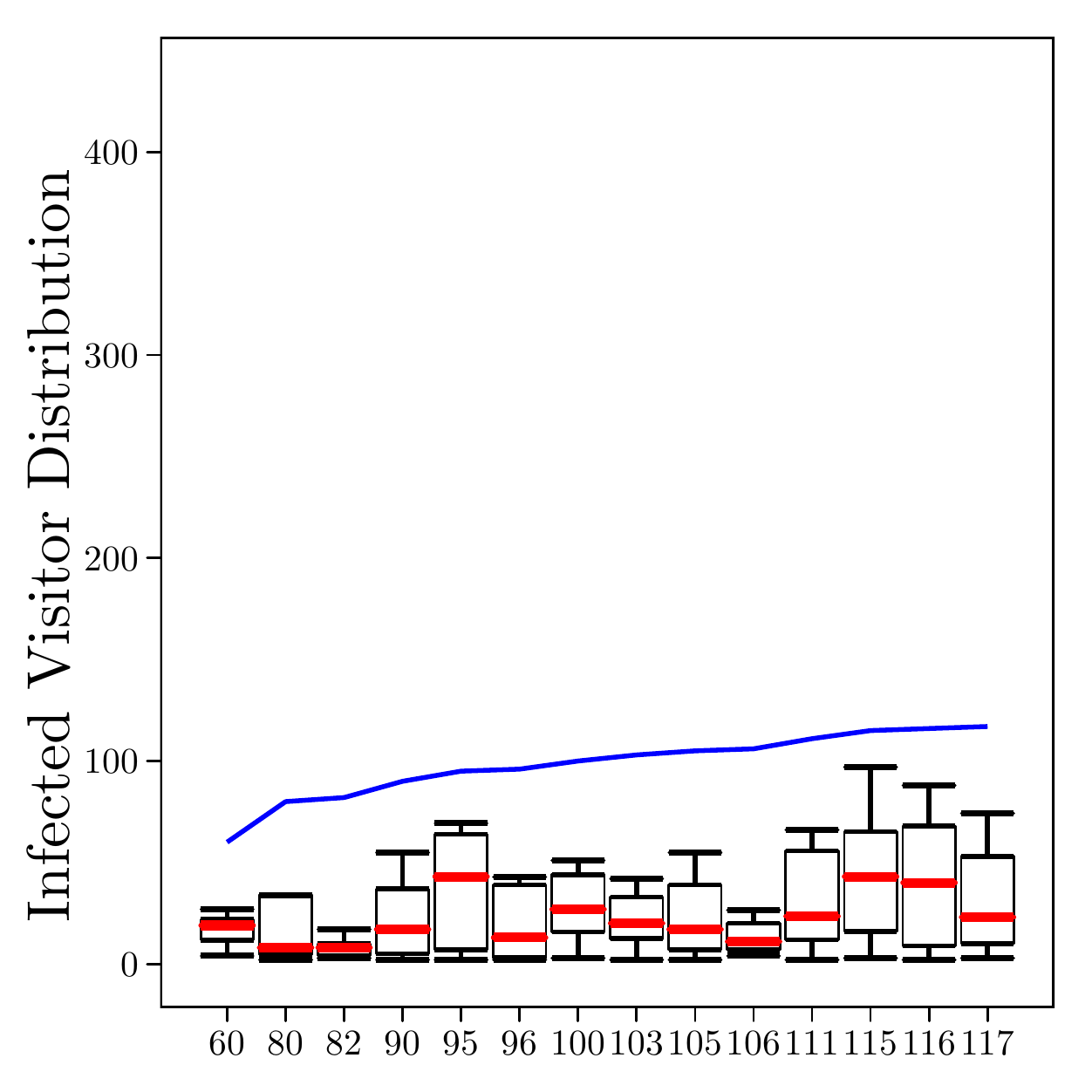}
\includegraphics[width=0.45\columnwidth,height=0.45\columnwidth]{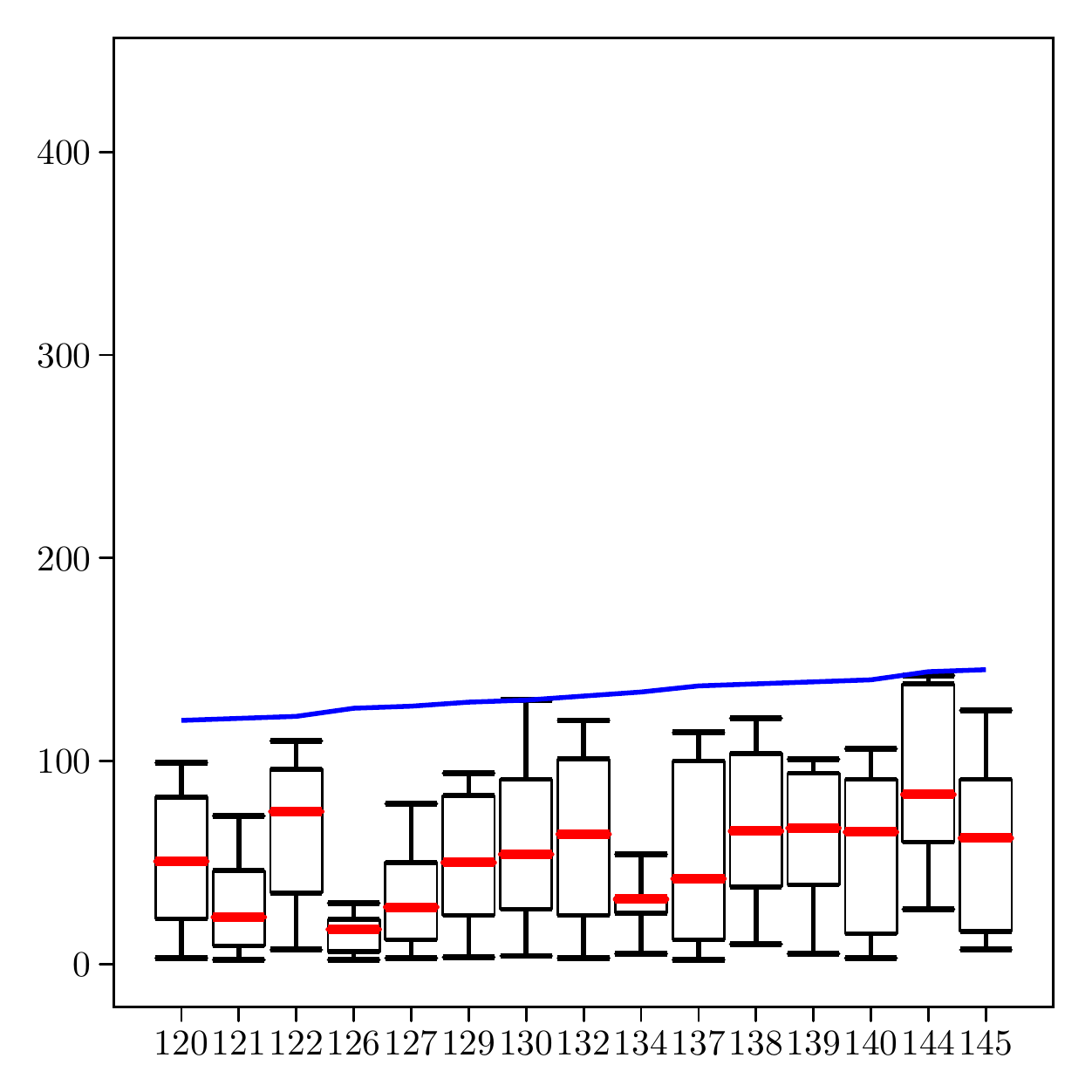}
\includegraphics[width=0.45\columnwidth,height=0.45\columnwidth]{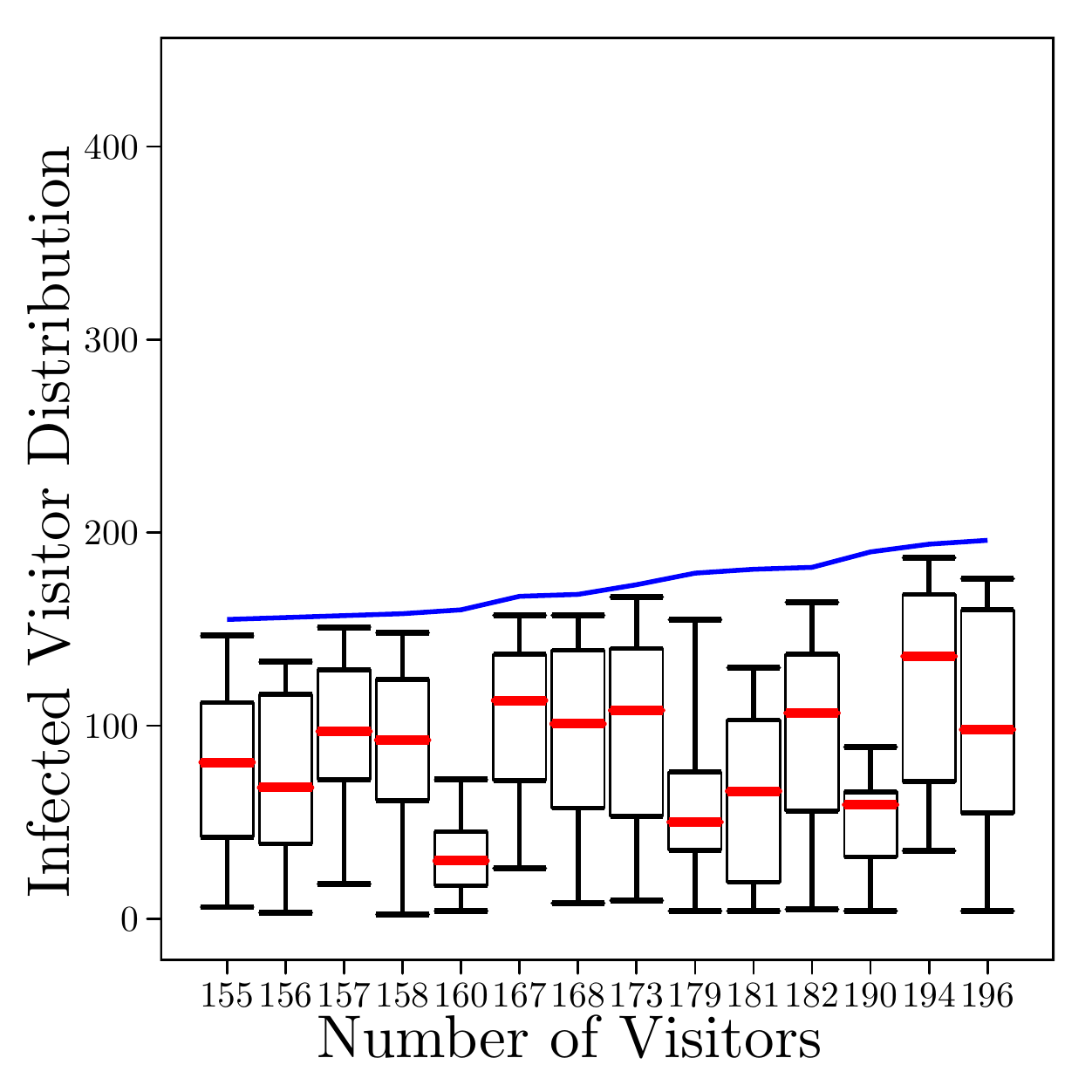}
\includegraphics[width=0.45\columnwidth,height=0.45\columnwidth]{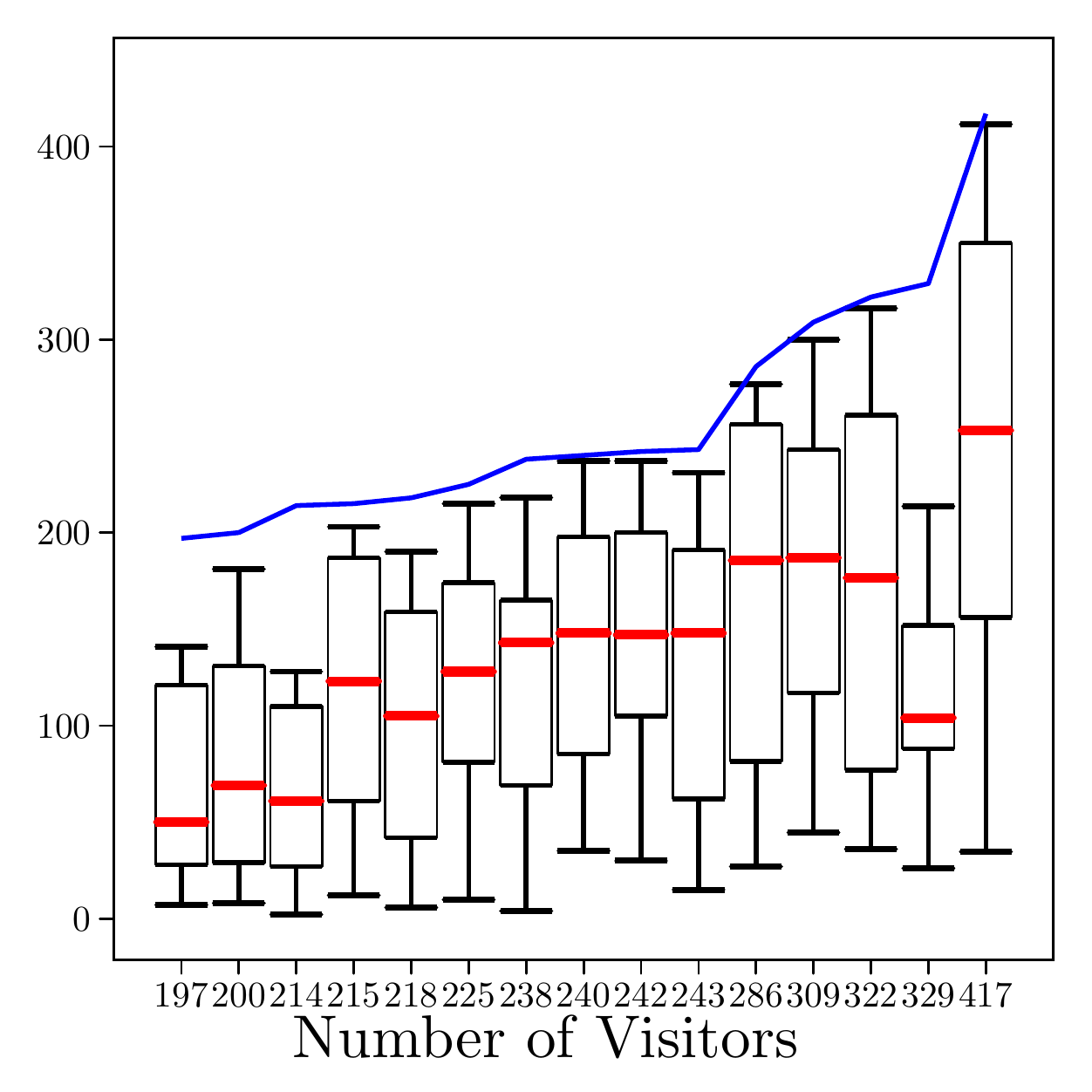}
\caption{(Color online) Results of spreading dynamics in the SG data: the figure
  shows panel boxplots of the final number of infected individuals in
  one day, versus the number of visitors in that day. The blue line
  represents the total number of daily visitors, giving an upper bound
  for the number of infected. The bottom and top of the rectangular
  boxes correspond to the $25^{th}$ and $75^{th}$ quantile of the
  distribution of infected individuals at the end of each day, and the
  red lines correspond to the median ($50^{th}$ quantile). The
  $5^{th}$ and $95^{th}$ are also shown (black horizontal lines).
}\label{boxplot-number-infected}
 \end{figure}

Let us now consider some other quantitative properties of the
spreading process, in particular the number of individuals reached by
the infection/information at the end of one day. In the SG case
Fig.~\ref{boxplot-number-infected} shows the distributions for each day, as boxplots, displaying the median together with the $5^{th}$, $25^{th}$, $75^{th}$ and $95^{th}$ percentiles. Days are arranged horizontally
from left to right, in increasing number of visitors.
A high degree of heterogeneity is visible.
The blue line corresponds to the number of daily visitors, that is
the maximum number of individuals who can potentially be infected.
We observe that the number of infected individuals is usually well
below this limit. The number of reached individuals also depends
on the number of CC in the aggregated network,
as the spreading process cannot propagate from one CC to another.
In fact, the limit for which all visitors are infected can be reached 
only if the aggregated network is globally connected, that occurs
only when the global number of visitors is large enough.
These results hint at the high intrinsic variability of the final outcome
of an epidemic-like process in a situation where individuals stream
through a building. A totally different picture emerges for
the HT09 conference, where the infection is almost always able to
reach all the participants. 

\begin{figure}
\includegraphics[width=0.45\columnwidth, height=0.45\columnwidth]{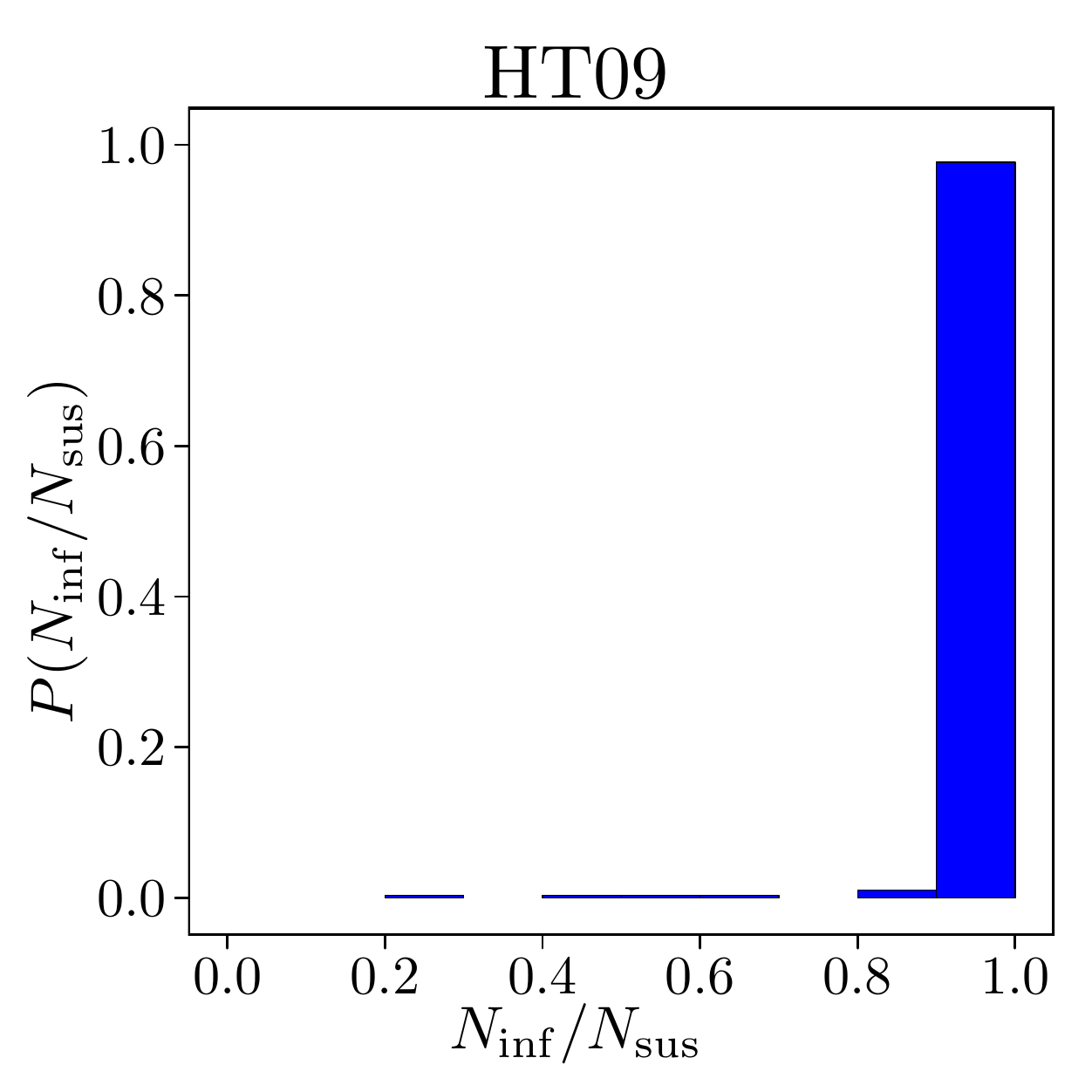}
 \includegraphics[width=0.45\columnwidth,height=0.45\columnwidth]{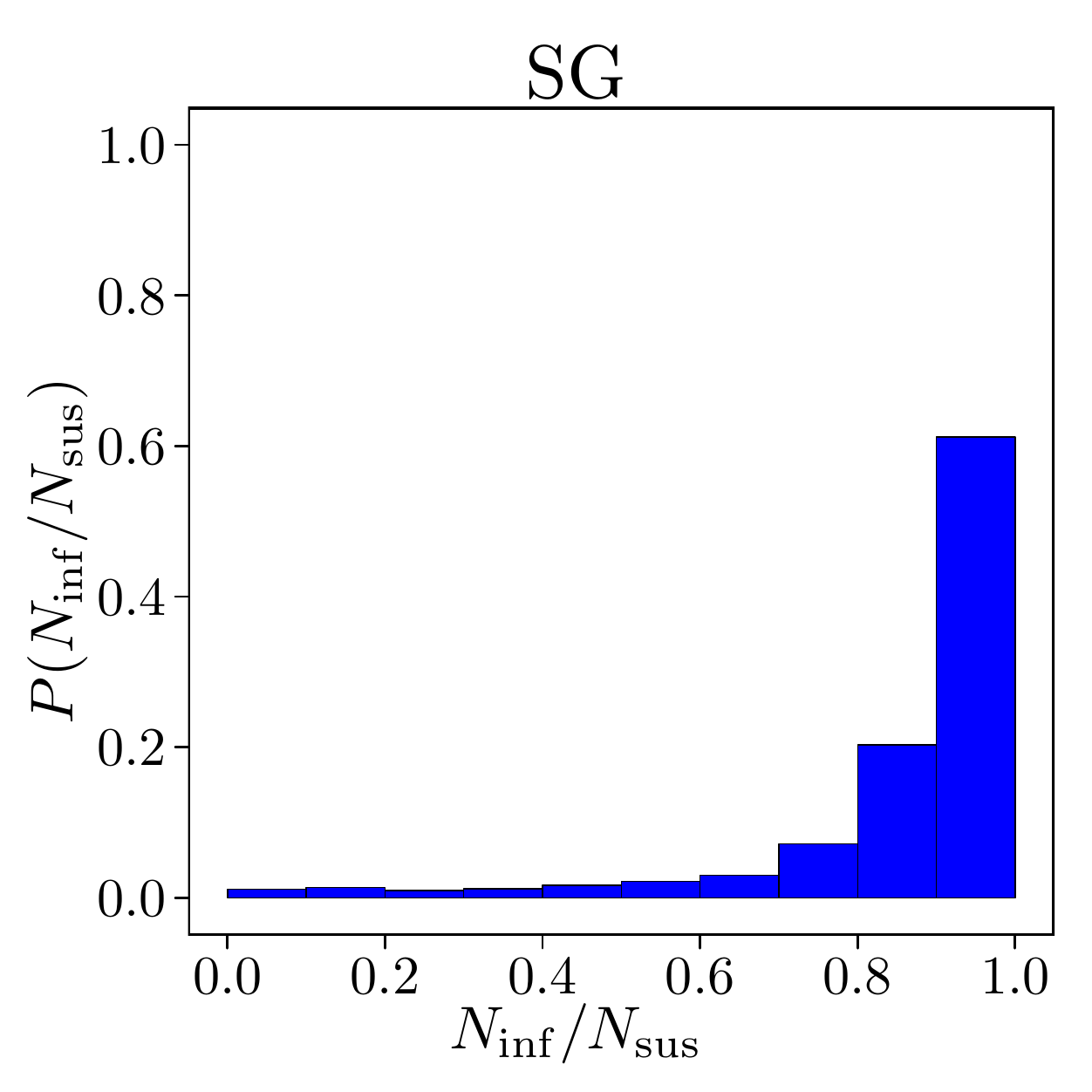}
\caption{Distribution of the ratio $N_{\rm inf}/N_{\rm sus}$ for the
  HT09 (left) and the SG (right) data, averaged over all potential
  seeds. $N_{\rm inf}$ is the final number of infected individuals at
  the end of one day, while $N_{\rm sus}$ is the number of individuals
  that could potentially be reached by a causal transmission path
  starting at the seed.  $N_{\rm sus}$ is given by the number of
  individuals visiting the premises in the same day, from the time the
  seed enters the premises, and belonging to the same CC as the seed.
}\label{ratio-ninf-nsus}
\end{figure}

As mentioned previously, the spreading process cannot reach individuals
who have left the venue before the seed enters, or the individuals
who belong to a CC different from that of the seed.
Therefore, we consider the ratio of the final number of infected individuals,
$N_{\rm inf}$ to the number $N_{\rm sus}$ of individuals who can be potentially reached through causal transmission paths starting at the seed.
The distributions of this ratio is reported in Fig.~\ref{ratio-ninf-nsus}.
We observe that in the case of HT09 (left) almost all
the potentially infected individuals will be infected by the end of
the day, whereas the distribution of $N_{\rm inf}/N_{\rm sus}$
is broader in the SG case (right).
We notice that a static network description would inevitably lead to all
individuals in the seed's CC catching the infection, a fact that
can be a severe (and misleading) approximation of reality.

\begin{figure}
\includegraphics[width=0.45\columnwidth, height=0.45\columnwidth]{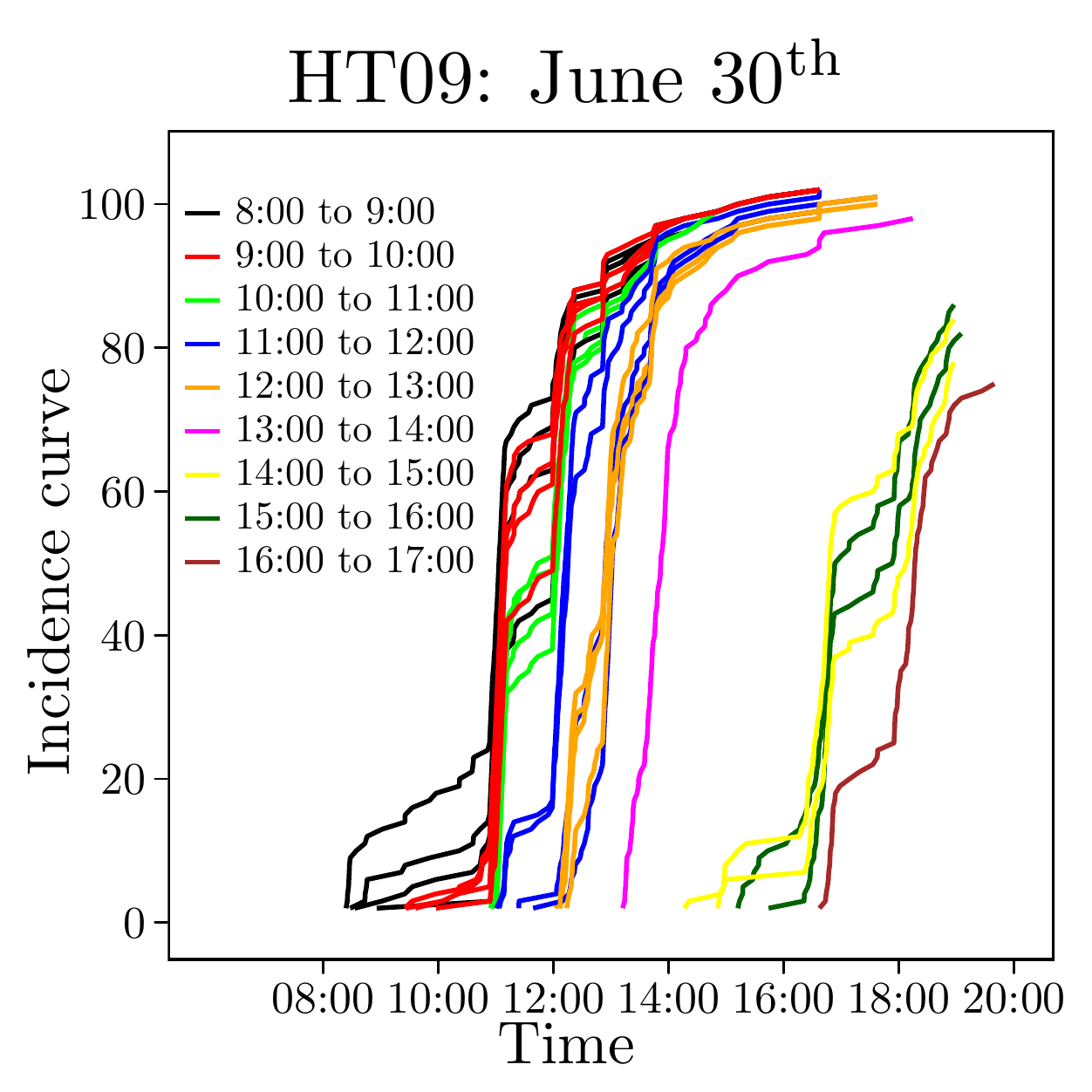}
\includegraphics[width=0.45\columnwidth,height=0.45\columnwidth]{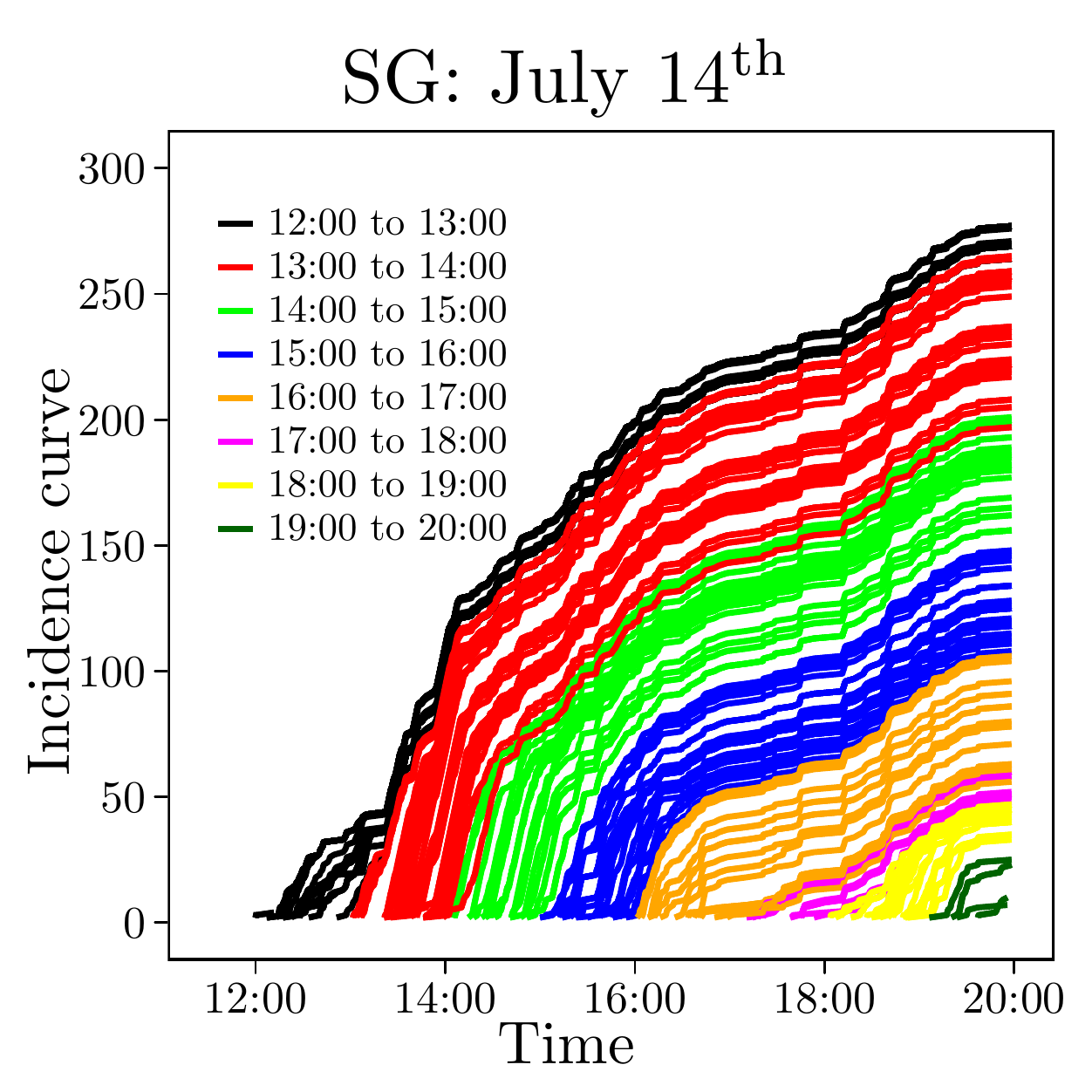}
\includegraphics[width=0.45\columnwidth,height=0.45\columnwidth]{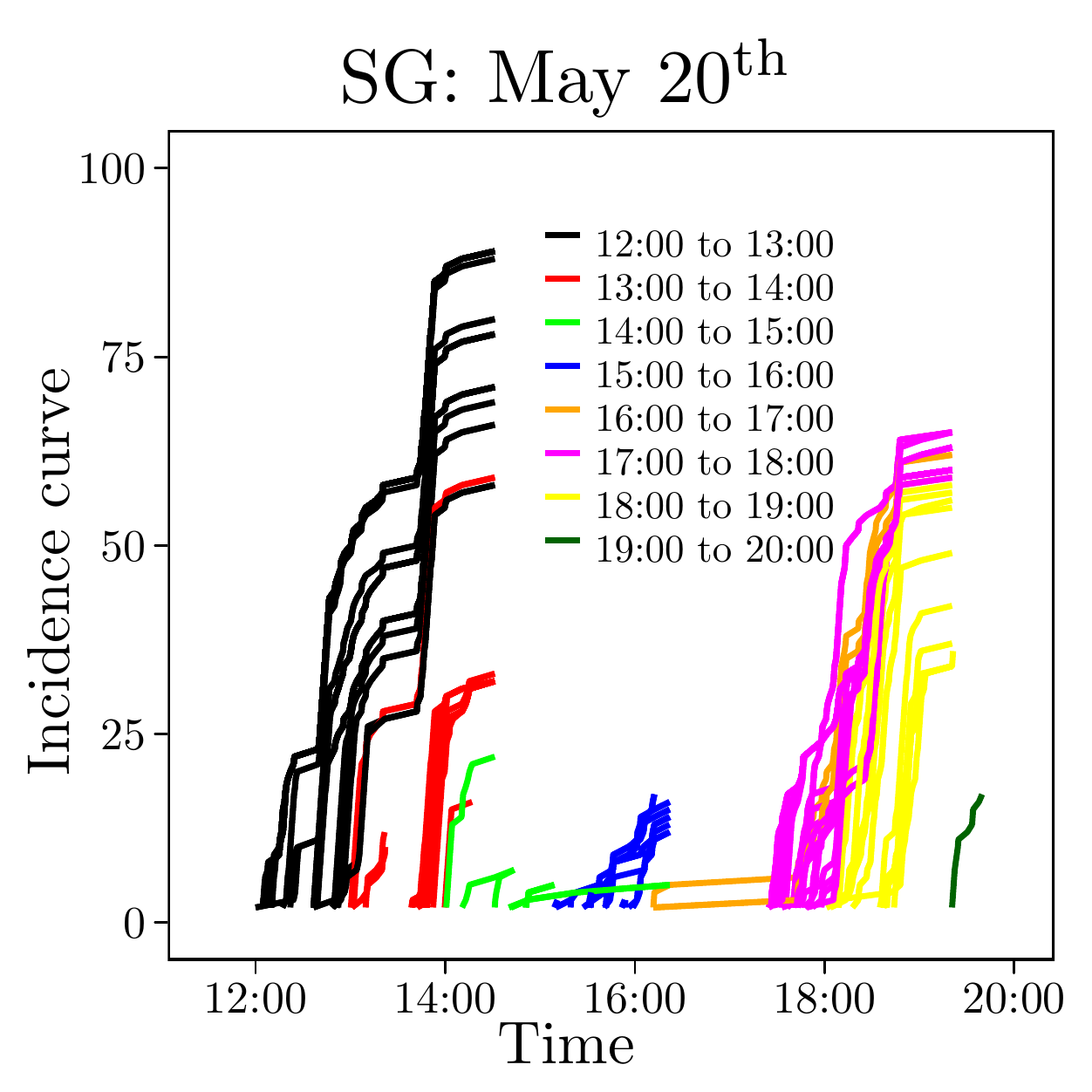}
\includegraphics[width=0.45\columnwidth,height=0.45\columnwidth]{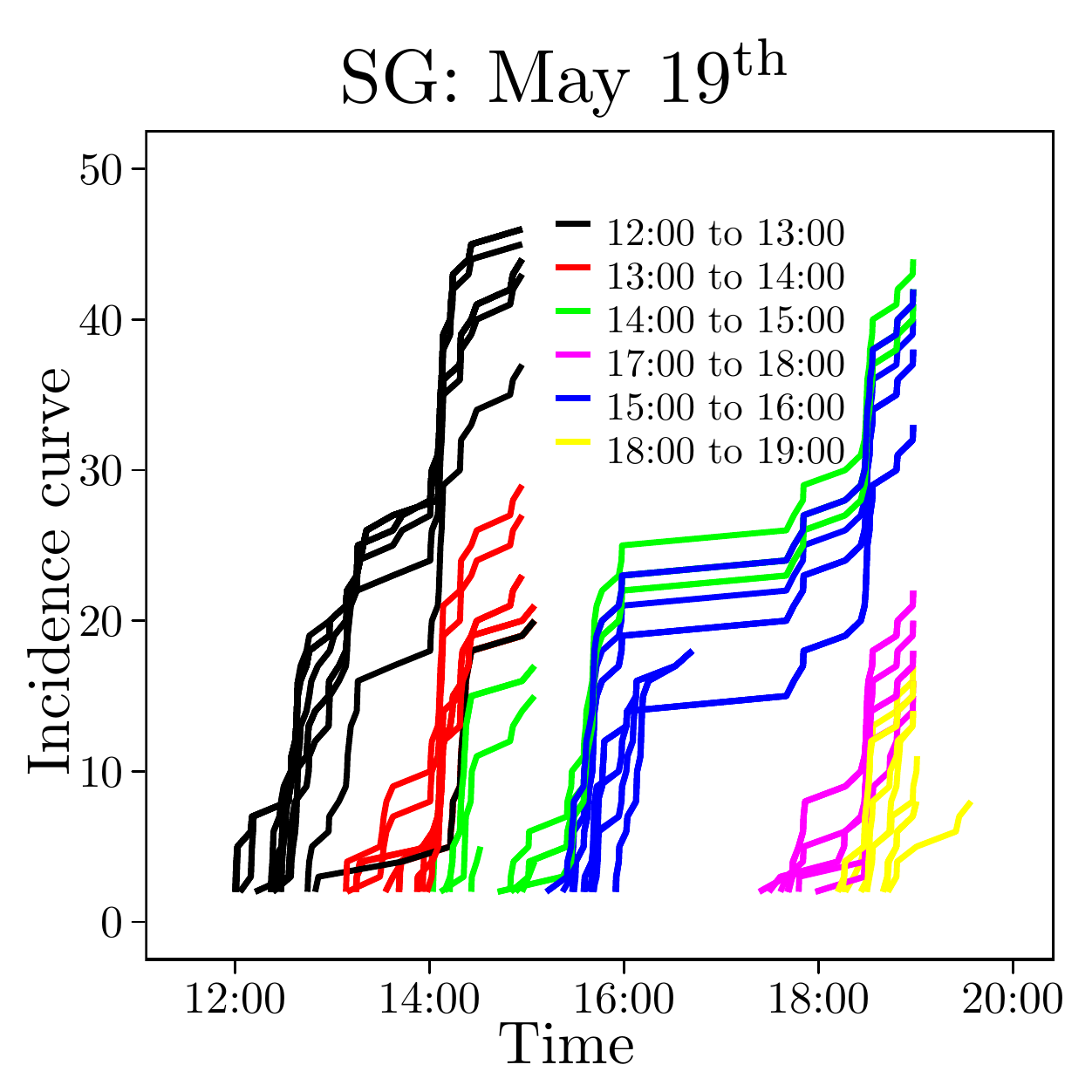}
\caption{(Color online) Incidence curves, giving the number of
  infected versus time for a spreading phenomenon simulated in the
  HT09 and SG data.  Clockwise from top: HT09, June 30$^{\rm th}$
  (aggregated network consisting of a single CC with $N_{1}=102$
  individuals); SG network for July 14$^{\rm th}$ (one CC,
  $N_{1}=282$ individuals), May 19$^{\rm th}$ (two CCs,
  $N_{1}= N_{2}=49$ individuals) and May 20$^{\rm th}$
  (two CCs, $N_{1}= N_{2}=89$ individuals). Each curve
  corresponds to a different seed, and is color-coded according to the
  starting time of the spreading.  }\label{epidemics-within-day}
\end{figure}

For each day the chosen seed generates a deterministic spreading
process for which we can compute the cumulative number of infected
individuals as a function of time, a quantity hereafter referred to as
an incidence curve. Figure~\ref{epidemics-within-day}
shows the results for a selected day of the
HT09 conference and for three different days of the SG data.

In the case of the HT09 conference, the earliest possible seeds
are the conference organizers, but little happens until conference participants gather for the coffee break and/or meet up
at the end of the first talk, between $10$:$00$ and $11$:$00$.
A strong increase in the number of
infected individuals is then observed, and a second strong increase occurs
during the lunch break. Due to the concentration in time of transmission
events, spreading processes reach very similar (and high) incidence
levels after a few hours, regardless of the initial seed or its arriving time.
Even processes started after $15$:$00$ can reach about $80\%$ of
the conference participants. Thus, the crucial point for the spreading
process does not consist in knowing where and when the epidemic trajectory
has started, but whether the seed or any other subsequently infected
individual attend the coffee break or not.

A different picture is obtained in the SG case:
First, in order to reach almost all participants the epidemics must
spread on a globally connected network and start early (black curves
for July $14^{th}$ data). Even in such a favorable setting for spreading,
the incidence curves do not present sharp gradients, and later
epidemics are unable to infect a large fraction of daily visitors.
The incidence curves for May $19^{th}$ and $20^{th}$ of Fig.~\ref{epidemics-within-day} show that different scenarios
can also occur: due to the fragmented nature of the
network, the final fraction of infected individuals can fluctuate greatly,
and sharp increases of the incidence can be observed
when dense groups such as those visible in
Fig.~\ref{aggregated-networks} are reached.\\

%
%
\section{Conclusions}
\label{conclusions}

In this paper we have shown that the analysis of time-resolved
network data can unveil interesting
properties of behavioral networks of face-to-face interaction between
individuals. We considered data collected in two very different
settings, representative of two types of social gatherings: the HT09
conference is a ``closed'' systems in which a group of individuals
gathers and interacts in a repeated fashion, while the SG museum
deployment is an ``open'' environment with a flux of individuals
streaming through the premises. 

We took advantage of the accurate time-resolved nature
of our data sources to build dynamically evolving behavioral networks.
We analyzed aggregated networks, constructed by
aggregating the face-to-face interactions during time intervals of one day,
and provided a comparison of their properties in both
settings. We assessed the role of network dynamics on the outcome
of dynamical processes such as spreading processes of informations
or of an infectious agent.

Our analysis shows that the behavioral networks of individuals
in conferences and in a museum setting exhibit both 
similarities and important differences. The topologies of the aggregated
networks are widely different: the conference networks are rather
dense small-worlds, while the SG networks have a larger diameter and are
possibly made of several connected components --- they do not form
small-worlds, and their ``elongated'' shape can be put in relation
with the fact that individuals enter the premises at different times
and remain there only for a limited amount of time. The networks'
differences are also unveiled by a
percolation analysis, which reveals how the SG aggregated networks
can easily be dismantled by removing links that act as ``bridges''
between groups of individuals; on the contrary,
aggregated networks at a conference are more ``robust'',
even with respect to targeted link removal.

Interestingly, some important similarities are also observed:
the degree distributions of aggregated networks, for example,
are short-tailed in both cases. Moreover, despite the
higher social activity at a conference, both the distribution of the
contact event durations and the distribution of the total time spent
in face-to-face interactions by two individuals
are very similar.

The study of simple spreading processes unfolding on the dynamical
networks of interaction between individuals allowed us to delve
deeper into the time-resolved nature of our data. Comparison of
the spreading dynamics on the time-dependent networks
with the corresponding dynamics on the aggregated networks shows
that the latter easily yields erroneous conclusions.
In particular, our results highlight the strong impact of causality
in the structure of transmission chains, that can differ significantly
from those obtained on a static network. The temporal properties
of the contacts are crucial in determining the spreading patterns
and their properties. Studies about the role of the
initial seed and its properties on the spreading patterns,
or the determination of the most crucial nodes for propagation,
can be misleading if only the static aggregated network
is considered. In more realistic dynamics, the fastest path
is typically not the shortest path of the aggregated network,
and the role of causality is clearly visible in the analysis
of the seed-to-last-infected paths.

Spreading phenomena unfold in very different ways in the two settings
we investigated: at a conference, people interact repeatedly and with
bursts of activity, so that transmission events also occur in a bursty
fashion, and most individuals are reached at the end of the day; in a
streaming situation, instead, the fraction of reached individuals can
be very small due to either the lack of global connectivity or the
late start of the spreading process.  Detailed information on the
temporal ordering of contacts is therefore crucial. We also note that
in more realistic settings with non-deterministic spreading,
information about the duration of contacts, and not only their
temporal ordering, would also turn out to be very relevant
and lead to an interesting interplay between the contact timescale
and the propagation timescale~\cite{stehle:2010}. Future work will also
address the issue of sampling effects: the fact that not all 
the conference attendees participated to the data collection
may lead to an underestimation of spreading, since 
spreading paths between sampled attendees involving unobserved 
persons may have existed, but are not taken into account.

We close by stressing that as the data sources on person-to-person
interactions become richer and ever more pervasive,
the task of analyzing networks of interactions is unavoidably shifting
away from statics towards dynamics, and a pressing need is building up
for theoretical frameworks that can appropriately deal with streamed
graph data and large scales. At the same time, we have shown that
access to these data sources challenges a number of assumptions
and poses new questions on how well-known dynamical processes unfold
on dynamic graphs.

\section*{Acknowledgments}
Data collections of the scale reported in this manuscript are only possible
with the collaboration and support of many dedicated individuals.
We gratefully thank the Science Gallery in Dublin for inspiring ideas
and for hosting our deployment. Special thanks go to Michael John Gorman,
Don Pohlman, Lynn Scarff, Derek Williams, and all the staff members
and facilitators who helped to communicate the experiment
and engage the public.
We thank the organizers of the ACM Hypertext 2009 conference
and acknowledge the help of Ezio Borzani, Vittoria Colizza,
Daniela Paolotti, Corrado Gioannini and the staff members
of the ISI Foundation, as well as the help of several Hypertext 2009 volunteers.
We also thank Harith Alani, Martin Szomszor and Gianluca Correndo
of the Live Social Semantics team.
We warmly thank Bitmanufaktur and the OpenBeacon project,
and acknowledge technical support from Milosch Meriac and Brita Meriac.
We acknowledge stimulating discussions
with Alessandro Vespignani and Vittoria Colizza.
This study was partially supported by the FET-Open project
DYNANETS (grant no. 233847) funded by the European Commission.
Finally, we are grateful for the valuable feedback, the patience
and the support of the tens of thousands of volunteers
who participated in the deployments.




\end{document}